\documentclass[twocolumn]{svjour3}
\smartqed
\usepackage{amsmath,amssymb}
\usepackage{color}
\usepackage{graphicx}

\usepackage{tikz}
\usetikzlibrary{shapes.geometric, arrows}
\usetikzlibrary{calc}
\tikzstyle{start} = [ellipse, text centered, draw=black]
\tikzstyle{process} = [rectangle, text width=4.5cm, text centered, draw=black]
\tikzstyle{decision} = [diamond, draw=black]
\tikzstyle{arrow} = [thick,->,>=stealth]

\usepackage{rotating}

\usepackage{makecell}

\usepackage{structmech}
\usetikzlibrary{backgrounds}

\newcommand{\boldface}[1]{\mathbf{#1}}  % italic (slanted)

\newcommand{\bfd}{\mathbf{d}}
\newcommand{\bfe}{\mathbf{e}}
\newcommand{\bff}{\mathbf{f}}

\newcommand{\bfj}{\mathbf{j}}

\newcommand{\bfn}{\mathbf{n}}

\newcommand{\bfq}{\mathbf{q}}

\newcommand{\bft}{\mathbf{t}}
\newcommand{\bfu}{\mathbf{u}}

\newcommand{\bfx}{\mathbf{x}}
\newcommand{\bfy}{\mathbf{y}}

\newcommand{\bfA}{\mathbf{A}}
\newcommand{\bfB}{\mathbf{B}}
\newcommand{\bfC}{\mathbf{C}}
\newcommand{\bfD}{\mathbf{D}}
\newcommand{\bfE}{\mathbf{E}}
\newcommand{\bfF}{\mathbf{F}}

\newcommand{\bfH}{\mathbf{H}}
\newcommand{\bfI}{\mathbf{I}}
\newcommand{\bfJ}{\mathbf{J}}
\newcommand{\bfK}{\mathbf{K}}

\newcommand{\bfM}{\mathbf{M}}
\newcommand{\bfN}{\mathbf{N}}

\newcommand{\bfR}{\mathbf{R}}
\newcommand{\bfS}{\mathbf{S}}

\newcommand{\bfvarepsilon}{\boldsymbol{\varepsilon}}

\newcommand{\bfkappa}{\boldsymbol{\kappa}}

\newcommand{\bfmu}{\boldsymbol{\mu}}

\newcommand{\bfxi}{\boldsymbol{\xi}}

\newcommand{\bfsigma}{\boldsymbol{\sigma}}

\newcommand{\bfvarphi}{\boldsymbol{\varphi}}

\newcommand{\calC}{\mathcal{C}}

\newcommand{\dsC}{\mathbb{C}}

\newcommand{\bfzero}{\mbox{\boldmath $0$}}
\newcommand{\dd}{\;\!\mathrm{d}} 
\newcommand{\be}{\begin{equation}}
\newcommand{\ee}{\end{equation}}
\newcommand{\bea}{\begin{eqnarray}}
\newcommand{\eea}{\end{eqnarray}}
\newcommand{\bes}{\begin{equation*}}
\newcommand{\ees}{\end{equation*}}
\newcommand{\ba}{\begin{align}}
\newcommand{\ea}{\end{align}}
\newcommand{\tr}{\mathrm{tr}}
\begin{document}

\title{Multiscale modeling of cancellous bone considering full coupling of mechanical, electrical and magnetic effects} 

\author{Mischa Blaszczyk \and Klaus Hackl}

\institute{M. Blaszczyk and K.Hackl \at
	Institute of Mechanics of Materials, Ruhr-Universität Bochum, 44801 Bochum, Germany \\
	\email{mischa.blaszczyk@rub.de, klaus.hackl@rub.de}}

\maketitle
%Abstract
\begin{abstract}
Modeling of cancellous bone has important applications in the detection and treatment of fatigue fractures and diseases like osteoporosis. In this paper, we present a fully coupled multiscale approach considering mechanical, electrical and magnetic effects by using the multiscale finite element method and a two-phase material model on the microscale. We show numerical results for both scales, including calculations for a femur bone, comparing a healthy bone to ones affected by different stages of osteoporosis. Here, the magnetic field strength resulting from a small mechanical impact decreases drastically for later stages of the disease, confirming experimental research.
\end{abstract}
%%%%%%%%%%%%%%%%%%%%%%%% 
\section{Introduction} 
In the present contribution, we develop a multiscale model for cancellous bone taking mechanical, electrical and magnetic effects into account. An important application of this model is the early detection of osteoporosis. This bone disease reduces the mass density of the bone, making it thinner and weaker, increasing the likelihood of fractures. Sonography is used as a cheap, fast and non-invasive early detection technique for osteoporosis \cite{PMID:18599409}. Material modeling and numerical simulations are helpful tools in order to understand and evaluate experimental measurements and enable medical diagnostics based on this method. 

Bone is a composite material with impressive properties, drawing the interest of researchers of many different fields. As a material, it is very strong and stiff and has a high fracture toughness, while also maintaining a light weight \cite{Hamed2010}. Thus in recent decades, a lot of different approaches to investigate and simulate the material behavior of bone have appeared. Many analytical solutions are based on Biot's famous theory \cite{biot1,biot2}. Examples include \cite{Buchanan2006}, \cite{Chen2018} and \cite{Steeb2010}. Here, cortical bone is modeled as a solid, while bone marrow is assumed to be a fluid. The acoustic properties of bone material are then used to obtain mechanical material parameters of bone and the parameters of Biot's model. Additionally, the results are compared with the findings of experiments.

In contrast to the analytical solutions, many numerical approaches exist in the scope of bone modeling as well. The finite difference method was used in \cite{PMID:18599409} to obtain numerical results of ultrasound propagation in bone. Applications of the finite element method (FEM) on the topic of bone modeling include the simulation of mechanical properties of bone \cite{gardner2000,miller2002} and the simulation of osteogenic effects \cite{Wang2017}. In \cite{doi:10.1098/rsta.2010.0041} patient-specific FEM simulations are proposed in order to estimate the likelihood of osteoporotic fractures.

Since the bone microstructure is very complex and heterogenous, material modeling should take place on different scales. Currently used single scale models are  criticized in \cite{doi:10.1098/rsta.2010.0041} as oversimplified and multiscale approaches proposed instead. In \cite{Hamed2010}, the mechanical properties of bone are modeled on five different length scales from the nanoscale to the macroscale. Multiscale approaches can also be combined with numerical methods. The finite element square method ($\text{FE}^2$) extends the standard FEM approach by applying the multiscale concept and solving the differential equation systems on two scales via the FEM. An overview of the method can be found in \cite{schroeder2000,schroeder2013plasticity}. Basic works on this method include for example \cite{WILLIS19811,Suquet1987ElementsOH,CASTANEDA1997171} and applications to different materials can be found for example in \cite{https://doi.org/10.1002/pamm.200410144,MIEHE20022123}. An application of the $\text{FE}^2$ within the scope of bone modeling can be found in \cite{URAL2013141,PODSHIVALOV2011888,Pahr2008}, proposing different models to capture the microstructure of bone, allowing to investigate mechanical effects. In \cite{Ilic2010} and \cite{Klinge2013}, macroscopic material parameters were recovered by simulations on the microscale. The results obtained were subsequently used for macroscale simulations of wave propagation.

So far, all presented contributions focus only on the mechanical effects of bone. However, cortical bone possesses the properties of a piezoelectric solid. After the discovery of this effect \cite{yasuda1957,SHAMOS1963}, research considering these coupled physical effects has started.
A review on computer modeling of bone piezoelectricity can be found in \cite{MOHAMMADKHAH2019544}. There, applications are discussed as well. Since electric and magnetic effects are coupled physically via the Maxwell equations, it may be necessary to include magnetic effects as well. In \cite{GUZELSU198119}, bone was modeled as a hollow cylinder and analytical solutions of the coupled equations of all three effects were studied.

In this work, we present a fully coupled multiscale approach for modeling cancellous bone considering mechanical, electrical and magnetic effects and using two scales, the macro- and microscale. At the microscale, we assume a heterogenous material consisting of two phases, cortical bone and bone marrow. Cortical bone is modeled as piezoelectric, insulating solid, bone marrow as viscoelastic, conducting solid. Electrical and magnetic effects are coupled via the Maxwell equations. Based on energy methods in mechanics, we establish a thermodynamically consistent material model and derive the weak and strong form of the corresponding boundary value problem. We apply the FEM to solve the problem numerically. For multiscale analysis, we resort to the $\text{FE}^2$ method. To apply this method, we constructed a periodic representative volume element (RVE) and discuss the transition between scales.

The article is structured as follows: in Section 2 we discuss the material structure of cancellous bone and the $\text{FE}^2$ method. Then, we introduce the microscopic material model and derive the weak and strong form of the corresponding variational problem. Additionally, we cover the macroscale boundary value problem. In Section 3 we present the FEM implementation of the model and show details regarding scale transition and programming. In Section 4, we present numerical results, starting with microscale calculations, on to multiscale simulations for a cylindrical body and finally a true to scale model of a human femur bone. To close this article, we draw a short conclusion and give an outlook to future research envisioned in Section 5.

%%%%%%%%%%%%%%%%%%%%%%%%
\section{Material model}

\subsection{Structure and properties of cancellous bone}
Our work focuses on the description of the internal structure of cancellous (spongy) bone, which consists of small beams or shells of interconnected cortical bone and interstitial bone marrow. Cortical bone is mainly composed of elastic collagen fibers, which act as charge carriers. When applying a shear stress, these collagen fibers slip past each other, thus producing a piezoelectric effect. This was first measured in \cite{yasuda1957} and later validated in \cite{SHAMOS1963}. This means that, whenever a mechanical strain is present in the bone, an electric field is generated due to the piezoelectric effect. A time-dependent fluctuation of the electric field then creates a magnetic field due to Ampère's circuital law, coupling mechanical, electrical and magnetic effects all together. 

\begin{figure}[h]
	\centering
  \includegraphics[width=0.48\textwidth]{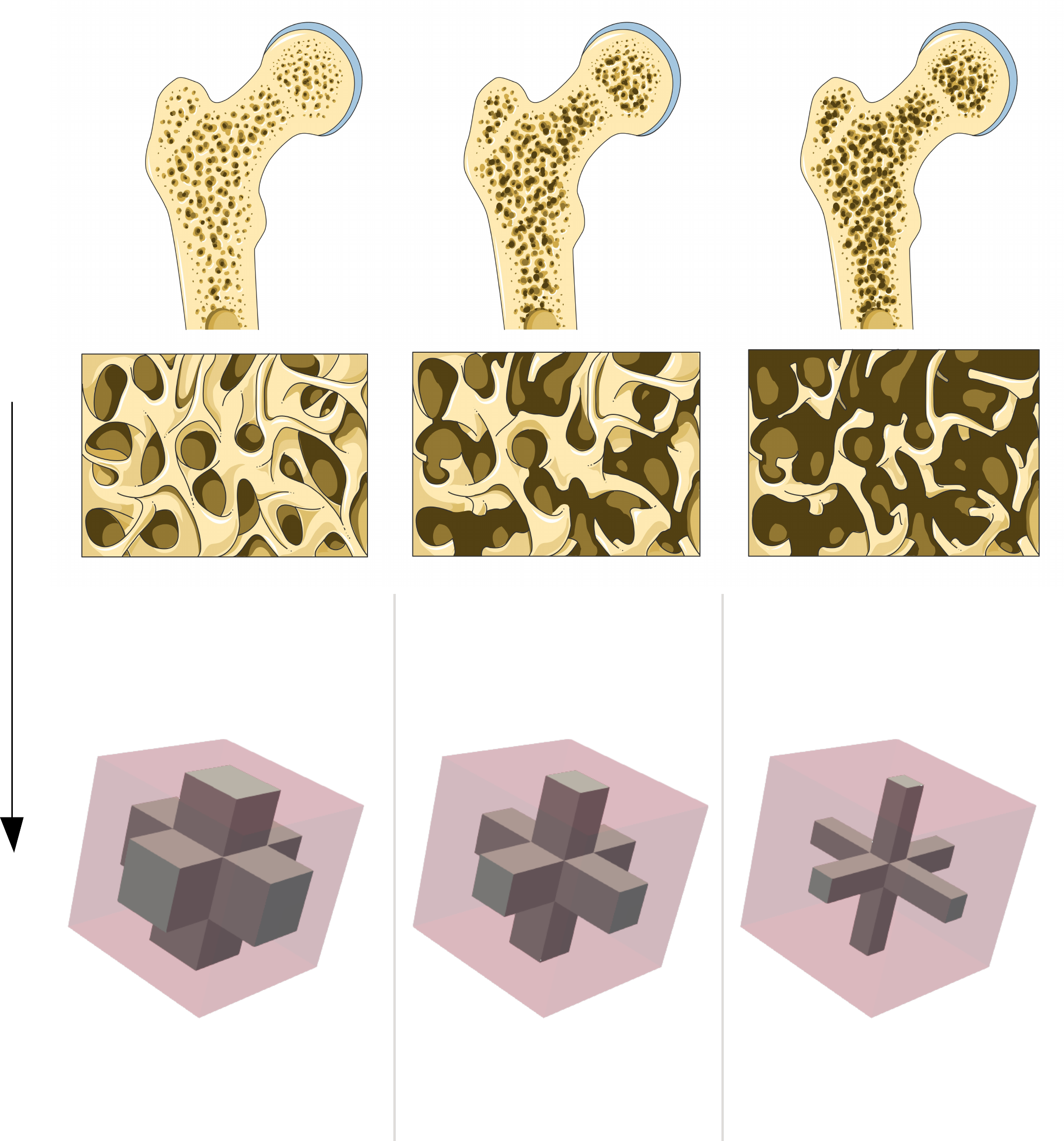}
	\caption{Bone phases depending on osteoporosis stage (cf. \cite{imageosteowikip}) and corresponding RVEs.}
	\label{fig1} 
\end{figure}

An important application of bone modeling is the early detection of osteoporosis, a bone disease, which manifests itself in the reduction of the cortical bone phase, thus reducing the strength of the bone and increasing the likelihood of fractures. Compared to a heal\-thy bone, the volume fraction of cortical bone for an affected bone can be reduced from $30\%$ to $5\%$ \cite{Steeb2010,Ilic2010}. Figure \ref{fig1} shows a comparison depending on the osteoporosis stage and illustrates the heterogeneity of the material. During the course of osteoporosis, the cortical bone (represented brighter) reduces and is replaced by bone marrow (represented in dark). Thus, we will employ different RVEs for the simulations. Here, the cortical bone phase is represented in gray, while the bone marrow phase is drawn in transparent red color.

Early detection of osteoporosis can be done via sono\-graphy: ultrasonic waves enter the bone and due to the described effects create a magnetic field, which can be measured \cite{GUZELSU198119} and - depending on the results - conclusions on the health status of the investigated bone can be drawn. In this contribution, we introduce a material model including all the described effects. It is important to note, that there are two different forms of coupling: while the piezoelectric coupling is captured via a suitable material model, the Maxwell coupling is of physical (electrodynamical) nature. 

%%%%%%%%%%%%%%%%%%%%%%%%%%
\subsection{Concept of the $\text{FE}^2$ method}
\begin{figure}[h]
	\centering
  \includegraphics[width=0.48\textwidth]{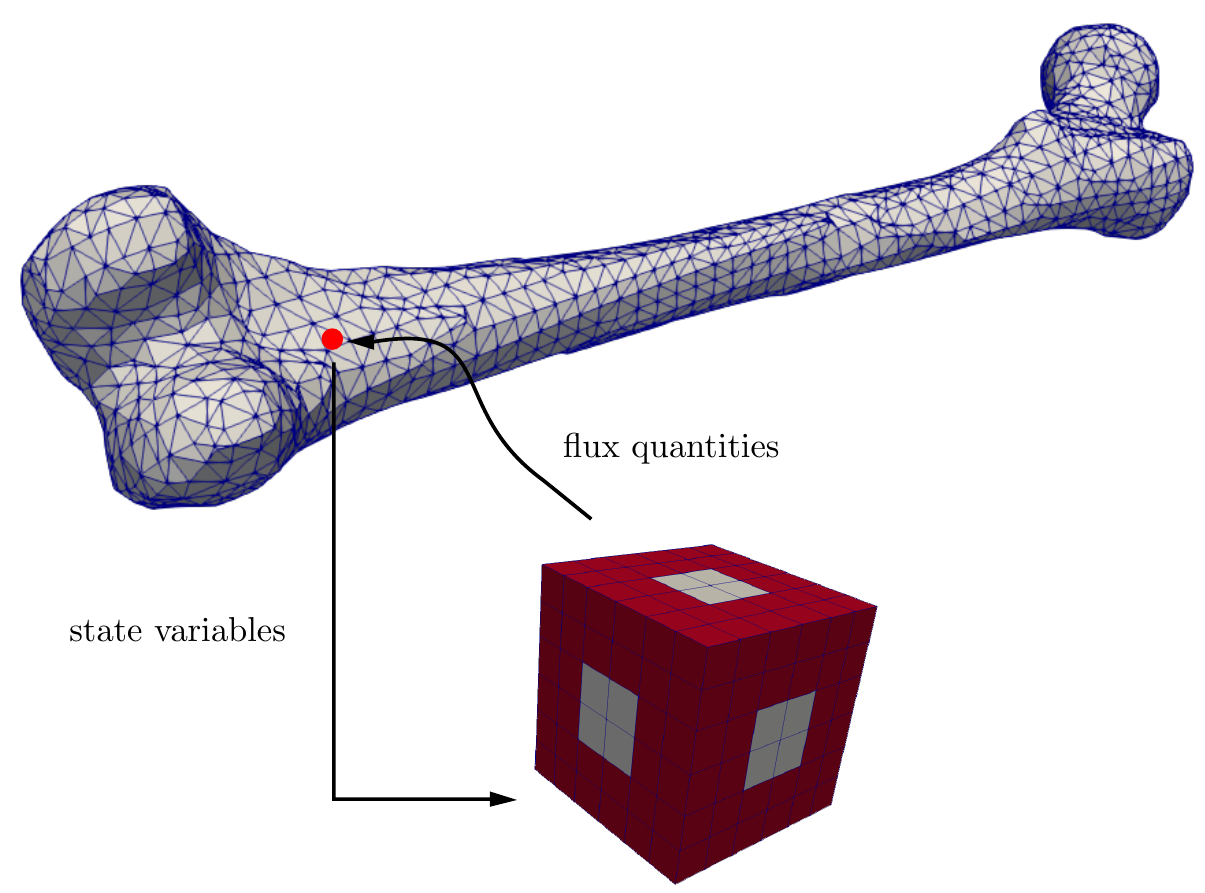}
	\caption{Transition between macro- and microscale. State variables enter as boundary conditions of the RVE problem. Flux quantities at the macroscale are calculated by averaging the RVE quantities.}
	\label{fig2} 
\end{figure}

To include micro heterogeneities directly, an extremely fine resolution of the problem would be necessary, resulting in a very high computation cost for the simulations. Alternatively, the $\text{FE}^2$ method is a homogenization technique, which captures the structure of micro heterogeneities by introducing a second - smaller - scale to the problem. If the material is statistically regular on the smaller scale, it can be modeled by a corresponding RVE \cite{schroeder2000,schroeder2013plasticity}. In this paper, we denote the larger scale as the macroscale and the smaller scale as the microscale. To obtain accurate results, the quotient of the characteristic lengths between micro- and macroscale should tend to zero, so the RVE has to be much smaller than the simulated macroscopic body. Figure \ref{fig2} illustrates this procedure: instead of using a material model on the macroscale, the state variables are linked to the microscale, where the RVE problem is solved. The microscale calculations yield average flux quantities and consistent tangent matrices, which then can be used for the solution of the macroscale problem, replacing a macroscopic material model. 

We denote spatial coordinates on the macroscale by $\bfx$ and on the microscale by $\bfy$. Quantities denoted as $\overline{(\cdot)}$ are affiliated to the macroscale. The transition between the scales regarding energy conservation and numerical treatment is discussed in Section 3.2.
%%%%%%%%%%%%%%%%%%%%%%%%%%%
\subsection{Variational formulation of the microscale problem}

The domain $\Omega := \Omega_{\bfy}$, representing the RVE of the micro problem, is split into a cortical bone part $\Omega_\mathrm{b}$ and a bone marrow part $\Omega_\mathrm{m}$. For any quantity, the indecies $(\cdot)_\mathrm{m}$ and $(\cdot)_\mathrm{b}$ are used to denote the affiliation to each phase. If no index is present, the quantity or equation is valid for both phases. We employ the following thermodynamic energy functional at the microscale: 
\begin{align}
\Pi &= \int\limits_{\Omega_\mathrm{b}} \Psi_{\mathrm{b}}(\bfvarepsilon,\bfE,\bfB) \dd V + \int\limits_{\Omega_\mathrm{m}} \Psi_{\mathrm{m}}(\bfvarepsilon,\bfvarepsilon^i,\bfE,\bfB) + \calC \dd V \nonumber \\
&+ \int\limits_{\Omega_\mathrm{m}} \int\limits_{t} \Delta(\dot{\bfvarepsilon}^i,\dot{\bfA}) \dd t \dd V + \int\limits_{\Omega} \Psi_{\mathrm{g}}(\nabla \cdot \bfA) \dd V - W_{\mathrm{ext}} \; \text{.}
\end{align} % energy potential
The functional contains the energy densities $ \Psi_{\mathrm{b}}$ and $ \Psi_{\mathrm{m}}$ of both phases, a volume constraint $\calC$, dissipation and gauge functionals ($\Delta$ and $\Psi_{\mathrm{g}}$) and the potential of the generalized external forces $W_{\mathrm{ext}}$. The main variables of the problem are then the mechanical displacements $\bfu$, the electric scalar potential $\varphi$ and the magnetic vector potential $\bfA$, yielding seven unknown variables for the three-dimensional model. The state variables are the mechanical strain $\bfvarepsilon$, the electric field $\bfE$ and the magnetic flux density $\bfB$, calculated as
\be
\bfvarepsilon = \frac{1}{2} \; (\nabla \bfu + \nabla^T \bfu) \; \text{,} \; \bfE = - \nabla \varphi - \dot{\bfA} \; \text{and} \; \bfB = \nabla \times \bfA \; \text{.}
\ee % state variables
This way, two of the four Maxwell equations are already satisfied:
\bea
\nabla \times \bfE &=& \nabla \times (- \nabla \varphi - \dot{\bfA}) = -\dot{\bfB} \quad \text{and} \nonumber \\
\nabla \cdot \bfB &=& \nabla \cdot (\nabla \times \bfA) = \bfzero \; \text{.}
\eea % Maxwell identities
For the mechanical strain, we use Voigt's notation \cite{10.1093/qjmam/43.1.15} as
\be \bfvarepsilon = \begin{pmatrix} \varepsilon_{xx} & \varepsilon_{yy} & \varepsilon_{zz} & 2 \; \varepsilon_{xy} & 2 \; \varepsilon_{yz} & 2 \; \varepsilon_{xz} \end{pmatrix}^T \; \text{.}
\ee
Then, the energy densities for both phases are
\bea
\Psi_\mathrm{b} &=& \frac{1}{2} \bigl( \bfvarepsilon \cdot \dsC_\mathrm{b} \cdot \bfvarepsilon - \bfE \cdot \bfxi_\mathrm{b} \cdot \bfE + \bfB \cdot \bfmu_\mathrm{b}^{-1} \cdot \bfB \bigr) \nonumber \\
& & - \bfe_\mathrm{b} \cdot \bfvarepsilon \cdot \bfE \quad \text{and} \nonumber \\
\Psi_\mathrm{m} &=& \frac{1}{2} \bigl( (\bfvarepsilon-\bfvarepsilon^i) \cdot \dsC_\mathrm{m} \cdot (\bfvarepsilon-\bfvarepsilon^i) - \bfE \cdot \bfxi_\mathrm{m} \cdot \bfE \nonumber \\
& & + \bfB \cdot \bfmu_\mathrm{m}^{-1} \cdot \bfB \bigr) \; \text{,} \label{eqdens}
\eea % energy densities
consisting of quadratic energies for mechanical, electrical and magnetic effects, resulting in a linear problem. We include a piezoelectric energy term for the cortical bone phase. For the bone marrow phase, an inelastic strain $\bfvarepsilon^i$ is introduced. Here, $\dsC$ is the mechanical stiffness tensor, $\bfxi$ is the permittivity tensor, $\bfmu^{-1}$ is the inverse permeability tensor and $\bfe_\mathrm{b}$ is the piezoelectric tensor. 
While it is possible to switch between state and flux variables via a Legendre transformation, the present formulation proves as the most suitable for our model, as it allows an easy inclusion of the Maxwell coupling and the electric dissipation. For linear problems, the transformation would change an extremal into a saddle point problem, thus excluding solvers, that require positive definiteness of the system matrix as a precondition. The constraint function reads
\be
\calC = \lambda \; \tr(\bfvarepsilon^i) \; \text{,} \label{eqcons}
\ee % constraint
enforcing volume conservation of the inelastic deformation. Here, $\lambda$ is a Lagrange multiplier. The dissipation function is
\be
\Delta = \frac{1}{2} \; (\mu_\mathrm{v}^{-1} |\dot{\bfvarepsilon}^i|^2 - \bfkappa \; \bfE^2) \; \text{,} \quad \text{with} \quad \bfJ = \bfkappa \; \bfE \; \text{.} \label{eqdiss}
\ee % dissipation
Thus, $\Delta$ governs the evolution of the inelastic strain and the energy loss due to conduction, which occur both only in the bone marrow phase. Here, the viscosity parameter $\mu_\mathrm{v}^{-1}$, the electric conductivity tensor $\bfkappa$ and the electric current density $\bfJ$ are introduced. The gauge function is
\be
\Psi_{\mathrm{g}}=\frac{\gamma}{2} (\nabla \cdot \bfA)^2 \label{eqgauge}
\ee % gauge
and ensures, that a unique solution for the magnetic vector potential $\bfA$ is obtained by penalizing its divergence, effectively requiring, that $\nabla \cdot \bfA$ vanishes and thus improving the numerical stability \cite{doi:10.1002/cnm.818}. The penalty parameter $\gamma$ is a numerical parameter used to control the gauge term. Finally, the potential of generalized external forces is
\begin{align}
W_{\mathrm{ext}} &= \int\limits_{\Omega} (\bff \cdot \bfu - q_{\mathrm{v}} \cdot \varphi + \bfj_{\mathrm{v}} \cdot \bfA) \dd V \nonumber \\
&+ \int\limits_{\partial \Omega} (\bft \cdot \bfu - q_{\mathrm{s}} \cdot \varphi + \bfj_{\mathrm{s}} \cdot \bfA) \dd A \; \text{.}
\end{align} % W_ext
Here, $\bff$ and $\bft$ are the mechanical volume and surface forces, $q_{\mathrm{v}}$ and $q_{\mathrm{s}}$ are the electric volume and surface charges and $\bfj_{\mathrm{v}}$ and $\bfj_{\mathrm{s}}$ are the volume and surface currents. 

By calculating the derivative of the energy density with respect to the state variables, we find the following constitutive equations for both phases:
\begin{align}
\bfsigma &:= \displaystyle\frac{\partial \Psi_\mathrm{b}}{\partial \bfvarepsilon} = \dsC_{\mathrm{b}} \cdot \bfvarepsilon - \bfe_{\mathrm{b}}^T \cdot \bfE \;\text{,} \nonumber \\
\bfD &:= -\displaystyle\frac{\partial \Psi_\mathrm{b}}{\partial \bfE} = \bfe_{\mathrm{b}} \cdot \bfvarepsilon + \bfxi_\mathrm{b} \cdot \bfE \;\text{,} \nonumber \\
\bfH &:= \displaystyle\frac{\partial \Psi_\mathrm{b}}{\partial \bfB} = \bfmu_{\mathrm{b}}^{-1} \cdot \bfB \;\text{,} \nonumber \\
& \nonumber \\
\bfsigma &:= \frac{\partial \Psi_\mathrm{m}}{\partial \bfvarepsilon} = \dsC_{\mathrm{m}} \cdot (\bfvarepsilon - \bfvarepsilon^i) \;\text{,} \nonumber \\
\bfD &:= -\frac{\partial \Psi_\mathrm{m}}{\partial \bfE} = \bfxi_\mathrm{m} \cdot \bfE \quad \text{and} \nonumber \\
\bfH &:= \frac{\partial \Psi_\mathrm{m}}{\partial \bfB} = \bfmu_{\mathrm{m}}^{-1} \cdot \bfB \; \text{.} \label{eqconslaws}
\end{align} % const laws
For the bone marrow, the additional constitutive equations are
\begin{align}
\bfsigma &:= -\displaystyle\frac{\partial \Psi_\mathrm{m}}{\partial \bfvarepsilon^i} = \dsC_{\mathrm{m}} \cdot (\bfvarepsilon - \bfvarepsilon^i) \quad \text{and} \\
\bfJ &:= \displaystyle\frac{\partial \Delta}{\partial \bfE} = -\frac{\partial \Delta}{\partial \dot{\bfA}} = \bfkappa_{\mathrm{m}} \; \bfE \quad \text{,}
\end{align} % const laws bone marrow
introducing the flux quantities mechanical stress $\bfsigma$, \\ electric displacement $\bfD$ and magnetic field strength $\bfH$. For the cortical bone phase the viscosity parameter $\mu_\mathrm{v}^{-1}$ and the electric conductivity tensor $\bfkappa$ vanish. The material tensors satisfy
\begin{align}
\dsC &:= \displaystyle\frac{\partial \bfsigma}{\partial \bfvarepsilon} \; \text{,} \quad & \bfxi &:= \frac{\partial \bfD}{\partial \bfE} \; \text{,} \quad & \bfe &:= \displaystyle\frac{\partial \bfD}{\partial \bfvarepsilon} = -(\displaystyle\frac{\partial \bfsigma}{\partial \bfE})^T \; \text{,} \nonumber \\
\bfmu^{-1} &:= \displaystyle\frac{\partial \bfH}{\partial \bfB} \; \text{,} \quad & \bfkappa &:= \frac{\partial \bfJ}{\partial \bfE} \; \text{.}
\end{align} % material tensors
%%%%%%%%%%%%%%%%%%%%%%%%%%%%
\subsection{Weak and strong form of the microscale problem}

To calculate the weak and strong form of the problem, the energy functional has to become stationary with respect to the main variables and internal variables, leading to
\be
\int\limits_{t_0}^{t_1} \Pi \dd t \rightarrow \underset{\bfu \text{,} \; \varphi \text{,} \; \bfA \text{,} \; \bfvarepsilon^i}{stat} \; \text{.} 
\ee % stat-Bedingung
The stationary condition of the first variation of the energy functional reads then 
\bea
\displaystyle\int\limits_{\Omega_b} \biggl( \frac{\partial \Psi_b}{\partial \bfvarepsilon} \delta \bfvarepsilon + \frac{\partial \Psi_b}{\partial \bfE} \delta \bfE + \frac{\partial \Psi_b}{\partial \bfB} \delta \bfB \biggr) \dd V & \nonumber \\ 
+ \displaystyle\int\limits_{\Omega_m} \biggl( \frac{\partial \Psi_m}{\partial \bfvarepsilon} \delta \bfvarepsilon + \frac{\partial \Psi_m}{\partial \bfvarepsilon^i} \delta \bfvarepsilon^i + \frac{\partial \Psi_m}{\partial \bfE} \delta \bfE & \nonumber \\
+ \displaystyle\frac{\partial \Psi_m}{\partial \bfB} \delta \bfB + \frac{\partial \Delta}{\partial \dot{\bfvarepsilon}^i} \delta \bfvarepsilon^i + \frac{\partial \Delta}{\partial \dot{\bfA}} \delta \bfA & \nonumber \\
+ \lambda \bfI \delta \bfvarepsilon^i \biggr) \dd V + \displaystyle\int\limits_{\Omega} \frac{\partial \Psi_{g}}{\partial (\nabla \cdot \bfA)} & \nonumber \\
\delta(\nabla \cdot \bfA) \dd V - \delta W_{ext} = \delta \Pi =& 0 \; \forall \delta \bfu \text{,} \delta \varphi \text{,} \delta \bfA \text{,} \delta \bfvarepsilon^i \text{.} \nonumber \\
& \label{eqfirstvar}
\eea % 1.Variation
The variation of the generalized external forces is
\begin{align}
\delta W_{\mathrm{ext}} = \int\limits_{\Omega} ( \bff \cdot \delta \bfu - q_{\mathrm{v}} \cdot \delta \varphi + \bfj_{\mathrm{v}} \cdot \delta \bfA ) \dd V \nonumber \\
+ \int\limits_{\partial \Omega} ( \bft \cdot \delta \bfu - q_{\mathrm{s}} \cdot \delta \varphi + \bfj_{\mathrm{s}} \cdot \delta \bfA ) \dd A \; \text{.}
\end{align} % delta W_ext
Using the introduced energy densities, constraint, dissipation and gauge functions, Eqs. \eqref{eqdens}, \eqref{eqcons}, \eqref{eqdiss}, \eqref{eqgauge}, and inserting the constitutive equations Eq. \eqref{eqconslaws}, Eq. \eqref{eqfirstvar} simplifies to
\begin{align}
\int\limits_{\Omega} \bigl( \bfsigma \cdot \delta \bfvarepsilon - \bfD \cdot \delta \bfE + \bfH \cdot \delta \bfB - \bfJ \cdot \delta \bfA \nonumber \\
+ \gamma (\nabla \cdot \bfA) \cdot \delta (\nabla \cdot \bfA) + (-\bfsigma + \mu_\mathrm{v}^{-1} \dot{\bfvarepsilon}^i + \lambda \bfI) \nonumber \\
\delta \bfvarepsilon^i \bigr) \dd V - \delta W_{ext} = 0 \; \forall \delta \bfu \text{,} \delta \varphi \text{,} \delta \bfA \text{,} \delta \bfvarepsilon^i \text{.}
\end{align} % weak form complete
Here, the identity vector is denoted as $\bfI$. We find the evolution equation of the inelastic strain:
\be
-\bfsigma + \mu_\mathrm{v}^{-1} \dot{\bfvarepsilon}^i + \lambda \bfI = \bfzero \; \text{.} \label{eqevoans}
\ee % Evogl. Ansatz
To calculate the Lagrange multiplier, the trace is applied to Eq. \eqref{eqevoans}:
\be
-\tr(\bfsigma) + \underbrace{\mu_\mathrm{v}^{-1} \; \tr(\dot{\bfvarepsilon}^i)}_{=0} + \lambda \; \tr(\bfI) = 0 \; \Rightarrow \; \lambda = \frac{1}{3} \; \tr(\bfsigma) \; \text{.} \label{eqevotr}
\ee % Trace
The second term in Eq. \eqref{eqevotr} must vanish because of the introduced volume constraint. This leads to the final evolution equation 
\be
\dot{\bfvarepsilon}^i = \mu_\mathrm{v} \; \bfsigma_{\mathrm{dev}} \; \text{,} \label{eqevogl}
\ee % Evogl final
with $\bfsigma_{\mathrm{dev}} = \bfsigma - \frac{1}{3} \; \tr(\bfsigma) \; \bfI$ denoting the deviatoric part of the mechanical stress $\bfsigma$. The time integration of the evolution equation is discussed in Section 3.
 
To calculate the strong form of the problem, the remaining variational equation is used:
\begin{align}
\int\limits_{\Omega} ( \bfsigma \cdot \delta \bfvarepsilon - \bfD \cdot \delta \bfE + \bfH \cdot \delta \bfB - \bfJ \cdot \delta \bfA
 & \nonumber \\ 
 + \gamma (\nabla \cdot \bfA) \cdot \delta (\nabla \cdot \bfA) ) \dd V - \delta W_{\mathrm{ext}} &= 0 \; \forall \delta \bfu \text{,} \; \delta \varphi \text{,} \; \delta \bfA \; \text{.} \label{eqwfred}
\end{align} % weak form reduced

This form is later used to insert a FEM ansatz. We apply partial integration to each term. Details can be found in Appendix A. We obtain
\begin{align}
\nabla \cdot \bfsigma + \bff &= \bfzero \quad \text{in} \; \Omega \nonumber \\
\bfsigma \cdot \bfn &= \bft \quad \text{on} \; \partial \Omega \nonumber \\ 
\nabla \cdot \bfD &= q_{\mathrm{v}} \quad \text{in} \; \Omega \nonumber \\
\bfD \cdot \bfn &= -q_{\mathrm{s}} \quad \text{on} \; \partial \Omega \nonumber \\
\nabla \times \bfH &= \dot{\bfD} + \bfJ + \gamma \nabla(\nabla \cdot \bfA) + \bfj_{\mathrm{v}} \quad \text{in} \; \Omega \nonumber \\
\bfH \times \bfn &= \bfj_{\mathrm{s}} - \gamma (\nabla \cdot \bfA) \bfn \quad \text{on} \; \partial \Omega \label{eqstrong}
\end{align} % Strong form
recovering the mechanical equilibrium condition, the two remaining Maxwell equations and boundary conditions, including the gauge. Here, $\bfn$ is the normal vector pointing outwards. Additionally, we receive the jump conditions between the phases
\bea
[[\bfsigma]]_{\mathrm{bm}} \cdot \bfn &=& \bft \quad \text{on} \; \partial \Omega_\mathrm{bm} \nonumber \\
\left[[\bfD]\right]_{\mathrm{bm}} \cdot \bfn &=& -q_{\mathrm{s}} \quad \text{on} \; \partial \Omega_\mathrm{bm} \nonumber \\
\left[[\bfH]\right]_{\mathrm{bm}} \times \bfn &=& \bfj_{\mathrm{s}} - \gamma \; [[\nabla \cdot \bfA]]_{\mathrm{bm}} \; \bfn \quad \text{on} \; \partial \Omega_\mathrm{bm} \label{eqmc}
\eea % jump conditions
on the interface $\partial \Omega_\mathrm{bm}$ and the evolution equation of the inelastic strain Eq. \eqref{eqevogl}
in $\Omega_\mathrm{m}$.
Here $[[\cdot]]_{12} := (\cdot)_1 - (\cdot)_2$ denotes the difference between the phases. It should be noted that the strong form is valid for both phases, but the calculation of the flux variables and the inelastic strain evolution depends on the specific material parameters and thus in which phase the calculation is done.
%%%%%%%%%%%%%%%%%%%%
\subsection{Macroscale problem}

For the macroscale, the following boundary value problem in the domain $\Omega_{\bfx}$ has to be solved: find the set $\{\overline{\bfu}, \overline{\varphi}, \overline{\bfA}\}$, such that
\begin{align}
\nabla \cdot \overline{\bfsigma} + \overline{\bff} &= \bfzero \quad \text{in} \; \Omega_{\bfx} \nonumber \\
\overline{\bfsigma} \cdot \bfn &= \overline{\bft} \quad \text{on} \; \partial \Omega_{\bfx} \nonumber \\ 
\nabla \cdot \overline{\bfD} &= \overline{q}_{\mathrm{v}} \quad \text{in} \; \Omega_{\bfx} \nonumber \\
\overline{\bfD} \cdot \bfn &= -\overline{q}_{\mathrm{s}} \quad \text{on} \; \partial \Omega_{\bfx} \nonumber \\
\nabla \times \overline{\bfH} &= \dot{\overline{\bfD}} + \overline{\bfJ} + \gamma \nabla(\nabla \cdot \overline{\bfA}) + \overline{\bfj}_{\mathrm{v}} \quad \text{in} \; \Omega_{\bfx} \nonumber \\
\overline{\bfH} \times \bfn &= \overline{\bfj}_{\mathrm{s}} - \gamma (\nabla \cdot \overline{\bfA}) \bfn \quad \text{on} \; \partial \Omega_{\bfx} \; \text{,}
\end{align} % strong form macro
with the state variables
\be
\overline{\bfvarepsilon} = \frac{1}{2} \; (\nabla \overline{\bfu} + \nabla^T \overline{\bfu}) \; \text{,} \; \overline{\bfE} = - \nabla \overline{\varphi} - \boldface{\dot{\overline{A}}} \; \text{and} \; \overline{\bfB} = \nabla \times \overline{\bfA} \; \text{,}
\ee % state variables macro
and the calculation of the fluxes depending on the microscale calculations
\be
(\overline{\bfsigma}, \overline{\bfD}, \dot{\overline{\bfD}}, \overline{\bfH}, \overline{\bfJ}) = f_{\mathrm{RVE}}(\overline{\bfvarepsilon}, \overline{\bfE}, \overline{\bfB}) \; \text{.}
\ee % macro fluxes with RVE
We transform the strong form into the weak form by multiplying with test functions of the main variables and again using partial integration:
\begin{align}
\int\limits_{\Omega_{\bfx}} ( \overline{\bfsigma} \cdot \delta \overline{\bfvarepsilon} - \overline{\bfD} \cdot \delta \overline{\bfE} + \overline{\bfH} \cdot \delta \overline{\bfB} - \overline{\bfJ} \cdot \delta \overline{\bfA} & \nonumber \\
+ \gamma (\nabla \cdot \overline{\bfA}) \cdot \delta (\nabla \cdot \overline{\bfA}) ) \dd V - \delta \overline{W}_{\mathrm{ext}}=& 0 \quad \forall \delta \overline{\bfu} \text{,} \; \delta \overline{\varphi} \text{,} \; \delta \overline{\bfA} \; \text{.}
\end{align} % weak form macro
Here, the variation of the macroscopic generalized external forces is
\begin{align}
\delta \overline{W}_{\mathrm{ext}} &= \int\limits_{\Omega_{\bfx}} ( \overline{\bff} \cdot \delta \overline{\bfu} - \overline{q}_{\mathrm{v}} \cdot \delta \overline{\varphi} + \overline{\bfj}_{\mathrm{v}} \cdot \delta \overline{\bfA} ) \dd V \nonumber \\
&+ \int\limits_{\partial \Omega_{\bfx}} ( \overline{\bft} \cdot \delta \overline{\bfu} - \overline{q}_{\mathrm{s}} \cdot \delta \overline{\varphi} + \overline{\bfj}_{\mathrm{s}} \cdot \delta \overline{\bfA} ) \dd A \; \text{.}
\end{align} % delta W_ext macro
This form is again used in the next section to formulate the FEM.
%%%%%%%%%%%%%%%%%%%%
\section{Numerical implementation}

%%%%%%%%%%%%%%%%%%%%
\subsection{Finite element method}

To solve the boundary value problems on both scales, we insert a standard finite element approach \cite{zienkiewiczfinite2005} into the weak form of the problem for all main variables. In this section, we derive the resulting system for the microscale. It should be noted that the same system has to be solved for the macroscale, but each quantity $(\cdot)$ has to be replaced by its macro average quantity $\overline{(\cdot)}$. The inelastic strain is only present on the microscale and vanishes on the macroscale. Its calculation is not done via the FEM, but directly by using the evolution equation Eq. \eqref{eqevogl} on the integration point level. Details regarding the calculation of macro fluxes and consistent material tensors are given in the next subsection. Here, we denote nodal FEM values by $\boldface{\hat{(\cdot)}}$. For the evolution equation of the inelastic strain on the micro scale, we apply an explicit Euler scheme, yielding:
\be
\bfvarepsilon^i_{n+1} = \bfvarepsilon^i_{n} + \Delta t \; \dot{\bfvarepsilon}^i \quad \text{with} \quad \dot{\bfvarepsilon}^i = \mu_\mathrm{v} \bfsigma_{\mathrm{dev}} \; \text{.}
\ee % Evo-Gl. Euler
Here, $\Delta t$ is the time increment between two time steps. The standard FEM approach for the remaining system is
\begin{align}
\bfu &\approx \bfN_\mathrm{u} \; \boldface{\hat{u}} & \bfvarphi &\approx \bfN_{\varphi} \; \boldface{\hat{\varphi}} & \bfA &\approx \bfN_\mathrm{A} \; \boldface{\hat{A}} \nonumber \\
\delta \bfu &\approx \bfN_\mathrm{u} \; \delta \boldface{\hat{u}} & \delta \bfvarphi &\approx \bfN_{\varphi} \; \delta \boldface{\hat{\varphi}} & \delta \bfA &\approx \bfN_\mathrm{A} \; \delta \boldface{\hat{A}}
\end{align} % FE ansatz
approximating the main variable and their variations by shape functions times the nodal values of the functions $(\boldface{\cdot}) \approx \bfN_{\boldface{\cdot}} \; \boldface{\hat{(\cdot)}}$. For the state variables and the gauge this approach yields
\begin{align}
\hat{\bfvarepsilon} &= \bfB_\mathrm{u} \boldface{\hat{u}} \; \text{,} & \boldface{\hat{E}} &= - \bfB_\mathrm{grad} \hat{\bfvarphi} - \bfN_A \dot{\hat{\bfA}} \; \text{,} \nonumber \\
\hat{\bfB} &= \bfB_\mathrm{curl} \hat{\bfA} \; \text{,} & \nabla \cdot \hat{\bfA} &= \bfB_\mathrm{div} \hat{\bfA} \; \text{.}
\end{align} % state variables FE
Here, the operator matrices are
\begin{align}
\bfB_\mathrm{u} &= \begin{pmatrix}
\frac{\partial}{\partial x} & 0 & 0 \\
0 & \frac{\partial}{\partial y} & 0 \\
0 & 0 & \frac{\partial}{\partial z} \\
\frac{\partial}{\partial y} & \frac{\partial}{\partial x} & 0 \\
0 & \frac{\partial}{\partial z} & \frac{\partial}{\partial y} \\
\frac{\partial}{\partial z} & 0 & \frac{\partial}{\partial x}
\end{pmatrix} \cdot \bfN_\mathrm{u} \nonumber \\
\bfB_\mathrm{grad} &= \begin{pmatrix}
\frac{\partial}{\partial x} \\ \frac{\partial}{\partial y} \\ \frac{\partial}{\partial z}
\end{pmatrix} \cdot \bfN_{\varphi} \nonumber \\
\bfB_\mathrm{curl} &= \begin{pmatrix}
0 & \frac{\partial}{\partial z} & -\frac{\partial}{\partial y} \\
-\frac{\partial}{\partial z} & 0 & \frac{\partial}{\partial x} \\
\frac{\partial}{\partial y} & -\frac{\partial}{\partial x} & 0
\end{pmatrix} \cdot \bfN_\mathrm{A} \nonumber \\
\bfB_\mathrm{div} &= \begin{pmatrix}
\frac{\partial}{\partial x} & \frac{\partial}{\partial y} & \frac{\partial}{\partial z}
\end{pmatrix} \cdot \bfN_\mathrm{A}
\end{align} % B matrices 
Inserting these equations into the reduced weak form of the micro problem Eq. \eqref{eqwfred} and by using the arbitrariness of the test functions, we find the final equation system in matrix form as follows (a detailed derivation is given in Appendix B):
\be
\bfR := \bfF - \bfM \ddot{\bfd} - \bfC \dot{\bfd} - \bfK \bfd \overset{!}{=} \bfzero \label{eqfemmat}
\ee % matrix form FE
with the residual vector $\bfR = \begin{pmatrix}
	\bfR_\mathrm{u} & \bfR_\mathrm{\varphi} & \bfR_\mathrm{A}
\end{pmatrix}^T =$
\be
\begin{pmatrix}\boldface{\hat{f}} + \int\limits_{\Omega} -\bfB_\mathrm{u}^T \boldface{\hat{\sigma}} \dd V \\
\boldface{\hat{q}} + \int\limits_{\Omega} -\bfB_{\mathrm{grad}}^T \boldface{\hat{D}} \dd V \\
\boldface{\hat{j}} + \int\limits_{\Omega} (\bfN_\mathrm{A}^T (\dot{\hat{\bfD}}+\hat{\bfJ}) -\bfB_{\mathrm{curl}}^T \boldface{\hat{H}} - \gamma \bfB_{\mathrm{div}} \bfB^T_{\mathrm{div}} \boldface{\hat{A}} ) \dd V \end{pmatrix}
\ee % residual
and the generalized force and displacement vectors together with the mass, damping and stiffness matrices as follows:
\begin{align}
\bfd &:= \begin{pmatrix}
\boldface{\hat{u}} \\ \boldface{\hat{\varphi}} \\ \boldface{\hat{A}}
\end{pmatrix} \text{,} \quad
\bfF = \begin{pmatrix}
\boldface{\hat{f}} \\ \boldface{\hat{q}} \\ \boldface{\hat{j}}
\end{pmatrix} \text{,} \nonumber \\
\bfM &= \begin{pmatrix}
\bfzero & \bfzero & \bfzero \\
\bfzero & \bfzero & \bfzero \\
\bfzero & \bfzero & \bfM_{\bfA\bfA}
\end{pmatrix} \text{,} \quad
\bfC = \begin{pmatrix}
\bfzero & \bfzero & \bfC_{\bfu \bfA} \\
\bfzero & \bfzero & \bfC_{\boldface{\varphi} \bfA}\\
\bfC_{\bfA \bfu} & \bfC_{\bfA \boldface{\varphi}} & \bfC_{\bfA \bfA}
\end{pmatrix} \text{,} \nonumber \\
\bfK &= \begin{pmatrix}
\bfK_{\bfu \bfu} & \bfK_{\bfu \boldface{\varphi}} & \bfzero \\
\bfK_{\boldface{\varphi} \bfu} & \bfK_{\boldface{\varphi} \boldface{\varphi}} & \bfzero \\
\bfzero & \bfK_{\bfA \boldface{\varphi}} & \bfK_{\bfA \bfA}
\end{pmatrix} \text{.} \nonumber
\end{align} %matrix system
\begin{align}
\bfK_{\bfu \bfu} &:= -\frac{\partial \bfR_\mathrm{u}}{\partial \boldface{\hat{u}}} = \int\limits_{\Omega} \bfB_\mathrm{u}^T \; \dsC_\mathrm{tang} \; \bfB_\mathrm{u} \; \dd V \text{,} \nonumber \\
\displaystyle\bfK_{\bfu \boldface{\varphi}} &:=  -\frac{\partial \bfR_\mathrm{u}}{\partial \boldface{\hat{\varphi}}} = \int\limits_{\Omega} \bfB_\mathrm{u}^T \; \bfe^T \; \bfB_{\mathrm{grad}} \; \dd V \text{,} \nonumber \\
\bfK_{\boldface{\varphi} \bfu} &:=  -\frac{\partial \bfR_\varphi}{\partial \boldface{\hat{u}}} = \int\limits_{\Omega} \bfB_{\mathrm{grad}}^T \; \bfe \; \bfB_\mathrm{u} \; \dd V \text{,} \nonumber \\
\displaystyle \bfK_{\boldface{\varphi} \boldface{\varphi}} &:=  -\frac{\partial \bfR_\varphi}{\partial \boldface{\hat{\varphi}}} = \int\limits_{\Omega} - \bfB_{\mathrm{grad}}^T \; \bfxi \; \bfB_{\mathrm{grad}} \; \dd V \text{,} \nonumber \\
\bfK_{\bfA \bfA} &:=  -\frac{\partial \bfR_\mathrm{A}}{\partial \boldface{\hat{A}}} = \int\limits_{\Omega} \bfB_{\mathrm{curl}}^T \bfmu^{-1} \bfB_{\mathrm{curl}} + \gamma \bfB_{\mathrm{div}} \bfB_{\mathrm{div}}^T \dd V \text{,} \nonumber \\
\displaystyle \bfK_{\bfA \boldface{\varphi}} &:=  -\frac{\partial \bfR_\mathrm{A}}{\partial \boldface{\hat{\varphi}}} = \int\limits_{\Omega} \bfN_\mathrm{A}^T \; \bfkappa \; \bfB_{\mathrm{grad}} \; \dd V \text{,} \nonumber \\
\bfC_{\bfu \bfA} &:= -\frac{\partial \bfR_\mathrm{u}}{\partial \dot{\hat{\bfA}}} = \int\limits_{\Omega} \bfB_\mathrm{u}^T \; \bfe^T \; \bfN_\mathrm{A} \; \dd V \text{,} \nonumber \\
\displaystyle \bfC_{\bfA \bfu} &:= -\frac{\partial \bfR_\mathrm{A}}{\partial \dot{\hat{\bfu}}} = \int\limits_{\Omega} - \bfN_\mathrm{A}^T \; \bfe \; \bfB_\mathrm{u} \; \dd V \text{,} \nonumber \\
\bfC_{\boldface{\varphi} \bfA} &:= -\frac{\partial \bfR_\varphi}{\partial \dot{\hat{\bfA}}} = \int\limits_{\Omega} - \bfB_{\mathrm{grad}}^T \; \bfxi \; \bfN_\mathrm{A} \; \dd V \text{,} \nonumber \\
\displaystyle \bfC_{\bfA \boldface{\varphi}} &:= -\frac{\partial \bfR_\mathrm{A}}{\partial \dot{\hat{\bfvarphi}}} = \int\limits_{\Omega} \bfN_\mathrm{A}^T \; \bfxi \; \bfB_{\mathrm{grad}} \; \dd V \text{,} \nonumber \\
\bfC_{\bfA \bfA} &:= -\frac{\partial \bfR_\mathrm{A}}{\partial \dot{\hat{\bfA}}} = \int\limits_{\Omega} \bfN_\mathrm{A}^T \; \bfkappa \; \bfN_\mathrm{A} \; \dd V \text{,} \nonumber \\ \displaystyle \bfM_{\bfA \bfA} &:= -\frac{\partial \bfR_\mathrm{A}}{\partial \boldface{\ddot{\hat{A}}}} = \int\limits_{\Omega} \bfN_\mathrm{A}^T \; \bfxi \; \bfN_\mathrm{A} \; \dd V \text{.}
\end{align} % stiffness / damping / mass matrices
The material tensors depend again on the phase. We calculate the mechanical stiffness tangent matrix $\dsC_\mathrm{tang}$ by introducing a time discretization as:
\be
\dsC_\mathrm{tang} = \frac{\partial \bfsigma_{n+1}}{\partial \bfvarepsilon_{n+1}} = \begin{cases} \dsC_{\mathrm{b}} \quad \text{in } \Omega_\mathrm{b} \\ 
\dsC_{\mathrm{m}} - \dsC_{\mathrm{m}} \; \frac{\partial \bfvarepsilon_{n+1}^i}{\partial \bfvarepsilon_{n+1}} \quad \text{in } \Omega_\mathrm{m}\end{cases}
\ee % tangent ansatz
For the bone marrow phase, the calculation depends on the inelastic strain $\bfvarepsilon_{n+1}^i$:
\be
\frac{\partial \bfvarepsilon_{n+1}^i}{\partial \bfvarepsilon_{n+1}} = \mu_\mathrm{v} \frac{\partial \bfsigma_\mathrm{dev}}{\partial \bfsigma} \underbrace{\frac{\partial \bfsigma}{\partial \bfvarepsilon_{n+1}}}_{\dsC_\mathrm{m}} \; \text{,}
\ee % tangent bone marrow calculation
with 
\begin{align}
\bfD_{\mathrm{devd}} &:= \frac{\partial \bfsigma_\mathrm{dev}}{\partial \bfsigma} 
&=& \bfI - \frac{1}{3} \begin{pmatrix}1 & 1 & 1 & 0 & 0 & 0 \\ 
1 & 1 & 1 & 0 & 0 & 0 \\
1 & 1 & 1 & 0 & 0 & 0 \\ 
0 & 0 & 0 & 0 & 0 & 0 \\
0 & 0 & 0 & 0 & 0 & 0 \\
0 & 0 & 0 & 0 & 0 & 0 \end{pmatrix} \nonumber \\
& &=& \frac{1}{3} \begin{pmatrix}2 & -1 & -1 & 0 & 0 & 0 \\ 
-1 & 2 & -1 & 0 & 0 & 0 \\
-1 & -1 & 2 & 0 & 0 & 0 \\ 
0 & 0 & 0 & 3 & 0 & 0 \\
0 & 0 & 0 & 0 & 3 & 0 \\
0 & 0 & 0 & 0 & 0 & 3 \end{pmatrix}\text{.}
\end{align} % devd matrix
The mechanical stiffness tangent matrix for the bone marrow phase is then
\be
\dsC_\mathrm{tang} = \dsC_m - \dsC_m (\mu_\mathrm{v} \bfD_{\mathrm{devd}} \dsC_m) \quad \text{in } \Omega_\mathrm{m} \; \text{.}
\ee % tangent bone marrow
In order to solve the resulting second order differential equation system, a suitable time integration scheme is necessary. Here we use a JWH-$\alpha$-scheme introduced in \cite{kadapagenalpha}, where also details regarding advantages and implementation of this method can be found. For the time integration, the time increment $\Delta t$ and the additional numerical parameter $\rho_{\infty}$ are needed. By combining the method with a regular Newton-Raphson scheme, we transform the matrix system of Eq. \eqref{eqfemmat} to
\be
\bfR^{i+1} := \bfR^i(d_{n+\alpha_f}, v_{n+\alpha_f}, \dot{v}_{n+\alpha_m}) - \bfS  \; \Delta \bfd \overset{!}{=} \bfzero \; \text{,}
\ee % newton raphson
with the index denoting the iteration and the generalized tangent matrix
\be
\bfS = \frac{\alpha_\mathrm{m}^2}{\alpha_\mathrm{f} \gamma_{\mathrm{a}}^2 \Delta t^2} \bfM + \frac{\alpha_\mathrm{m}}{\gamma_{\mathrm{a}} \Delta t} \bfC + \alpha_\mathrm{f} \bfK \; \text{,}
\ee % S matrix
which is the Jacobian of the system. Here $\Delta \bfd$ is the increment of the solution vector and $\alpha_\mathrm{f}$, $\alpha_\mathrm{m}$ and $\gamma_{\mathrm{a}}$ are numerical parameters depending on $\rho_{\infty}$ \cite{kadapagenalpha}. The residual $\bfR^i(d_{n+\alpha_f}, v_{n+\alpha_f}, \dot{v}_{n+\alpha_m})$ is calculated from either initial conditions for the first iteration of the first time step or else from the previous increment \cite{kadapagenalpha}. The resulting tangent matrix $\bfS$ is neither symmetric nor positive definite, limiting the choices for a suitable solver of the linear system.
%%%%%%%%%%%%%%%%%%%%
\subsection{Transition between the scales}
To connect the macro- and microscale in $\text{FE}^2$, it is important to discuss the transition between the scales. The Hill-Mandel conditions \cite{Hill1963,Hill1972,Sch:2009:dot,schroeder2016,LabSchLup:2019:ath,PhysRevE.99.022120} have to be fulfilled, guaranteeing energy conservation during the scale transition. Thus, the virtual work on the macroscale has to be equal to the virtual work on the microscale:
\begin{align}
\overline{\bfsigma} \cdot \delta \overline{\bfvarepsilon} &= \frac{1}{\Omega} \int\limits_{\Omega} \bfsigma \cdot \delta \bfvarepsilon \dd V \text{,} & \overline{\bfD} \cdot \delta \overline{\boldface{E}} &= \frac{1}{\Omega} \int\limits_{\Omega} \bfD \cdot \delta \bfE \dd V \text{,} \nonumber \\
\boldface{\overline{B}} \cdot \delta \boldface{\overline{H}} &= \frac{1}{\Omega} \int\limits_{\Omega} \bfB \cdot \delta \bfH \dd V \; \text{.}
\end{align} % Hill Mandel
For the macro-to-micro transition, these conditions can be fulfilled by three different types of boundary conditions on the microscale: Dirichlet, Neumann and periodic boundary conditions 
\cite{Ilic2010,schroeder2000,schroeder2013plasticity}. Here we chose periodic boundary conditions, as they are the only type of boundary condition, where the results on the microscale are independent from the relative geometry of the RVE \cite{schroeder2000,schroeder2013plasticity}. Additionally, as the RVE is periodic in space, this type of boundary condition is the most suitable. In the program, the periodic boundary conditions were applied by fixing all degrees of freedom at all corner nodes, preventing rigid body motions, and linking all degrees of freedom at opposite faces of the RVE, ensuring the periodicity. The micro state variables consist then of two parts: a term resulting from the microscopic main variables (denoted by $\boldface{\tilde{(\cdot)}}$), whose fluctuations are calculated, and a term contributed by the macroscale:
\begin{align}
\bfvarepsilon &= \tilde{\bfvarepsilon}(\bfy) + \overline{\bfvarepsilon}(\bfx) \text{,} & \bfE &= \boldface{\tilde{E}}(\bfy) + \boldface{\overline{E}}(\bfx) \text{,} \nonumber \\
\bfB &= \boldface{\tilde{B}}(\bfy) + \boldface{\overline{B}}(\bfx) \text{.}
\end{align} % micro state vars depending on macro state
This way, we calculate the flux variables on the microscale. For the micro-to-macro transition, the volume average of these flux quantities is sent back to the macroscale:
\begin{align}
\overline{\bfsigma}(\bfx) &= \frac{1}{\Omega} \int\limits_{\Omega} \bfsigma(\bfy) \dd V \text{,} & \overline{\bfD}(\bfx) &= \frac{1}{\Omega} \int\limits_{\Omega} \bfD(\bfy) \dd V \text{,} \nonumber \\ 
\dot{\overline{\bfD}}(\bfx) &= \frac{1}{\Omega} \int\limits_{\Omega} \dot{\bfD}(\bfy) \dd V \text{,} & \overline{\bfH}(\bfx) &= \frac{1}{\Omega} \int\limits_{\Omega} \bfH(\bfy) \dd V \text{,} \nonumber \\
\overline{\bfJ}(\bfx) &= \frac{1}{\Omega} \int\limits_{\Omega} \bfJ(\bfy) \dd V \text{.}
\end{align} % volaver fluxes
In this model, energy dissipation is considered in two ways. For the electric current $\bfJ$, the average is calculated and included in the scale transition, resulting in no energy loss during the scale transition. For the inelastic strain $\bfvarepsilon^i$, the complete state in every point and for every RVE is saved. Thus, the dissipation occurs only on the microscale and the energy conservation is fullfilled, as the virtual work send to the microscale is equal to the virtual work send back added to the energy dissipation on the microscale.
With the flux variables available on the macroscale, it is now possible to obtain the macro residual for the Newton-Raphson method and the calculation of consistent macro tangent moduli remains, which are needed for the iteration. The definitions of those moduli read
\begin{align}
\overline{\dsC} &:= \frac{\partial \overline{\bfsigma}}{\partial \overline{\bfvarepsilon}} \; \text{,} & \overline{\bfxi} &:= \frac{\partial \overline{\bfD}}{\partial \overline{\bfE}} \; \text{,} \nonumber \\
\overline{\bfe} &:= \frac{\partial \overline{\bfD}}{\partial \overline{\bfvarepsilon}} \; \text{,} & \overline{\bfmu^{-1}} &:= \frac{\partial \overline{\bfH}}{\partial \overline{\bfB}} \; \text{,} \nonumber \\ \overline{\bfkappa} &:= \frac{\partial \overline{\bfJ}}{\partial \overline{\bfE}} \; \text{.} 
\end{align}
The calculation can be done by applying a small numerical perturbation $\Delta_{\mathrm{tol}} = 10^{-8}$ to each entry of the corresponding state variable 
\begin{align}
\overline{\bfvarepsilon}^{p_i} &= \overline{\bfvarepsilon} + \Delta_{\mathrm{tol}} e_i \; \text{,} & \overline{\bfE}^{p_i} &= \overline{\bfE} + \Delta_{\mathrm{tol}} e_i \; \text{,} \nonumber \\
\overline{\bfB}^{p_i} &= \overline{\bfB} + \Delta_{\mathrm{tol}} e_i \; \text{,}
\end{align}
with the i-th unit vector $e_i$, and then calculating each entry of the macroscopic tangent tensors by evaluating the perturbated fluxes $\overline{\bfsigma}^{p_i}, \overline{\bfD}^{p_i}, \overline{\bfH}^{p_i}, \overline{\bfJ}^{p_i}$ by means of the RVE as
\begin{align}
\overline{\dsC}_{ji} &= \frac{\overline{\sigma}^{p_i}_j - \overline{\sigma}_{j}}{\overline{\varepsilon}^{p_i}_j - \overline{\varepsilon}_{j}} \; \text{,} & \overline{\xi}_{ji} &= \frac{\overline{D}^{p_i}_j - \overline{D}_{j}}{\overline{E}^{p_i}_j - \overline{E}_{j}} \; \text{,} \nonumber \\
\overline{e}_{ji} &= \frac{\overline{D}^{p_i}_j - \overline{D}_{j}}{\overline{\varepsilon}^{p_i}_j - \overline{\varepsilon}_{j}} & \overline{\mu^{-1}_{ji}} &= \frac{\overline{H}^{p_i}_j - \overline{H}_{j}}{\overline{B}^{p_i}_j - \overline{B}_{j}} \; \text{,} \nonumber \\
\overline{\kappa}_{ji} &= \frac{\overline{J}^{p_i}_j - \overline{J}_{j}}{\overline{E}^{p_i}_j - \overline{E}_{j}} \; \text{.}
\end{align}
Since for our model the same RVE is used everywhere and the non-linearity from the inelastic strain is very small, this calculation has to be done only once for all RVEs and all time steps, making this approach very efficient. Together with the calculated macro state variables, this allows to solve the macroscopic FE problem.
%%%%%%%%%%%%%%%%%%%%
\subsection{Implementation}

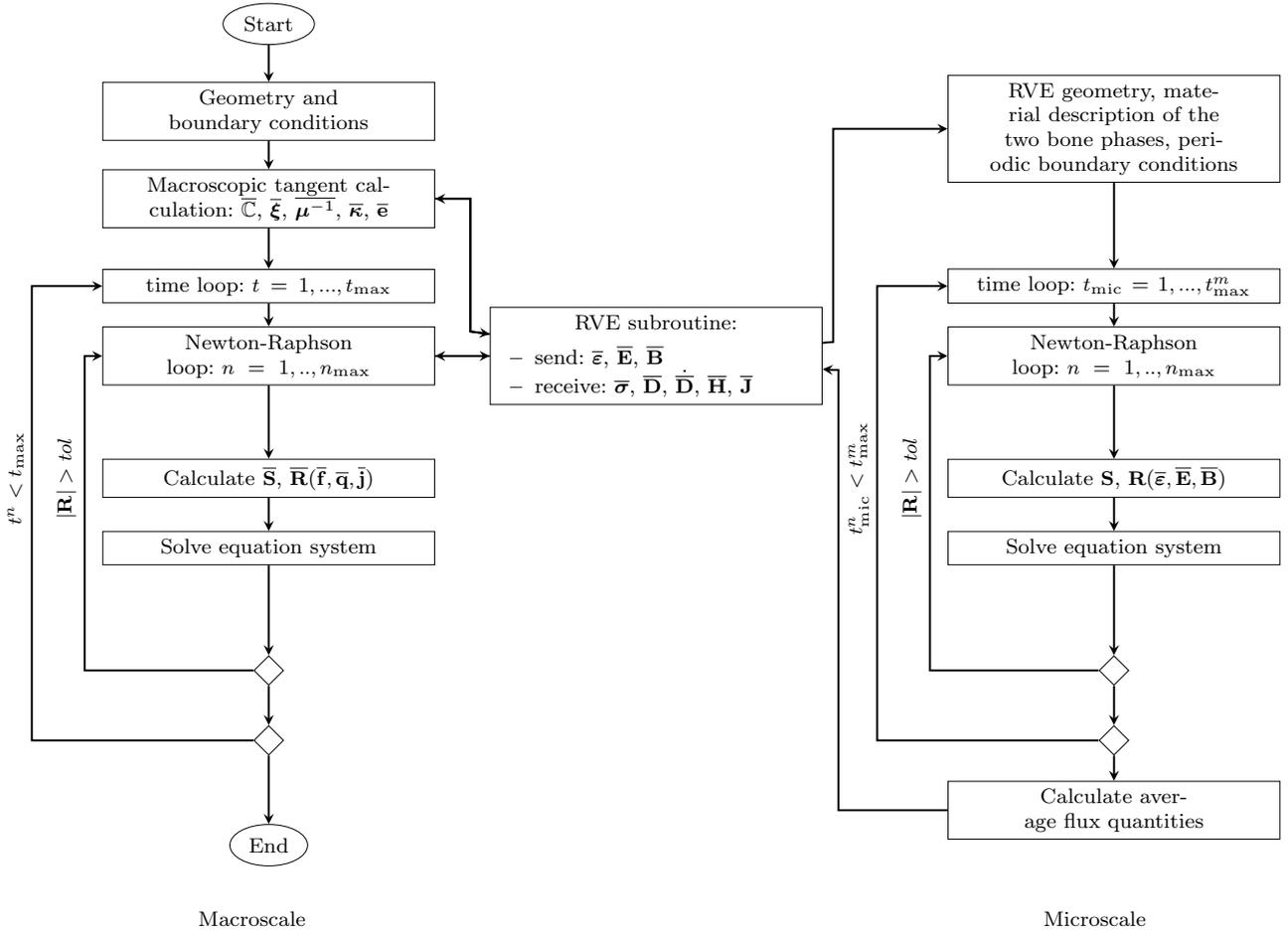
\begin{figure*}[h!]
	\centering
	\resizebox{0.98\textwidth}{!}{%
		\begin{tikzpicture}[node distance=1cm]
			\node (start) [start] {Start};
			\node (pro1) [process, below of=start, yshift =-0.25cm] {Geometry and boundary conditions};
			\node (pro2) [process, below of=pro1, yshift =-0.25cm] {Macroscopic tangent calculation: $\overline{\dsC}$, $\overline{\bfxi}$, $\overline{\bfmu^{-1}}$, $\overline{\bfkappa}$, $\overline{\bfe}$};
			
			\node (pro3) [process, below of=pro2, yshift =-0.25cm] {time loop: $t=1,...,t_\mathrm{max}$};
			\node (pro4) [process, below of=pro3] {Newton-Raphson loop: $n=1,..,n_\mathrm{max}$};
			
			\node (pro5) [process, below of=pro4, yshift =-0.75cm] {Calculate $\boldface{\overline{S}}$, $\boldface{\overline{R}}(\overline{\bff}, \overline{\bfq}, \overline{\bfj})$};
			\node (pro6) [process, below of=pro5] {Solve equation system};
			\node (in1) [decision, below of=pro6, yshift=-0.75cm] {};
			\node (in2) [decision, below of=in1] {};
			\node (end) [start, below of=in2, yshift=-0.5cm] {End};
			
			\node (mpro1) [process, below of=start, xshift=12cm, yshift =-0.5cm]{RVE geometry, material description of the two bone phases, periodic boundary conditions};
			\node (mpro3) [process, below of=mpro1, yshift =-1.25cm]{time loop: $t_\mathrm{mic}=1,...,t^{m}_\mathrm{max}$};
			\node (mpro4) [process, below of=mpro3] {Newton-Raphson loop: $n=1,..,n_\mathrm{max}$};
			
			\node (mpro5) [process, below of=mpro4, yshift =-0.75cm] {Calculate $\bfS$, $\bfR(\overline{\bfvarepsilon}, \overline{\bfE}, \overline{\bfB})$};
			\node (mpro6) [process, below of=mpro5] {Solve equation system};
			\node (min1) [decision, below of=mpro6, yshift=-0.75cm] {};
			\node (min2) [decision, below of=min1] {};
			\node (mpro7) [process, below of=min2] {Calculate average flux quantities};
			
			\node (promid) [process, below of=pro4, xshift=5.5cm, yshift=1cm]{RVE subroutine: 
				\begin{itemize}
					\item send: $\overline{\bfvarepsilon}$, $\overline{\bfE}$, $\overline{\bfB}$
					\item receive: $\overline{\bfsigma}$, $\overline{\bfD}$, $\dot{\overline{\bfD}}$, $\overline{\bfH}$, $\overline{\bfJ}$
			\end{itemize}};
			
			\coordinate (left2) at ($(pro2.east)+(0.5cm,0)$);
			\coordinate (leftsub2) at ($(pro2.east)+(0.5cm,-1.9cm)$);
			\coordinate (left3) at ($(pro3.west)+(-1cm,0)$);
			\coordinate (left4) at ($(pro4.west)+(-0.25cm,0)$);
			\coordinate (left5) at ($(pro5.west)+(-0.25cm,0)$);
			\coordinate (left6) at ($(pro6.west)+(-1cm,0)$);
			
			\coordinate (mleft4) at ($(mpro4.west)+(-0.25cm,0)$);
			\coordinate (mleft5) at ($(mpro5.west)+(-0.25cm,0)$);
			\coordinate (mleft3) at ($(mpro3.west)+(-1cm,0)$);
			\coordinate (mleft6) at ($(mpro6.west)+(-1cm,0)$);
			
			\coordinate (mleft1) at ($(mpro1.west)+(-1.55cm,0)$);
			\coordinate (mleft7) at ($(mpro7.west)+(-1.55cm,0)$);
			\coordinate (mrightmid1) at ($(promid.east)+(0.15cm,0.2cm)$);
			\coordinate (mrightmid2) at ($(promid.east)+(0.15cm,-0.2cm)$);
			
			\draw [arrow] (start) -- (pro1);
			\draw [arrow] (pro1) -- (pro2);
			\draw [arrow] (pro2) -- (pro3);
			\draw [arrow] (pro3) -- (pro4);
			\draw [arrow] (pro4) -- (pro5);
			\draw [arrow] (pro5) -- (pro6);
			\draw [arrow] (pro6) -- (in1);
			\draw [arrow] (in1) -- (in2);
			\draw [arrow] (in2) -- (end);
			
			\draw [arrow] (in1) -| (left5) node[anchor=south, rotate=90] {$|\bfR| > tol$} |- (left4) -- (pro4);
			\draw [arrow] (in2) -| (left6) node[anchor=south, yshift=1cm, rotate=90] {$t^n < t_{\mathrm{max}}$} |- (left3) -- (pro3);
			
			\draw [arrow] (mpro1) -- (mpro3);
			\draw [arrow] (mpro3) -- (mpro4);
			\draw [arrow] (mpro4) -- (mpro5);
			\draw [arrow] (mpro5) -- (mpro6);
			\draw [arrow] (mpro6) -- (min1);
			\draw [arrow] (min1) -- (min2);
			\draw [arrow] (min2) -- (mpro7);
			
			\draw [arrow] (min1) -| (mleft5) node[anchor=south, rotate=90] {$|\bfR| > tol$} |- (mleft4) -- (mpro4);
			\draw [arrow] (min2) -| (mleft6) node[anchor=south, yshift=1cm, rotate=90] {$t_\mathrm{mic}^n < t^{m}_{\mathrm{max}}$} |- (mleft3) -- (mpro3);
			
			\draw [arrow] (pro4) -- (promid);
			\draw [arrow] (promid) -- (pro4);
			
			\draw [arrow] (pro2) -| (left2) |- (leftsub2) -- (promid);
			\draw [arrow] (promid) -- (leftsub2) |- (left2) |- (pro2);
			
			\draw [arrow] (promid) -- (mrightmid1) |- (mleft1) -- (mpro1);
			\draw [arrow] (mpro7) -| (mleft7) |- (mrightmid2) -- (promid);
			
			\node[text width=3cm] at (0.5,-12.8) {Macroscale};
			\node[text width=3cm] at (12.5,-12.8) {Microscale};
			
		\end{tikzpicture}
	}%
	\caption{Program flow of the multiscale simulations.}
	\label{fig3}
\end{figure*}

For the simulations, we implemented a computer program in the language \textsc{Julia} \cite{juliawebsite}, using mainly the packages \textsc{JuAFEM} \cite{juafemwebsite} and \textsc{CoherentStructures} \cite{cohestrwebsite}. As the microscale calculations are not dependent on each other, we have parallelized the macroscale element routine, increasing the speed of the computations drastically. As the inelastic strain $\varepsilon^i$ is only present in the microscale, we used HDF5 files to store the complete state of the inelastic strain for every RVE for the previous and current time step. Thus, for the inelastic strain evolution no information is lost. In order to solve the linear systems, we used the BiCGStab(l) method of the package \textsc{KrylovMethods} \cite{krylovwebsite}, as it is stable, fast even without preconditioning the problem and can be used for any matrix type. Regarding the structure of the program, Figure \ref{fig3} shows the procedure.
%%%%%%%%%%%%%%%%%%%%
\section{Simulation results}

\subsection{Parameters and material tensors}

In this subsection, we discuss the numerical and material parameters employed. Unless explicity statet otherwise, the parameters from this subsection are used in all simulations. Regarding the numerical parameters, we use the same parameters for both scales. Here, the time integration parameter is $\rho_{\infty} = 0.5$, the time step increment is $\Delta_\mathrm{t} = 1 \cdot 10^{-3}\;\text{s}$, the Newton-Raphson tolerance is $\mathrm{tol}_{\mathrm{N}} = 1 \cdot 10^{-8}$ and the gauge penalty parameter is $\gamma = 1.0\; \text{s}^2\;\text{A}^2/(\text{kg\;m})$.

\begin{table*}[h]
\begin{tabular*}{\textwidth}{l@{\extracolsep{\fill}}lllll}
\hline
Material parameter &   & \multicolumn{2}{l}{Cortical bone} & \multicolumn{2}{l}{Bone marrow} \\ \hline
Young's modulus    & $E$ & $22.0$             & GPa            & $2.0$            & GPa            \\
Poisson's ratio                   & $\nu$ &         $0.32$         &      -         &   $0.3$             &       -        \\
Permittivity                   & $\xi_1$ &        $8.85 \cdot 10^{-12}$          &     F/m           &    $8.85 \cdot 10^{-12}$            &         F/m     \\
Permeability                  & $\mu_\mathrm{c}$ & $1.257 \cdot 10^{-6}$ & H/m & $1.257 \cdot 10^{-6}$ & H/m \\
Piezoelectric coefficient                   & $e_{15}$ & $3.0 \cdot 10^{-3}$ & $\text{A\;s}/\text{m}^2$ & $0$ & $\text{A\;s}/\text{m}^2$ \\
Electric conductivity                   & $\kappa_1$ & $0$ & S/m & $1.0 \cdot 10^{4}$ & S/m\\
Viscosity parameter                   & $\mu_\mathrm{v}$ & $0$ & s/GPa & $5.0 \cdot \Delta_t$ & s/GPa   \\
\hline         
\end{tabular*}
\caption{Default material parameters.}
\label{tab1}
\end{table*}

The used default material parameters are shown in Table \ref{tab1}. Young's modulus and Poisson's ratio for both phases can be found in \cite{Steeb2010}. All other parameters are of rather academical nature and influence the results only marginally. The resulting material tensors read
\begin{align}
\dsC &= \frac{E}{(1 + \nu)(1 - 2 \nu)} \; \cdot \nonumber \\
& \begin{pmatrix}1 - \nu & \nu & \nu & 0 & 0 & 0 \\
\nu & 1-\nu & \nu & 0 & 0 & 0 \\
\nu & \nu & 1-\nu & 0 & 0 & 0 \\
0 & 0 & 0 & \frac{1-2\nu}{2} & 0 & 0 \\
0 & 0 & 0 & 0 & \frac{1-2\nu}{2} & 0 \\
 0 & 0 & 0 & 0 & 0 & \frac{1-2\nu}{2} \\
\end{pmatrix} \text{,} \nonumber
\end{align}
\begin{align}
\bfxi &= \begin{pmatrix} \xi_1 & 0 & 0 \\
0 & \xi_1 & 0 \\
0 & 0 & \xi_1 \\
\end{pmatrix} \text{,} &
\bfe &= \begin{pmatrix} 0 & 0 & 0 & 0 & e_{15} & 0\\
0 & 0 & 0 & 0 & 0 & -e_{15} \\
0 & 0 & 0 & 0 & 0 & 0 \\
\end{pmatrix} \text{,} \nonumber \\
\bfmu^{-1} &= \begin{pmatrix} \mu_\mathrm{c}^{-1} & 0 & 0 \\
0 & \mu_\mathrm{c}^{-1} & 0 \\
0 & 0 & \mu_\mathrm{c}^{-1} \\
\end{pmatrix} \text{,} & 
\bfkappa &= \begin{pmatrix} \kappa_1 & 0 & 0 \\
0 & \kappa_1 & 0 \\
0 & 0 & \kappa_1 \\
\end{pmatrix} \text{.}
\end{align}

\begin{figure}[h!]
	\centering
  \includegraphics[width=0.48\textwidth]{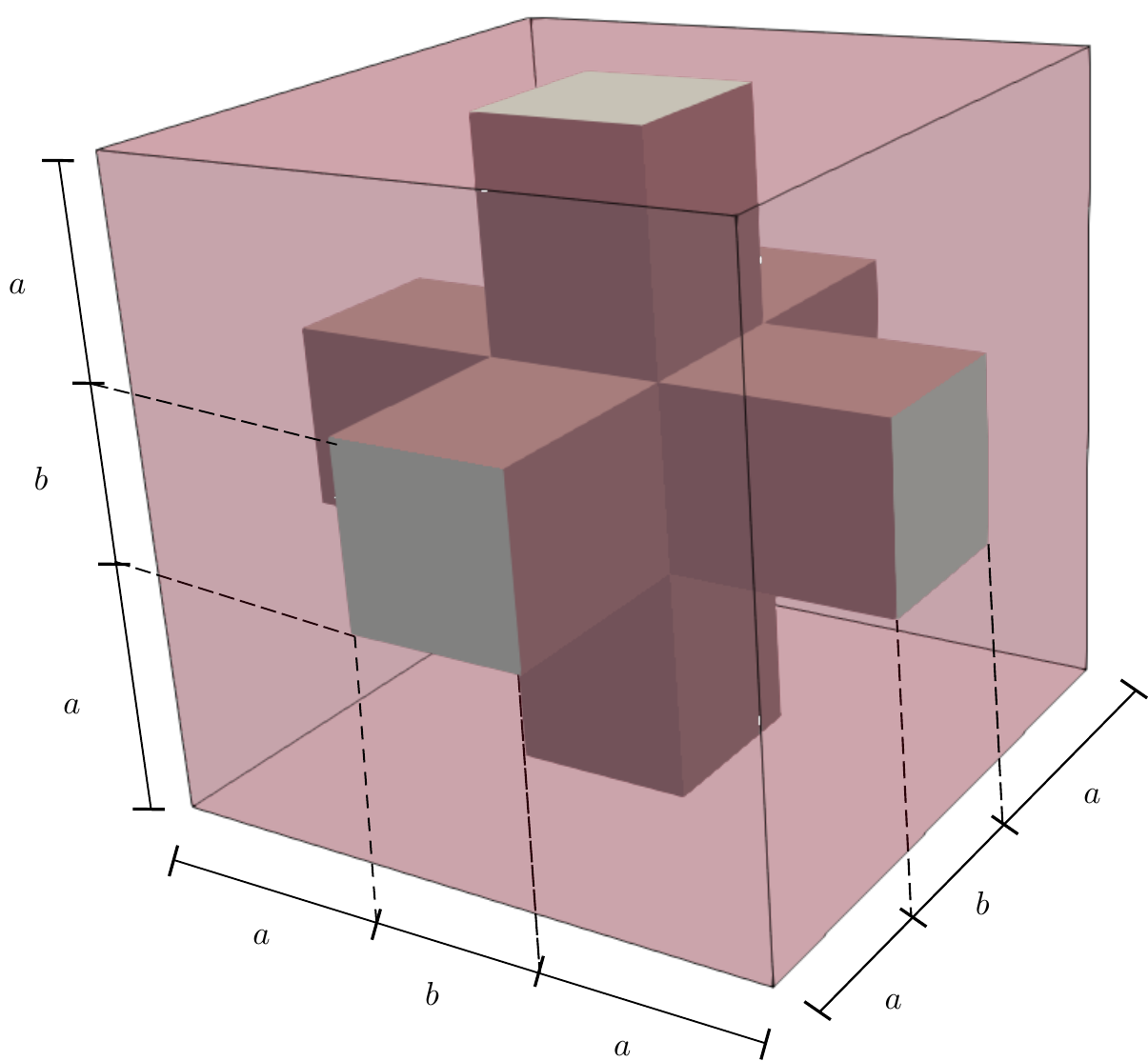}
	\caption{Periodic RVE with cortical bone phase (gray) and bone marrow phase (transparent red) and lengths parameters.}
	\label{fig4} 
\end{figure}

\begin{table}[h!]
\begin{tabular*}{0.48\textwidth}{l@{\extracolsep{\fill}}lll}
\hline
no. & $a\;[\text{mm}]$ & $b\;[\text{mm}]$ & $\rho_{\mathrm{b}}$ \\ \hline
$1$ & $0.43$ & $0.14$ & $5.3\%$ \\
$2$ & $0.40$ & $0.20$ & $10.4\%$ \\
$3$ & $0.38$ & $0.24$ & $14.5\%$ \\
$4$ & $0.36$ & $0.28$ & $19.1\%$ \\
$5$ & $0.34$ & $0.32$ & $24.2\%$ \\
$6$ & $0.32$ & $0.36$ & $29.5\%$ \\
\hline         
\end{tabular*}
\caption{Lengths parameter of the different RVEs.}
\label{tab2}
\end{table}
We assume linear isotropic material everywhere, excluding the piezoelectric tensor which is preferential in the $z$-axis due to the longitudinal orientation of the collagen fibers. It should be noted, that due to the form of the piezoelectric tensor, the material model as a whole is non-isotropic. 

For the generation of the meshes, we used the program \textsc{Gmsh} \cite{gmshwebsite}. We did the visualization of the results with \textsc{ParaView} \cite{paraviewwebsite} and \textsc{Julia}.

\subsection{Microscale model}

In this subsection, we restrict ourselves to microscale simulations. In order to compare periodic RVEs for different stages of osteoporosis, we introduce the lengths parameters $a$ and $b$ (Figure \ref{fig4}), which allow us to control the volume fractions of the phases. By using this convention, the total volume of the RVE is $V_{\mathrm{RVE}} = (2a+b)^3$. We only use RVEs with the same total volume of $V_{\mathrm{RVE}} = 1\;\text{mm}^3$, which is a suitable size for the microscale calculations \cite{Ilic2010}, making it easy to compare different RVEs. Thus, the choice of $a$ and $b$ is restricted by $2a+b=1\;\text{mm}$. The volume fraction of cortical bone for our RVE is $\rho_{\mathrm{b}}=(6 a b^2 + b^3)/(2a+b)^3$.

\begin{figure*}[h!]
\centering
  \includegraphics[width=0.48\textwidth]{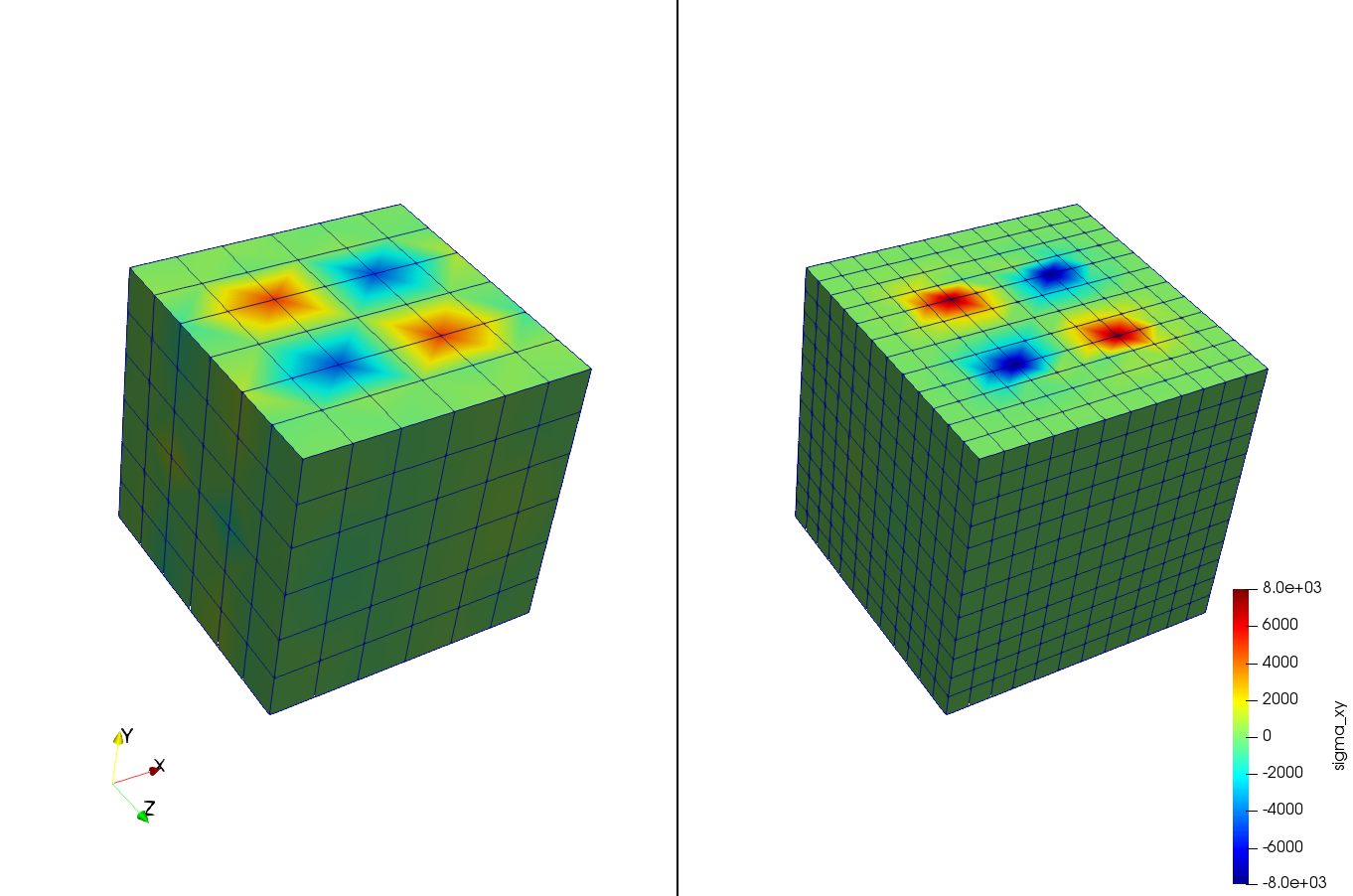} \quad \includegraphics[width=0.48\textwidth]{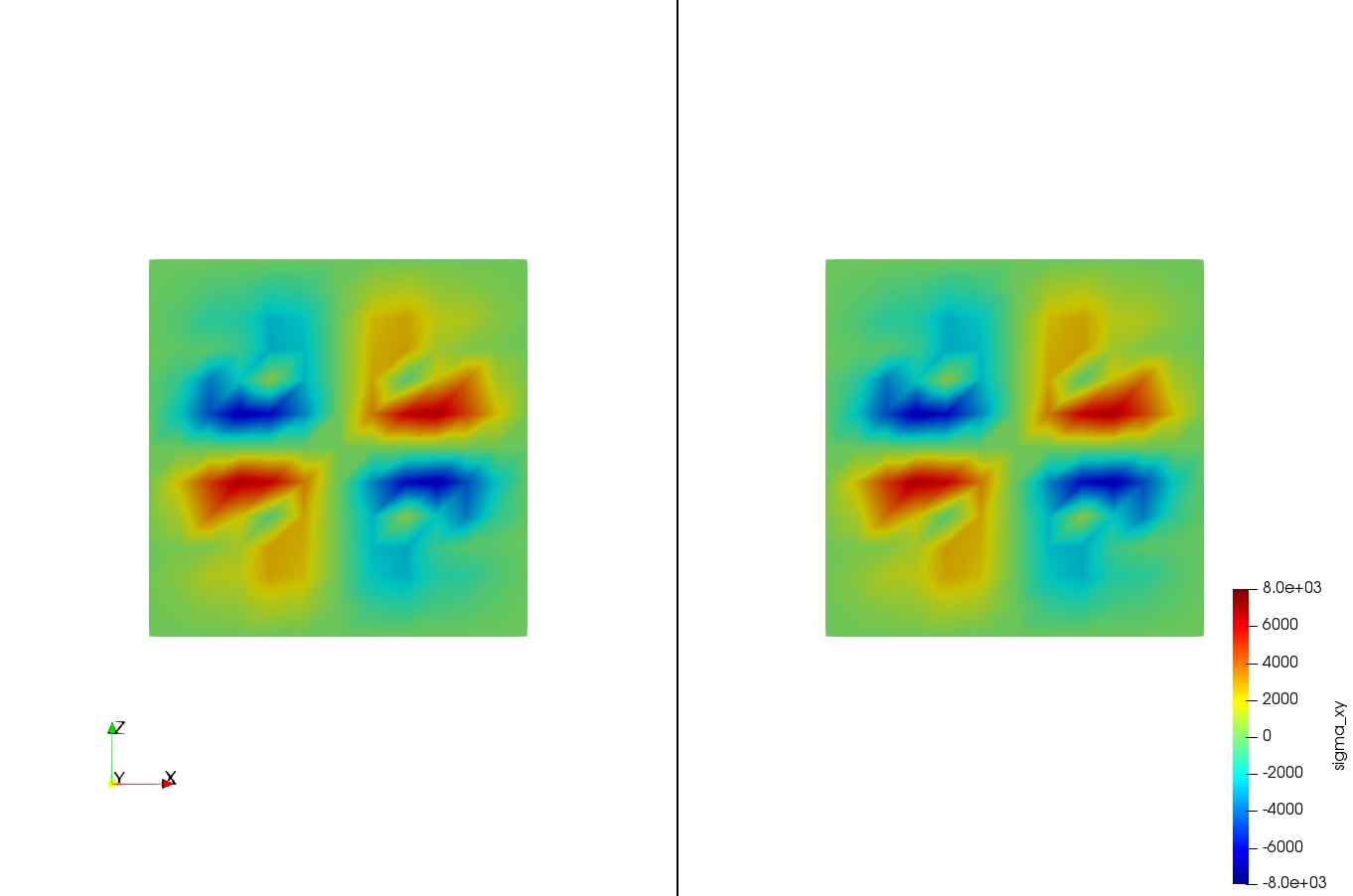}
  \hfil
  \includegraphics[width=0.48\textwidth]{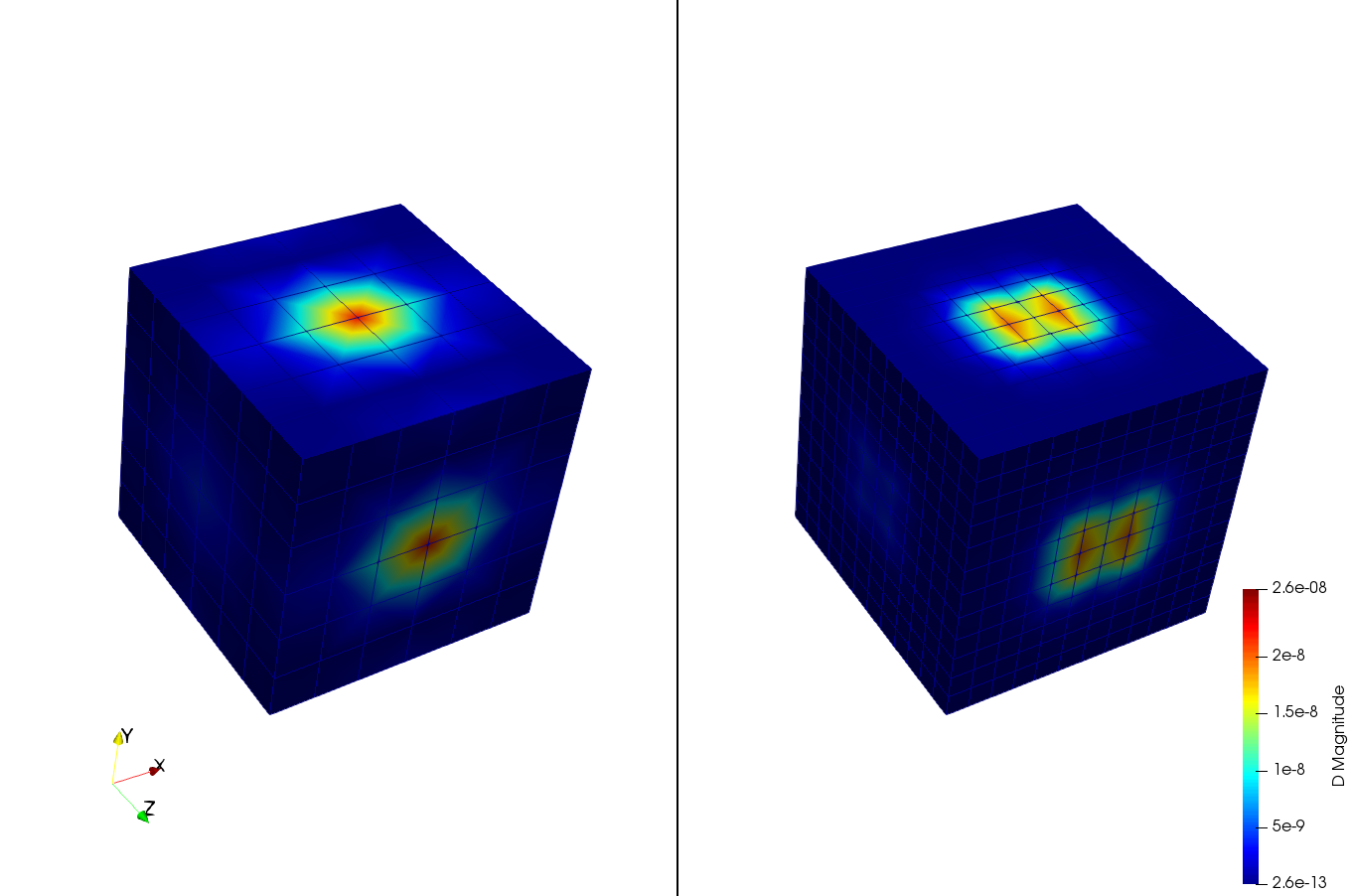}
  \quad
  \includegraphics[width=0.48\textwidth]{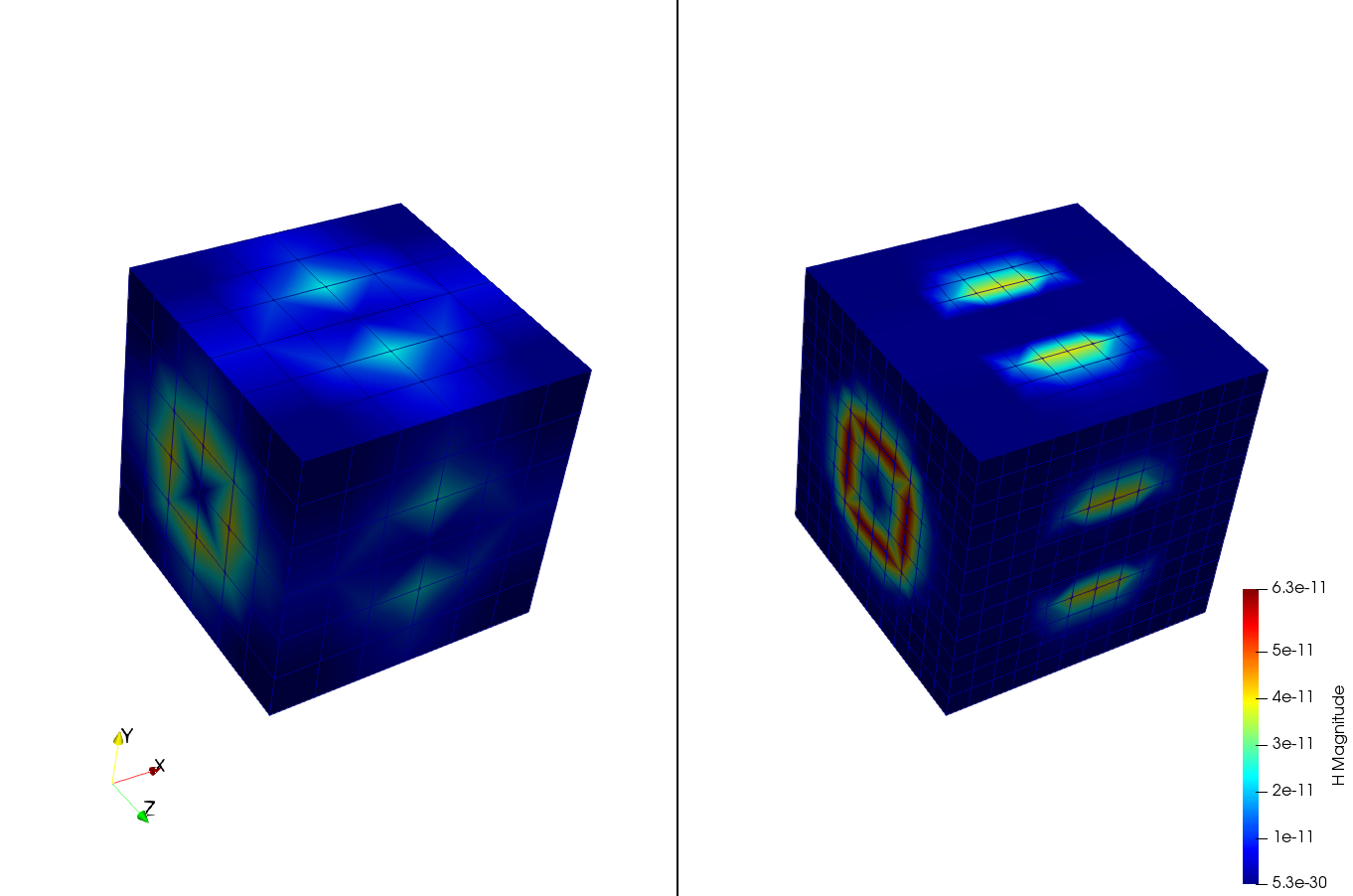}
\caption{Microscale simulation results of a coarse and fine mesh (left and right respectively) for all flux quantities. Top left: mechanical stress $\sigma_{xy}\;[\text{GPa}]$, top right: mechanical stress $\sigma_{xy}\;[\text{GPa}]$ in the $xz$-plane with $y=0$, bottom left: magnitude of the electric displacement field $\bfD\;[\text{A\;s}/\text{m}^2]$, bottom right: magnitude of the magnetic field stength $\bfH\;[\text{A/m}]$.}
\label{tab3}
\end{figure*}

In our first example, we use a healthy bone RVE with the parameters $a=0.32\;\text{mm}$ and $b=0.36\;\text{mm}$, resulting in $\rho_{\mathrm{b}}=29.5\%$. We compare different mesh resolutions. The first RVE consists of two elements in each phase block, resulting in six elements for each spatial direction. The second RVE consists of four elements in each block, resulting in twelve elements for each spatial direction. Here, all degrees of freedom for all corner nodes are restricted to zero and all opposite nodes are linked, to guarantee periodicity. We apply a macroscopic strain $\varepsilon_{yz} = 1 \cdot 10^{-5}$. Figure \ref{tab3} shows the results of the simulations.

Both simulations show quadratic convergence behavior and periodic results. For all quantities, the results between the two different used meshes look nearly identical confirming mesh independence of the results. This is not only fulfilled on the surface of the model, but also in the inner parts, as the slice (top right) shows. It should be noted that since the $\text{FE}^2$ method uses volume averaging, the coarse mesh with only six elements in each spatial direction is sufficient enough to create accurate results for the multiscale method and is exclusively used in the remaining examples of this paper.

\begin{figure}[h]
	\centering
  \includegraphics[width=0.48\textwidth]{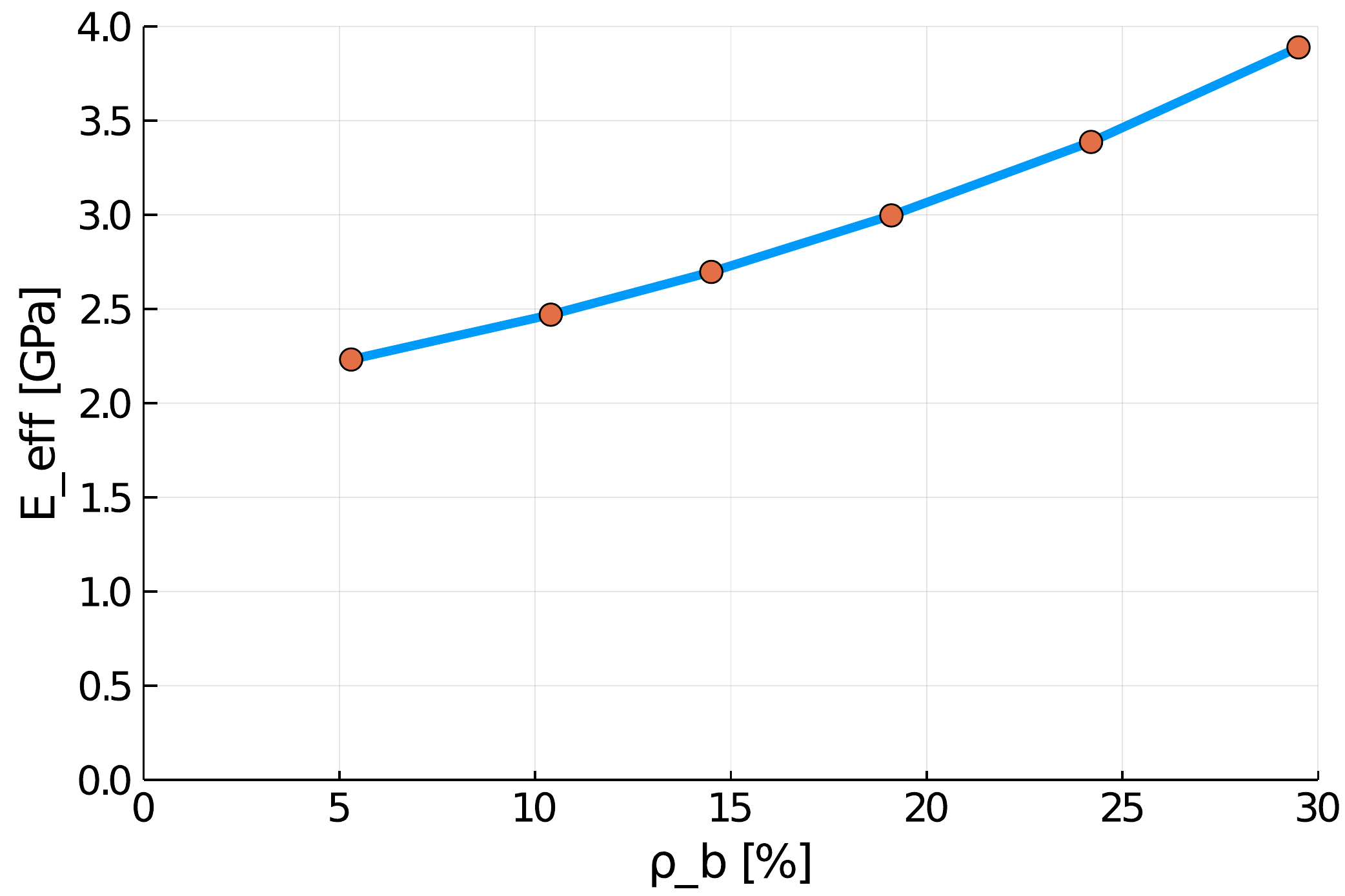}
	\caption{Effective Young's modulus $E_{\mathrm{eff}}$ against cortical bone volume fraction $\rho_{\mathrm{b}}$ for different RVEs.}
	\label{fig5} 
\end{figure}

To compare the model behavior for different stages of osteoporosis, we created RVEs with different volume fractions of cortical bone. Table \ref{tab2} shows the choice of the lengths parameters and the resulting volume fractions. The macroscopic mechanical stiffness tensor $\overline{\dsC} := \frac{\partial \overline{\bfsigma}}{\partial \overline{\bfvarepsilon}}$ was now evaluated for all RVEs by applying a small numerical perturbation as discussed in Section 3.2. We then calculate the effective Young's modulus as 
\be
E_{\mathrm{eff}}=\frac{\overline{\dsC}_{44} (3 \overline{\dsC}_{12} + 2 \overline{\dsC}_{44})}{(\overline{\dsC}_{12} +\overline{\dsC}_{44})} \; \text{.}
\ee
Figure \ref{fig5} shows a plot of the macroscopic Young's modulus against the volume fraction of cortical bone. Here, we observe a drastical reduction of the macroscopic Young's modulus with decreasing cortical bone fraction. Compared to a healthy bone ($\rho_{\mathrm{b}}=29.5\%$), the effective Young's modulus of the degenerated bone ($\rho_{\mathrm{b}}=5.3\%$) decreases to $57\%$ (from $3.89\;\text{GPa}$ to $2.32\;\text{GPa}$). Similar results can be found in \cite{Ilic2010}.

\subsection{Cylinder model}

\begin{figure}[h]
	\centering
  \includegraphics[width=0.48\textwidth]{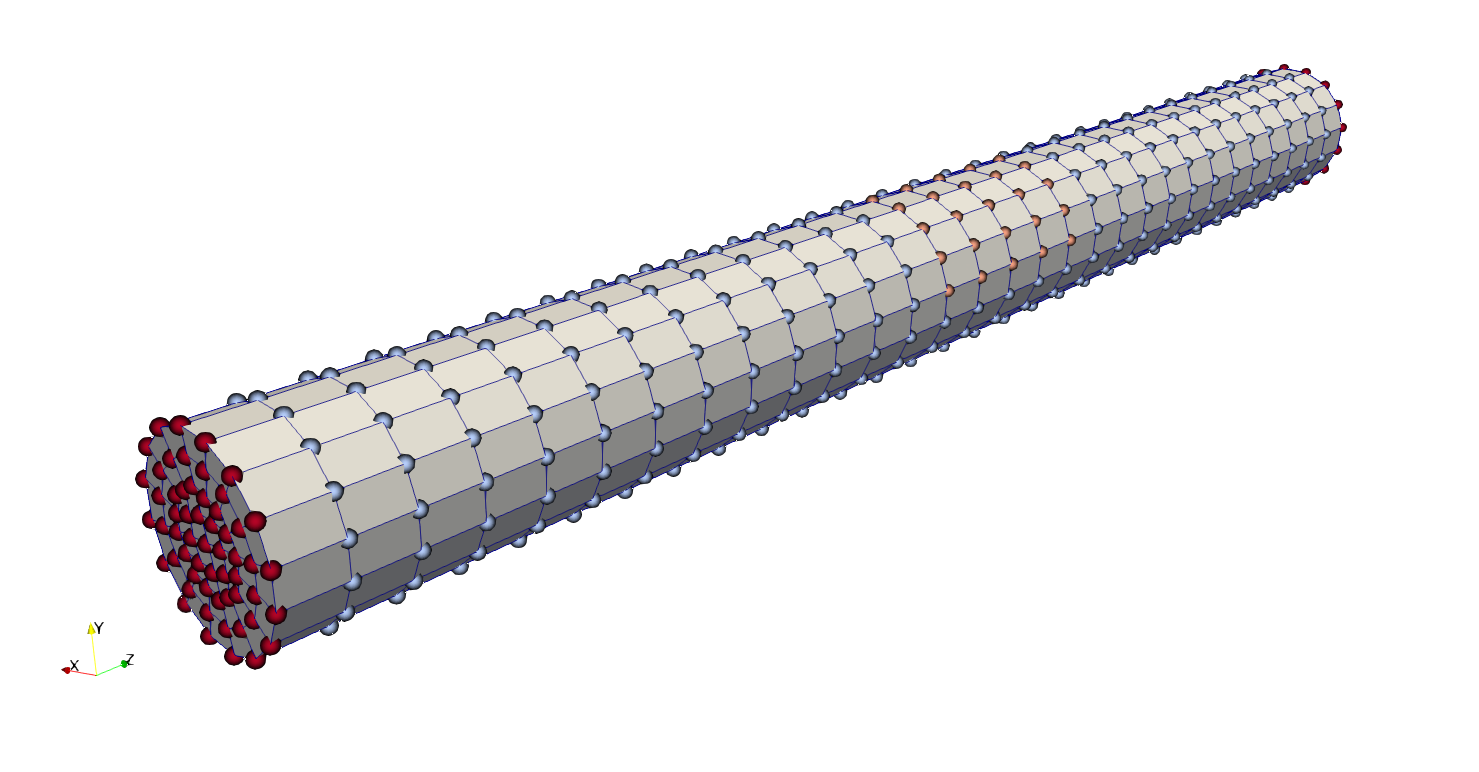}
	\caption{Cylinder mesh and displacement boundary conditions (red: all directions restricted, orange: only the x-direction restricted, blue-gray: no directions restricted).}
	\label{fig6} 
\end{figure}

\begin{figure}[h]
\centering
  \includegraphics[width=0.32\textwidth]{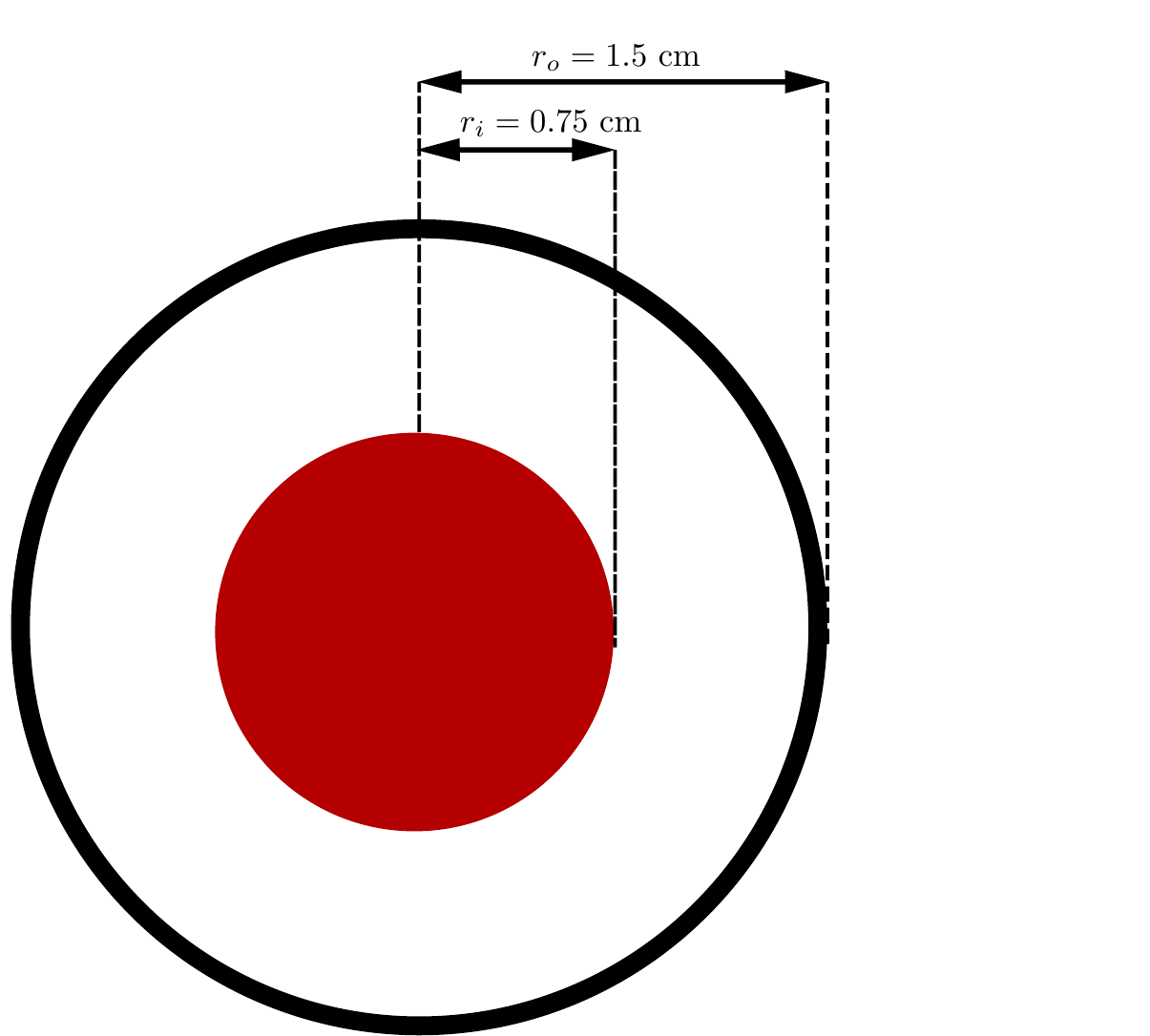}    
\caption{Cylinder front in the $xy$-plane for $z=0$ with grounded nodes in red.}
\label{tab4}
\end{figure}

\begin{figure}[h]
	\centering   \includegraphics[width=0.48\textwidth]{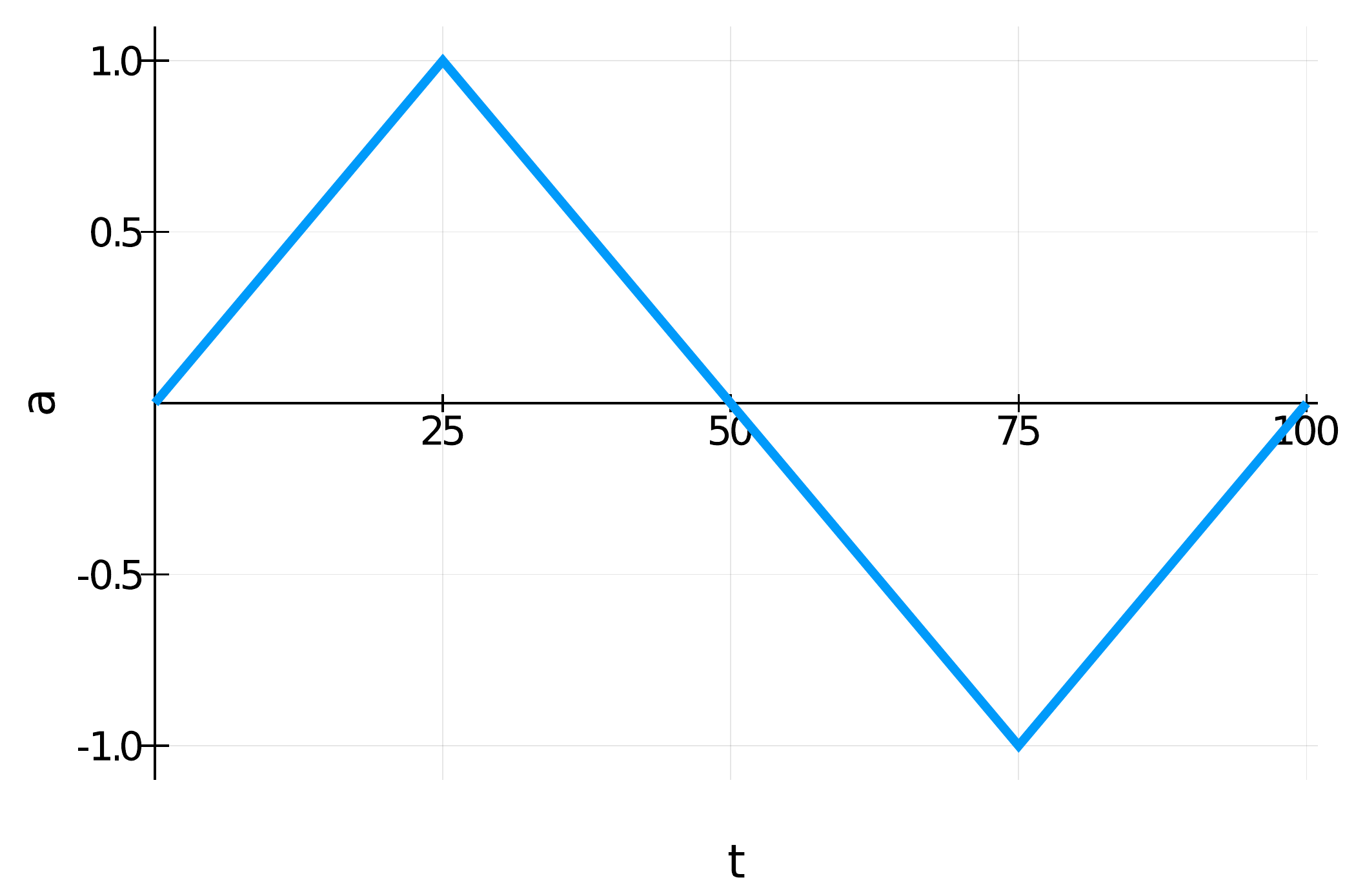} 
	\caption{Amplitude of the displacement function $a$ against the time step $t$. }
	\label{tab4a}
\end{figure}

In this section, we show results for a cylinder model, which has a length of $30\;\text{cm}$ and a diameter of $2 \; r_o = 3\;\text{cm}$. The mesh and the displacement boundary conditions are shown in Figure \ref{fig6}. The mesh consists of $1767$ nodes and $1440$ hexahedral elements. The left and right face is fixed, resulting in the boundary conditions $\bfu = \bfzero$ on the faces. Additionally, in the inner part of the left face ($r < r_i = 0.75 \; \text{cm}$) depicted in Figure \ref{tab4}, the cylinder is assumed to be grounded, resulting in $\varphi = 0$ and $\bfA =\bfzero$. We apply a time-dependent mechanical displacement in x-direction $u_x = u_{\max} \cdot a(t)$, with $u_{\max} = 2 \cdot 10^{-3}$, to the middle part of the cylinder and calculate $100$ time steps. Figure \ref{tab4a} shows the amplitude of the displacement function $a$ versus the time $t$. For the simulations, the time increment $\Delta_\mathrm{t} = 1 \cdot 10^{-4}\;\text{s}$ is used.

\begin{figure}[h!]
\centering
  \includegraphics[width=0.48\textwidth]{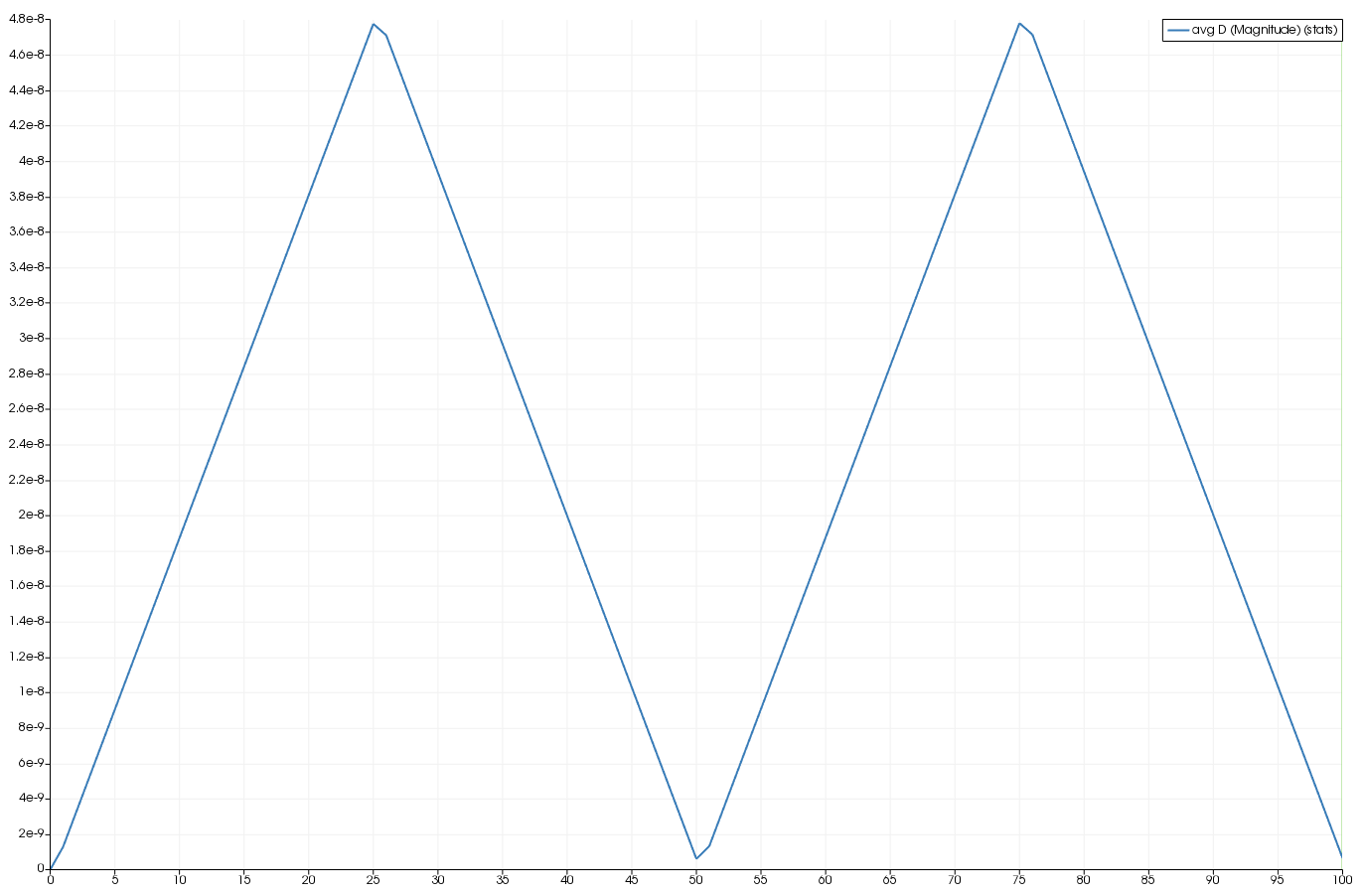}
\caption{Magnitude of the average electric displacement field $\bfD\;[\text{As}/\text{m}^2]$, plotted against the time $t$.}
\label{tab5}
\end{figure}

\begin{figure}[h!]
\centering 		
	\includegraphics[width=0.48\textwidth]{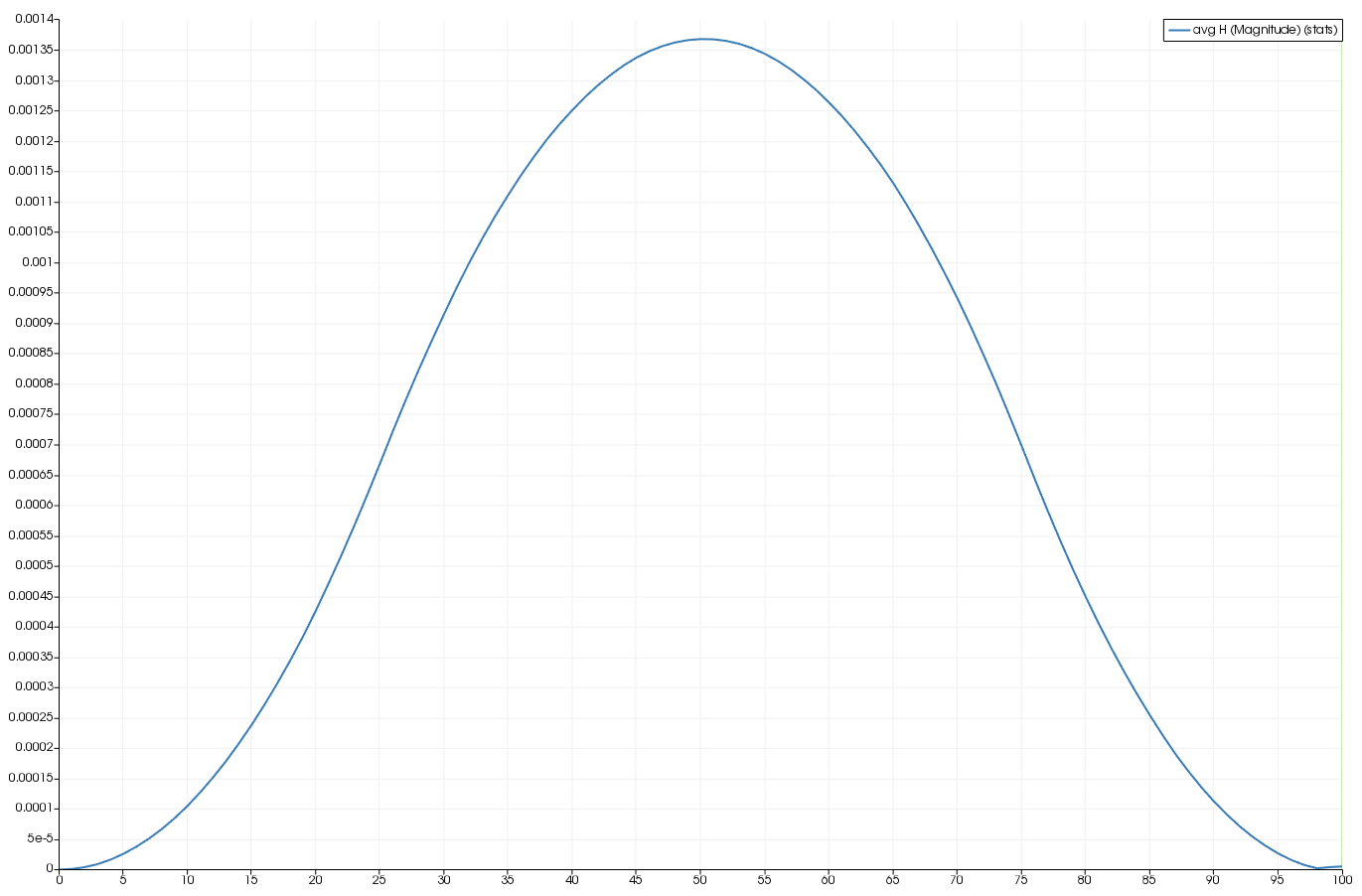}
\caption{Magnitude of the average magnetic field strength $\bfH\;[\text{A}/\text{m}]$, plotted against the time $t$.}
	\label{tab5a}
\end{figure}

First, we examine the simulation results for the heal\-thy bone (RVE 6, $\rho_{\mathrm{b}}=29.5\%$). Here, we observe qua\-dra\-tic convergence behavior for the macroscale as well. Figures \ref{tab5} and \ref{tab5a} show the magnitude of the average electric displacement field $\bfD$ and the magnitude of the average magnetic field stength $\bfH$, respectively, plotted against time $t$. The history of the average electric displacement field mimics the displacement boundary condition. Thus, the electric displacement field is caused mainly by the piezoelectric effect of the cortical bone material phase. In contrast, the magnitude of the average magnetic field strength increases until time $t=50$, where the maximum is reached. Then, the magnitude decreases again and at the end of the simulation, nearly no magnetic field is present. We conclude, that the magnetic field is caused mainly by the time change of the electric displacement field as described by the Maxwell equations.

\begin{figure*}[h!]
	\centering
  \includegraphics[width=\textwidth]{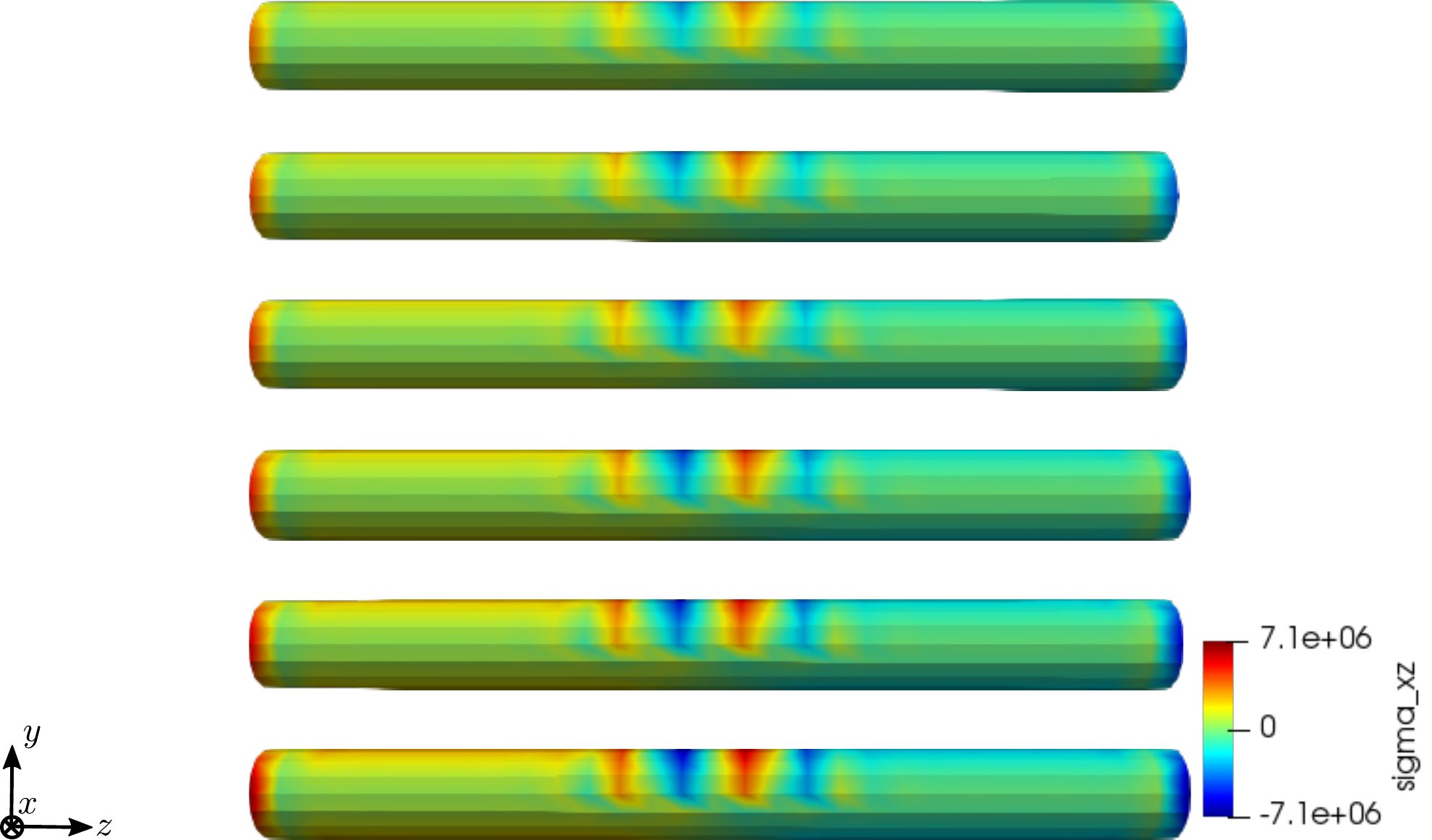}
	\caption{Simulation results for RVE 1 (top) to 6 (bottom): stress $\sigma_{xz}\;[\text{GPa}]$, $t=25$.}
	\vspace{0.5cm}
	\label{fig7} 
\end{figure*}

\begin{figure*}[h!]
	\centering
  \includegraphics[width=\textwidth]{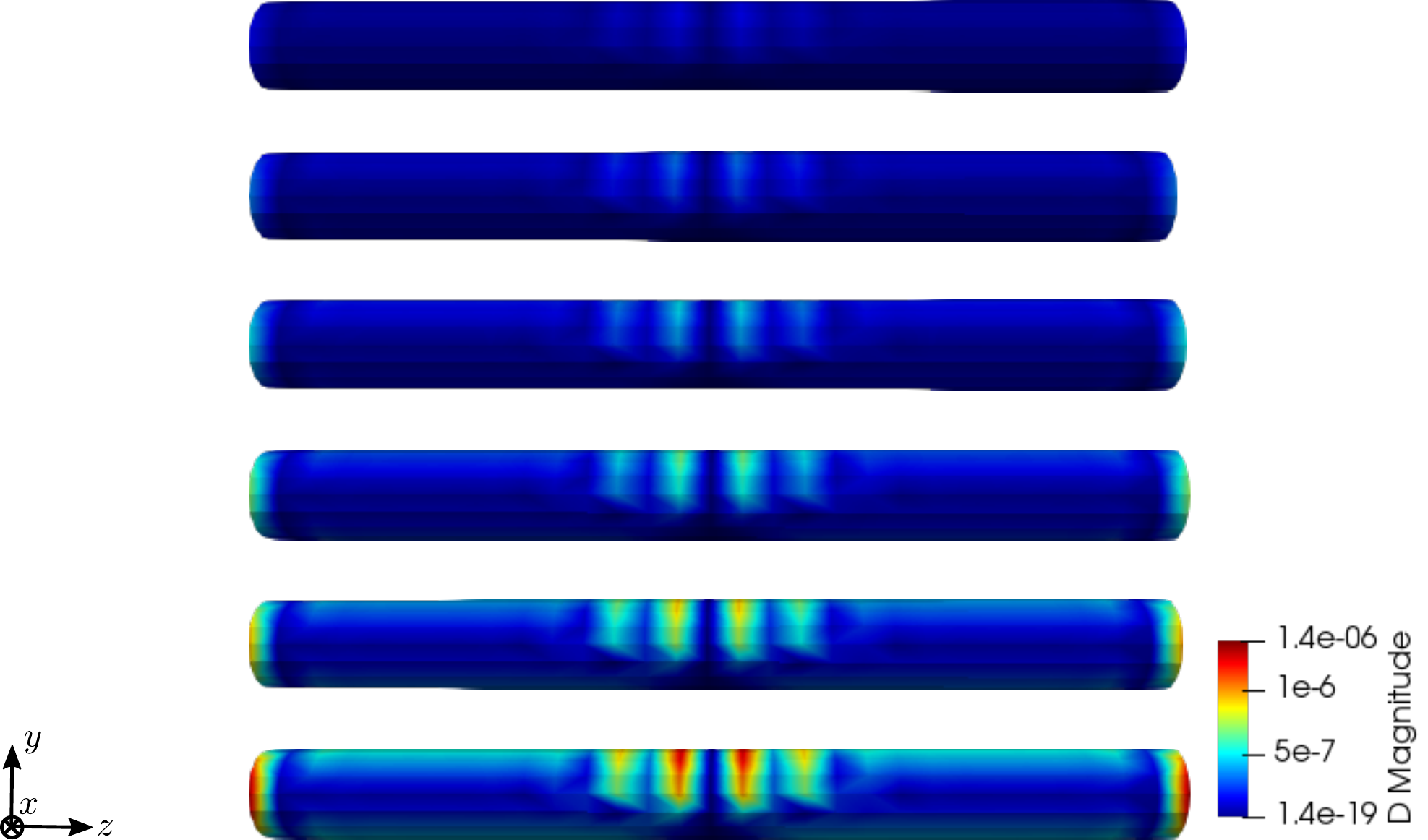}
	\caption{Simulation results for RVE 1 (top) to 6 (bottom): magnitude of the electric displacement field $\bfD\;[\text{As}/\text{m}^2]$, $t=25$.}
	\label{fig8} 
\end{figure*}

\begin{figure*}[h!]
	\centering
  \includegraphics[width=\textwidth]{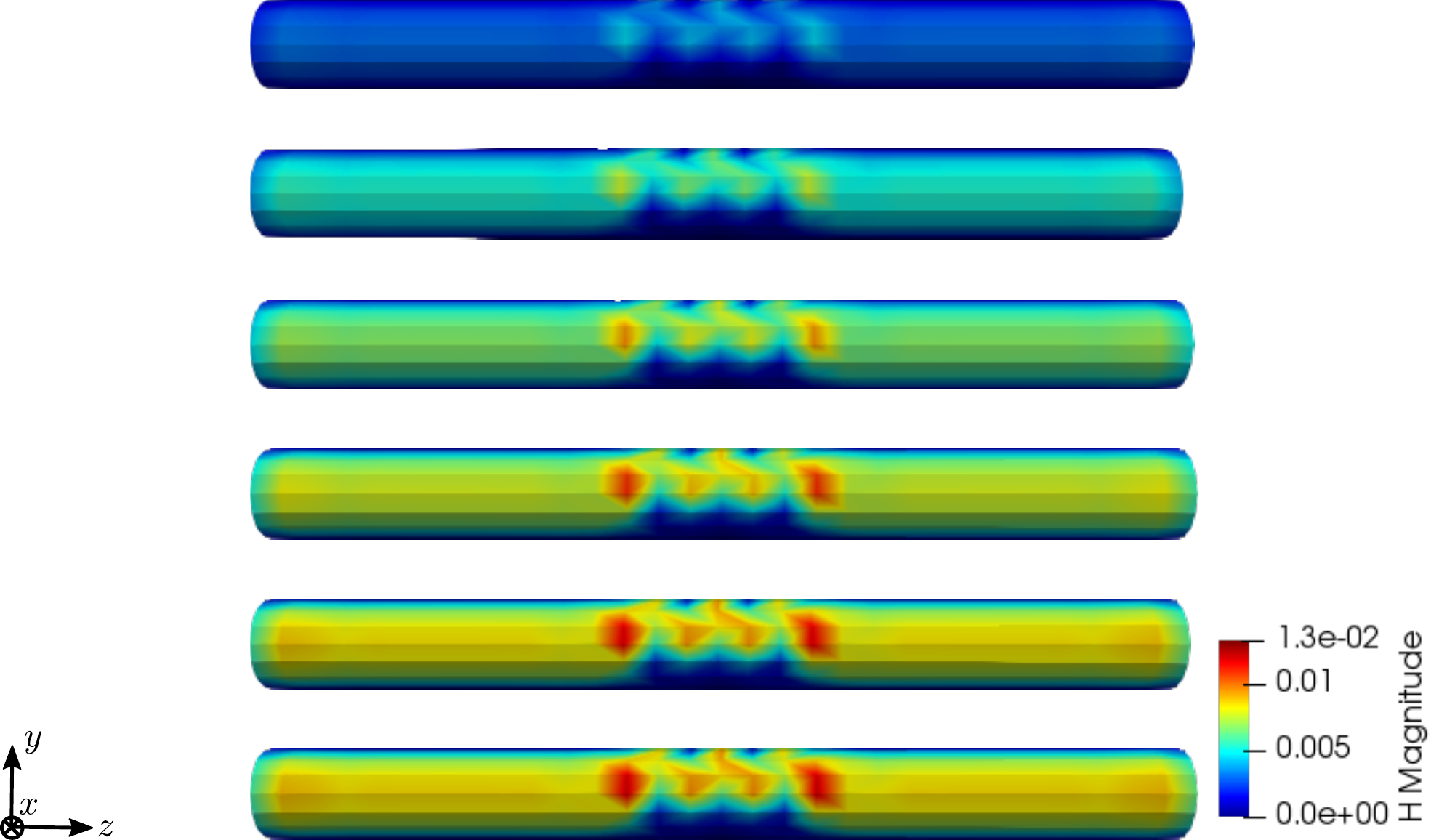}
	\caption{Simulation results for RVE 1 (top) to 6 (bottom): magnitude of the magnetic field strength $\bfH \; [\text{A}/\text{m}]$, $t=50$.}
	\vspace{0.5cm}
	\label{fig9} 
\end{figure*}

\begin{figure*}[h!]
	\centering
  \includegraphics[width=\textwidth]{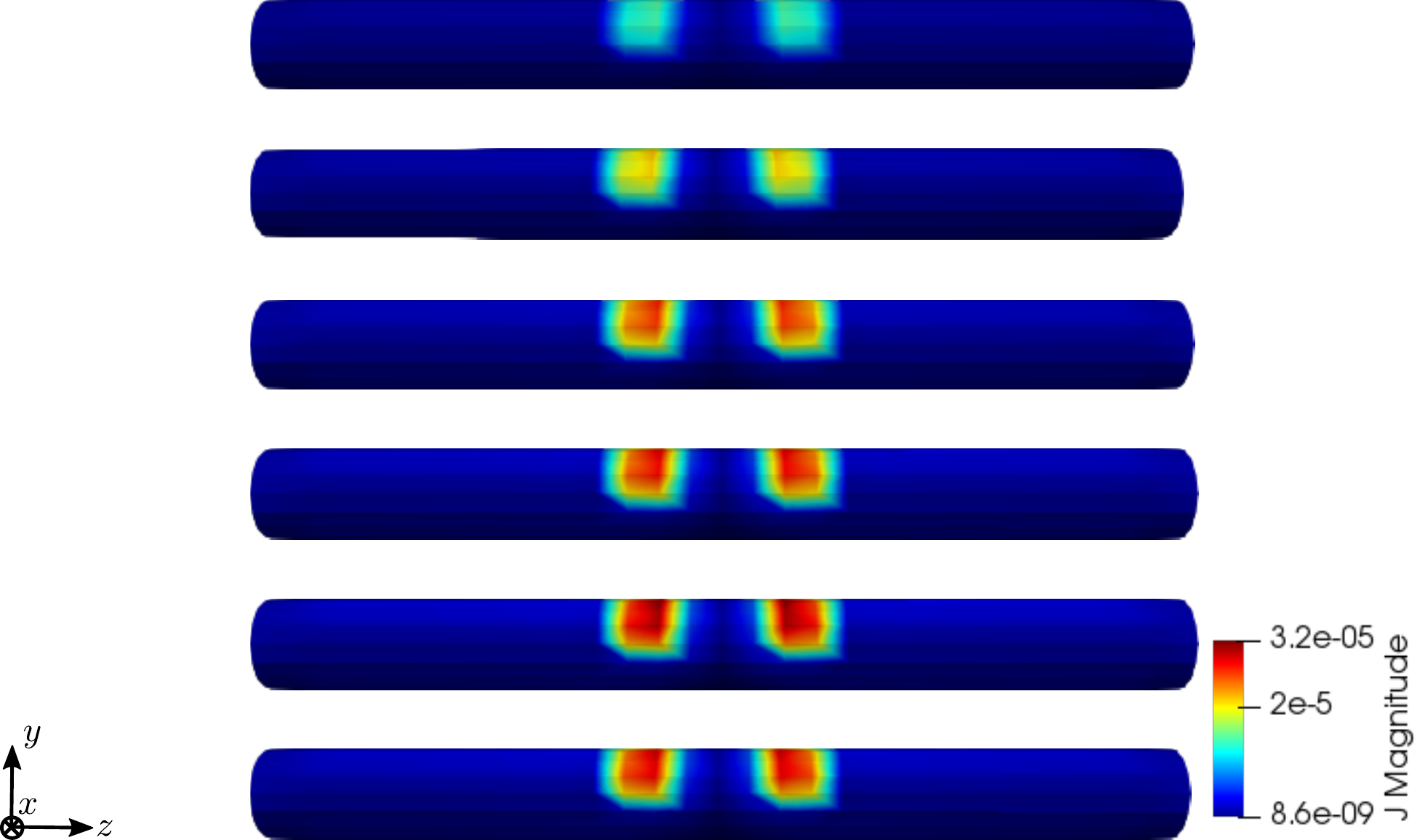}
	\caption{Simulation results for RVE 1 (top) to 6 (bottom): magnitude of the electric current density $\bfJ \; [\text{A}/\text{m}^2]$, $t=50$.}
	\label{fig10} 
\end{figure*}

To compare the different stages of osteoporosis, we use different RVEs (Table \ref{tab2}). The simulation results are shown in Figures \ref{fig7} to \ref{fig10}. Here, the number of the specific RVE increases from top to bottom.

\begin{figure*}[h!]
	\centering
	\includegraphics[width=\textwidth]{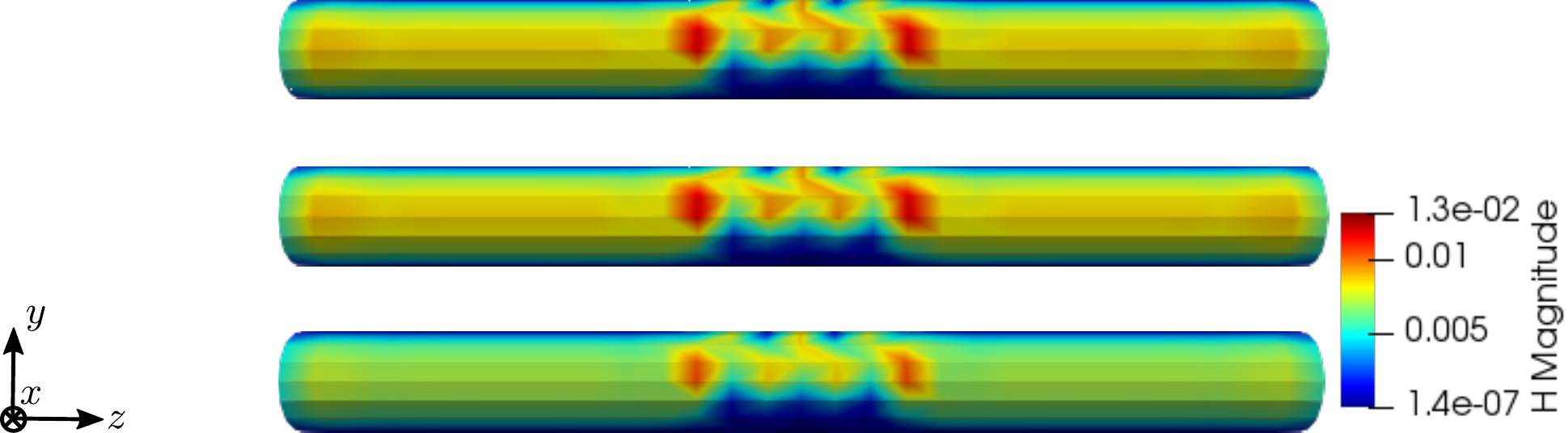}
	\caption{Simulation results for RVE 1 for the magnetic field strength $\bfH \; [\text{A}/\text{m}]$ with $\kappa_1  = 1 \cdot 10^{2} \; \text{S}/\text{m}$ (top), $\kappa_1  = 1 \cdot 10^{4} \; \text{S}/\text{m}$ (in the middle) and $\kappa_1  = 1 \cdot 10^{6} \; \text{S}/\text{m}$ (bottom), $t=50$.}
	\label{fig11} 
\end{figure*}

\begin{figure*}[h!]
	\centering
	\includegraphics[width=\textwidth]{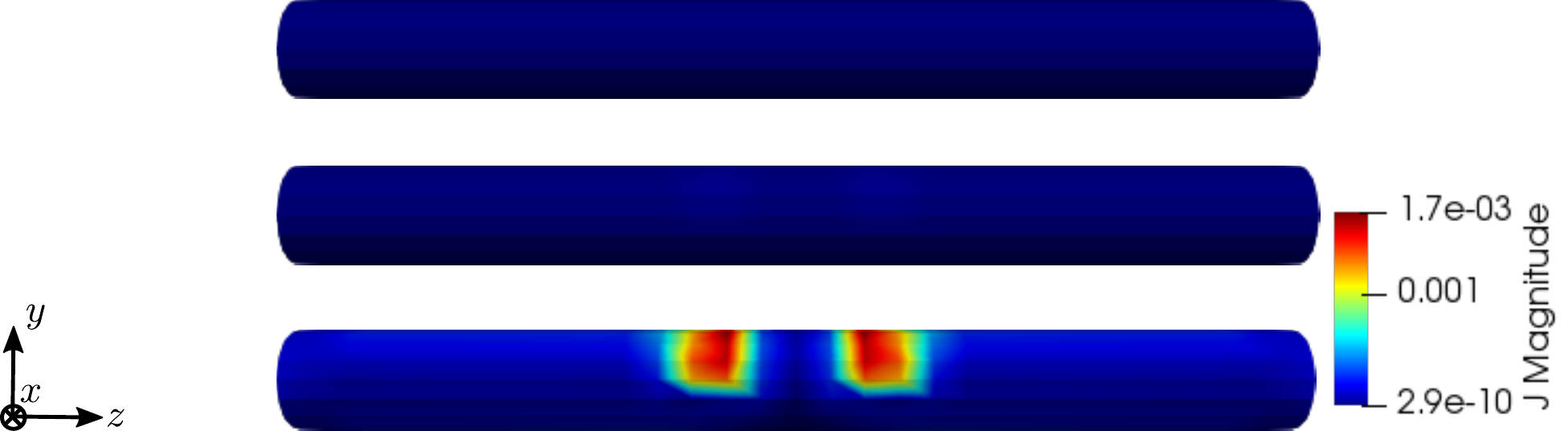}
	\caption{Simulation results for RVE 1 for the electric current density $\bfJ \; [\text{A}/\text{m}^2]$ with $\kappa_1  = 1 \cdot 10^{2} \; \text{S}/\text{m}$ (top), $\kappa_1  = 1 \cdot 10^{4} \; \text{S}/\text{m}$ (in the middle) and $\kappa_1  = 1 \cdot 10^{6} \; \text{S}/\text{m}$ (bottom), $t=50$.}
	\label{fig11a} 
\end{figure*}

As an additional example for the cylinder model, we performe a parameter study for the electric conductivity parameter $\kappa_1$, aiming to understand the interaction between the time derivative of the electric displacement field and the electric current density in the Maxwell equation. Figures \ref{fig11} and \ref{fig11a} show the results for RVE 1 and $\kappa_1 \in \{1 \cdot 10^{2} \; \text{S}/\text{m}, 1 \cdot 10^{4} \; \text{S}/\text{m}, 1 \cdot 10^{6} \; \text{S}/\text{m} \}$. 




For all quantities, we observe an increase for RVEs with higher volume fractions of cortical bone. Additionally, the difference between the RVEs is greater, the lower the volume fraction of cortical bone is. While the difference is barely noticable between RVE 5 and 6, the change of all quantities excluding the stress is distinct between RVEs 1 and 2. Qualitatively, we notice similar results between the different RVEs. 

Regarding the parameter study of the electric conductivity, we observe nearly identical results for the magnetic field strength $\bfH$ for the first two choices of $\kappa_1$, but a slight decrease for $\kappa_1 = 1 \cdot 10^{6} \; \text{S}/\text{m}$. The electric current density $\bfJ$ increases proportionally to the increase of the material parameter. Thus, for the first two choices of $\kappa_1$, nearly no electric current is visible. Regarding the magnetic field strength, the results again are qualitatively similar, suggesting that both parts of the Maxwell equation $\dot{\boldface{D}}$ and $\bfJ$ evolve in the same way. Here, the electric current density acts in the opposite direction of the time derivative of the electric displacement field, resulting in a slight decrease of the magnetic field strength. However, this effect is only relevant for very high values of the electric conductivity, as otherwise the contribution of $\dot{\boldface{D}}$ prevails.

\subsection{True to scale bone model}

\begin{figure}[h]
\centering
	\includegraphics[width=0.48\textwidth]{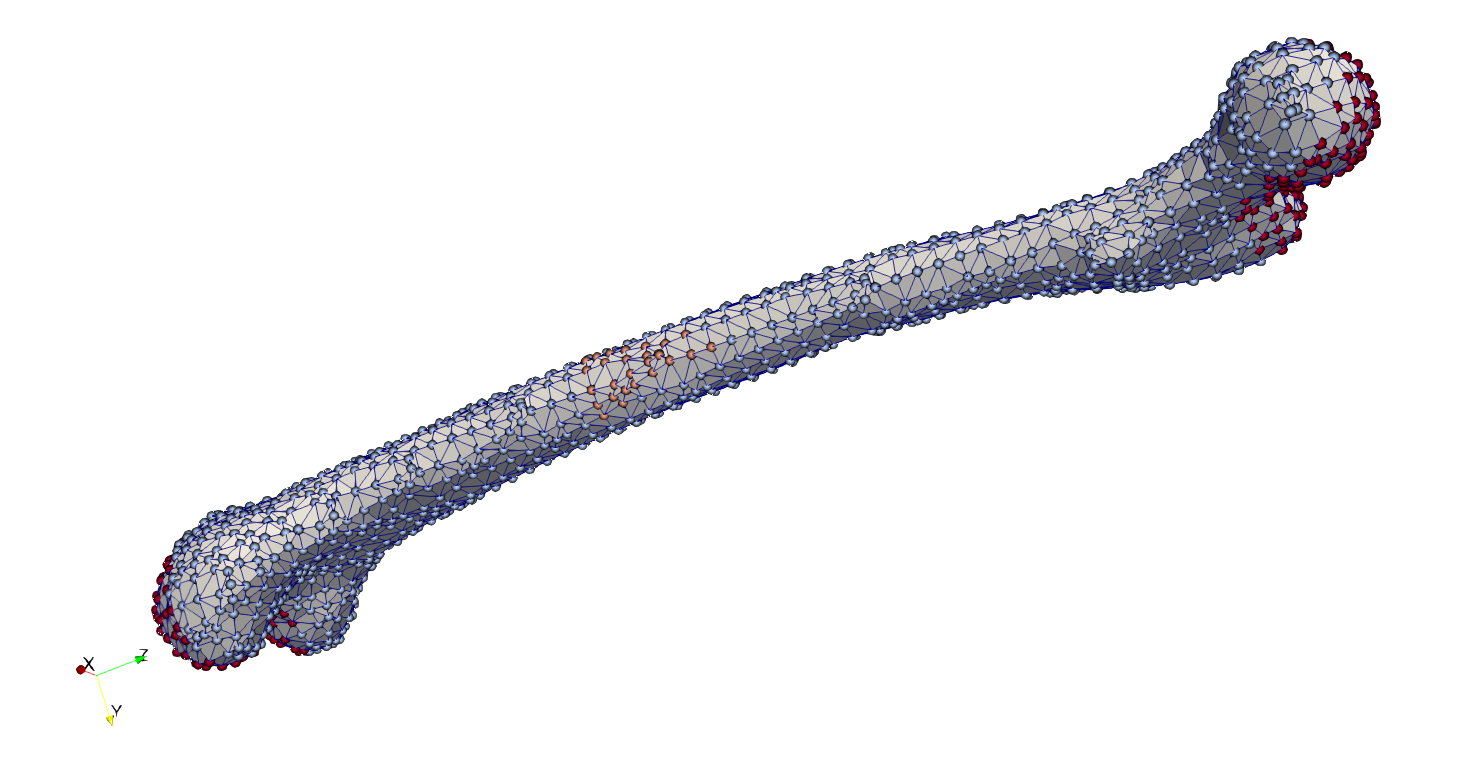}
\caption{Femur bone mesh and displacement boundary conditions (red: all directions restricted, orange: only the x-direction restricted, blue-gray: no directions restricted).}
\label{tab6} 
\end{figure}

\begin{figure}[h]
	\centering
		 \includegraphics[width=0.48\textwidth]{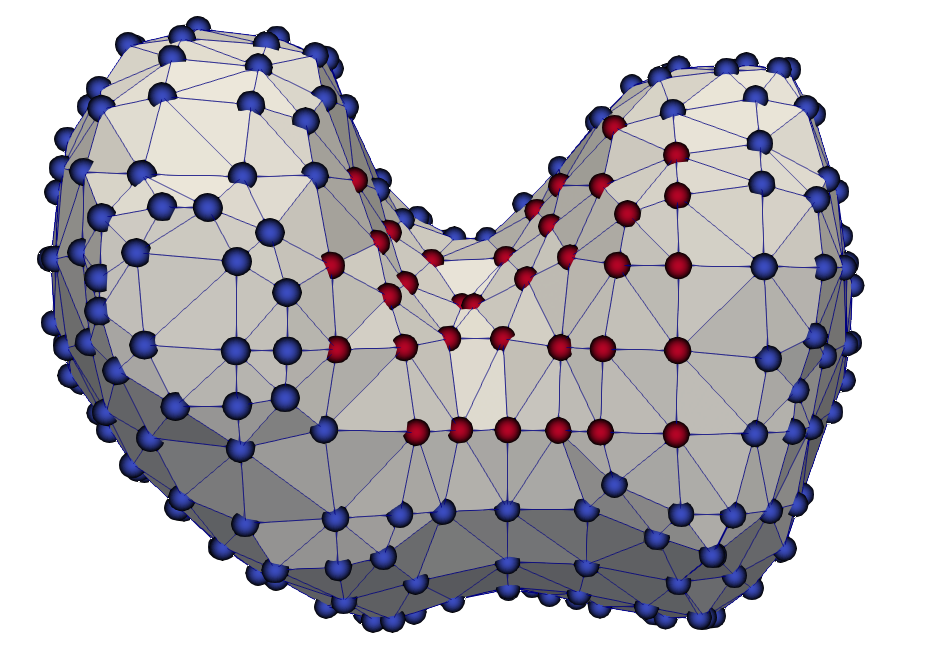} 
	\caption{Femur bone front with grounded nodes in red.}
	\label{tab6a}
\end{figure}

\begin{figure*}[h!]
	\centering
  \includegraphics[width=0.95\textwidth]{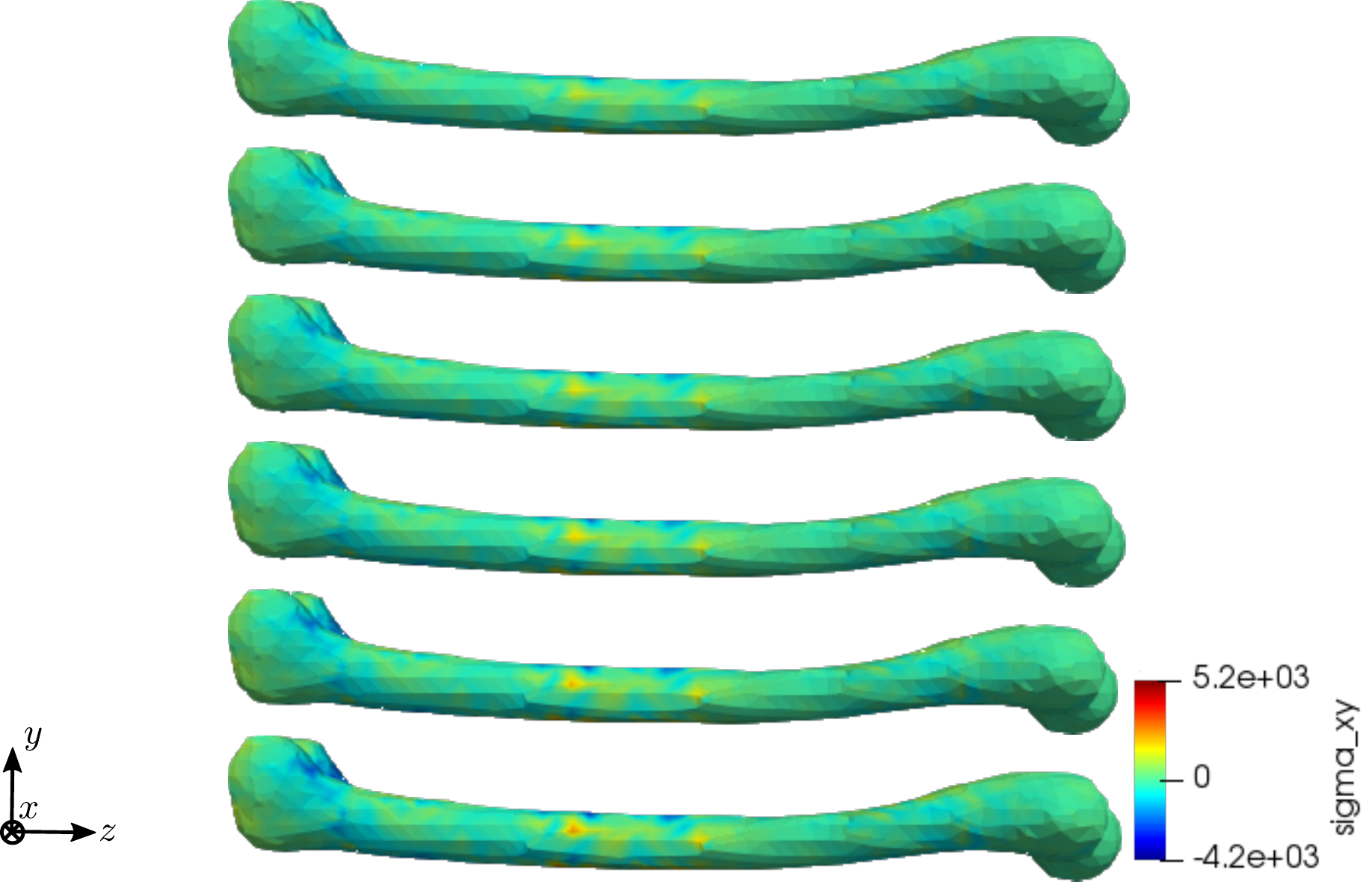}
	\caption{Simulation results for RVE 1 (top) to 6 (bottom): stress $\sigma_{xy} \; [\text{GPa}]$, $t=25$.}
	\label{fig12} 
\end{figure*}

\begin{figure*}[h!]
	\centering
  \includegraphics[width=0.95\textwidth]{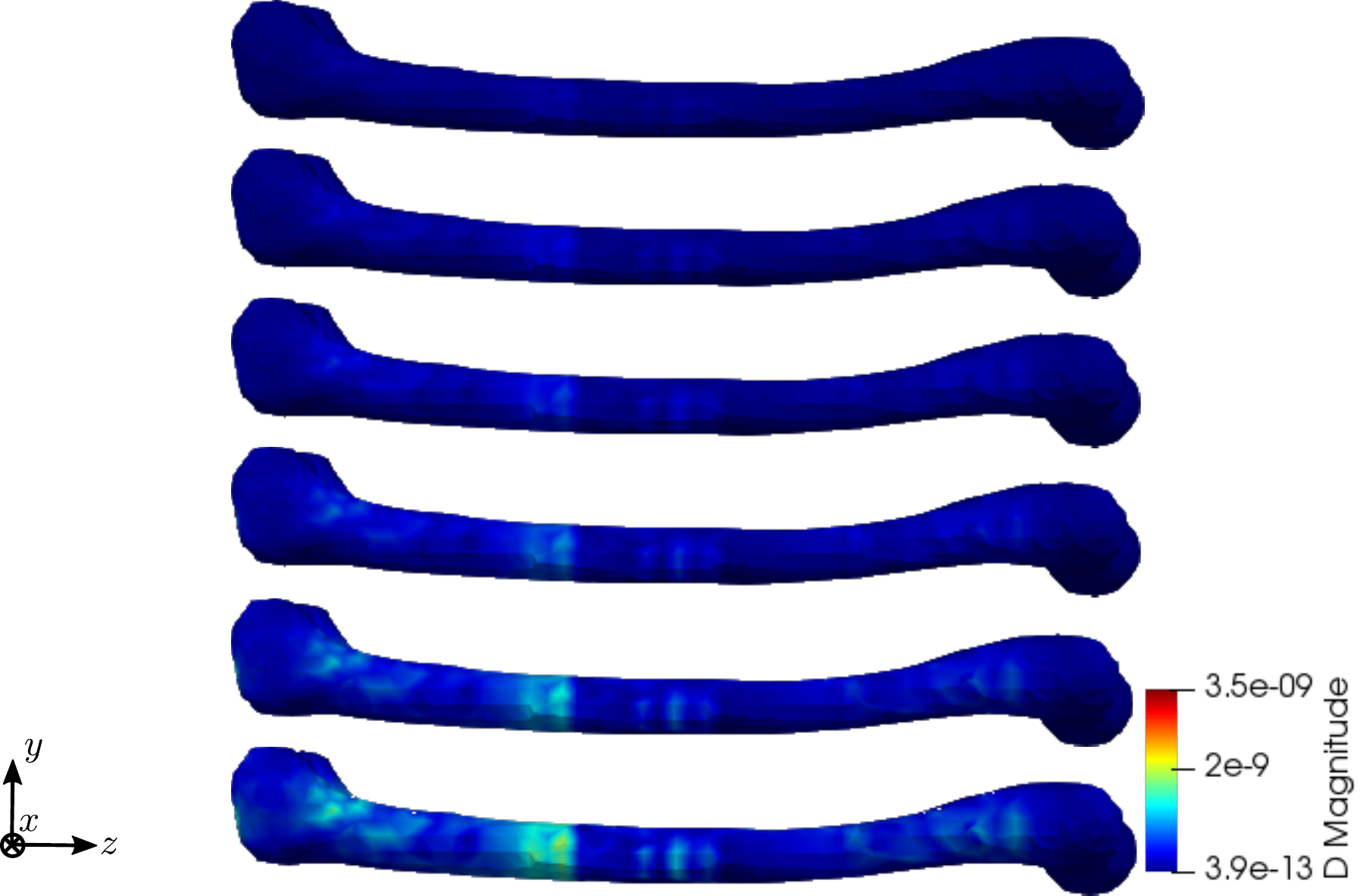}
	\caption{Simulation results for RVE 1 (top) to 6 (bottom): magnitude of the electric displacement field $\bfD \;  [\text{As}/\text{m}^2]$, $t=25$.}
	\label{fig13} 
\end{figure*}

\begin{figure*}[h!]
	\centering
  \includegraphics[width=0.95\textwidth]{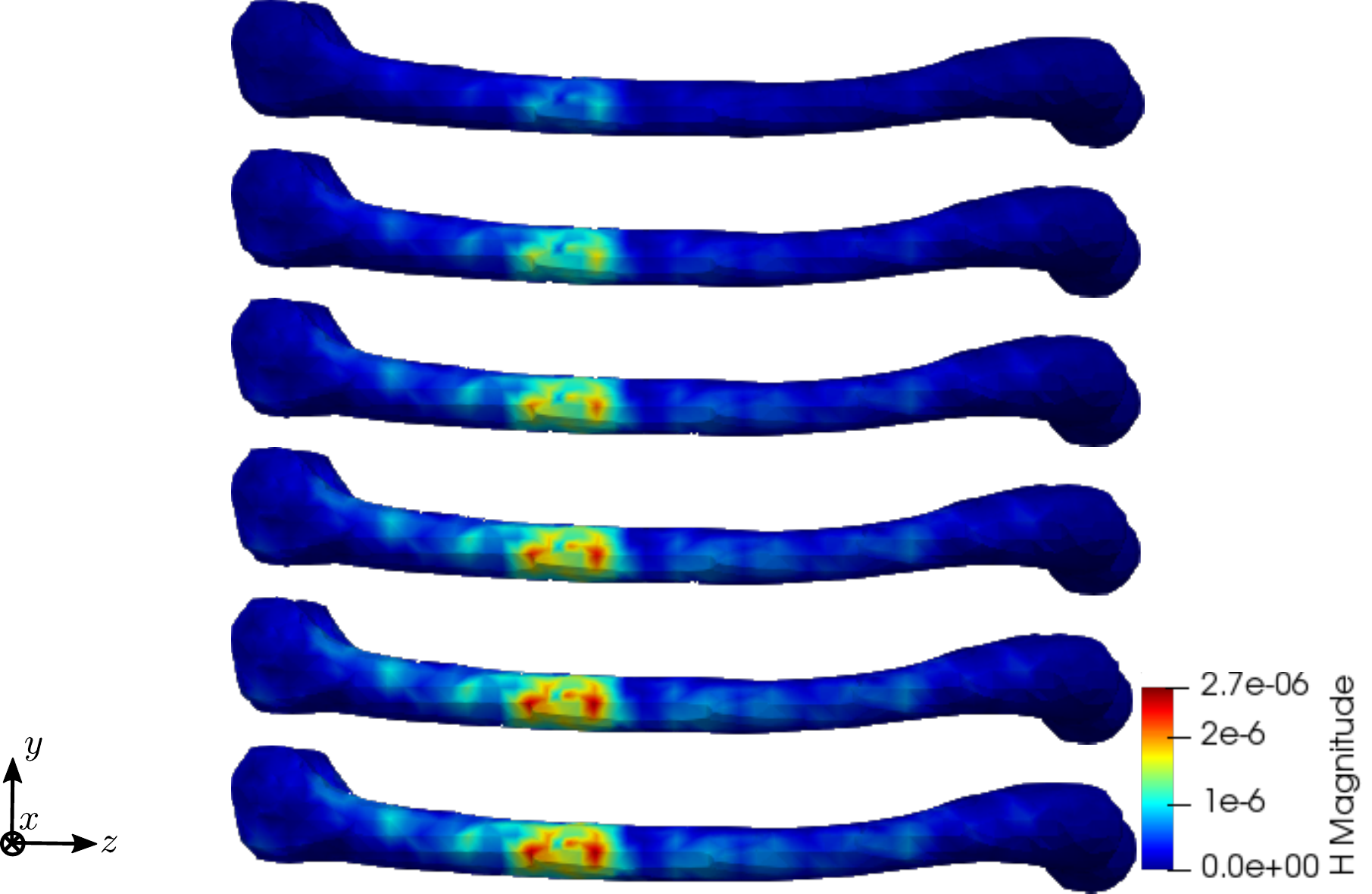}
	\caption{Simulation results for RVE 1 (top) to 6 (bottom): magnitude of the magnetic field strength $\bfH \; [\text{A}/\text{m}]$, $t=50$.}
	\label{fig14} 
\end{figure*}

\begin{figure*}[h!]
	\centering
  \includegraphics[width=0.95\textwidth]{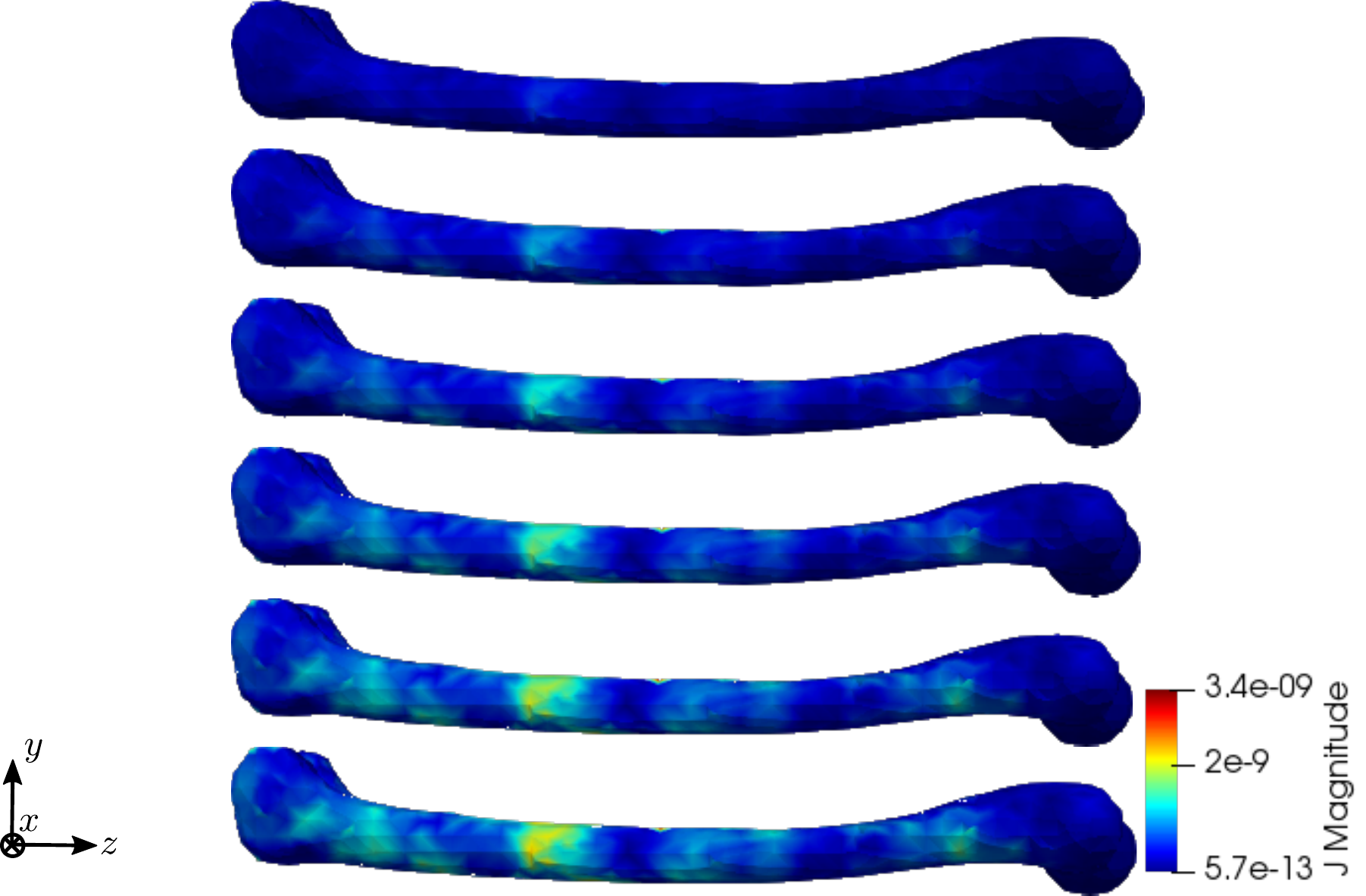}
	\caption{Simulation results for RVE 1 (top) to 6 (bottom): magnitude of the electric current density $\bfJ \; [\text{A}/\text{m}^2]$, $t=50$.}
	\label{fig14a} 
\end{figure*}

\begin{figure*}[h!]
	\centering
  \includegraphics[width=0.95\textwidth]{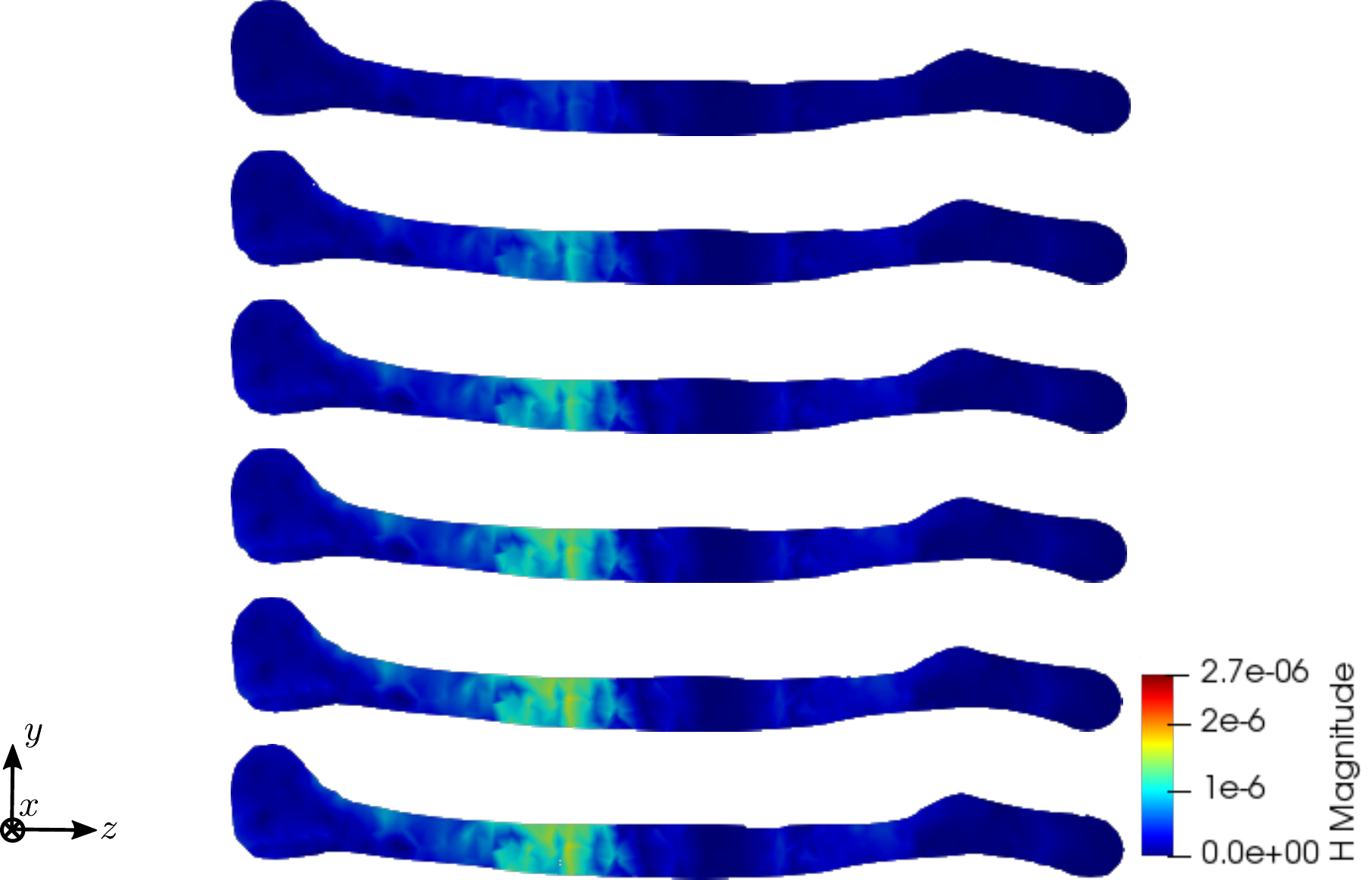}
	\caption{Simulation results for RVE 1 (top) to 6 (bottom): magnitude of the magnetic field strength $\bfH \; [\text{A}/\text{m}]$, slice, $t=50$.}
	\label{fig15} 
\end{figure*}

\begin{figure*}[h!]
	\centering
  \includegraphics[width=0.95\textwidth]{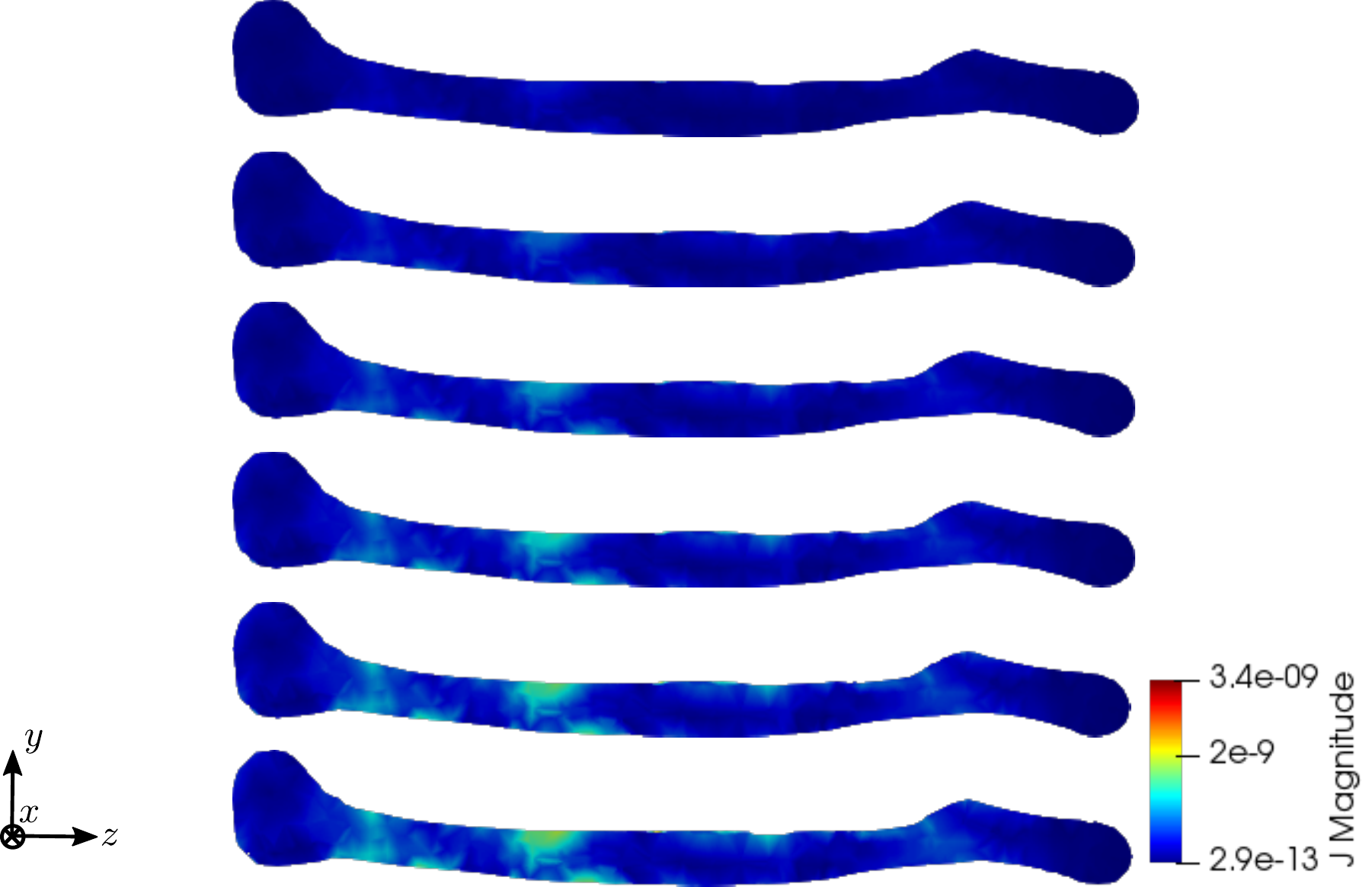}
	\caption{Simulation results for RVE 1 (top) to 6 (bottom): magnitude of the electric current density $\bfJ \; [\text{A}/\text{m}^2]$, slice, $t=50$.}
	\label{fig16} 
\end{figure*}

We examin a true to scale model of a human femur bone from \cite{lifesciencedbwebsite} and slightly modify it by using the software \textsc{Blender} \cite{blenderwebsite}, improving the mesh. Again the model has a length of about $30\;\text{cm}$. The mesh and the displacement boundary conditions are shown in Figure \ref{tab6}. The mesh consists of $1660$ nodes and $4944$ tetrahedral elements. The grounded nodes are shown in Figure \ref{tab6a}. Again, we apply the mechanical displacement depicted in (Figure \ref{tab4a}) with $u_{\max} = 2 \cdot 10^{-6}$ to the middle section and calculate $100$ time steps. Here, the time increment $\Delta_\mathrm{t} = 1 \cdot 10^{-2}\; \text{s}$ is used for the simulations. To compare different stages of osteoporosis, we use again different RVEs (Table \ref{tab2}). Figures \ref{fig12} to \ref{fig16} show the results.

\begin{figure}[h]
\centering
  \includegraphics[width=0.42\textwidth]{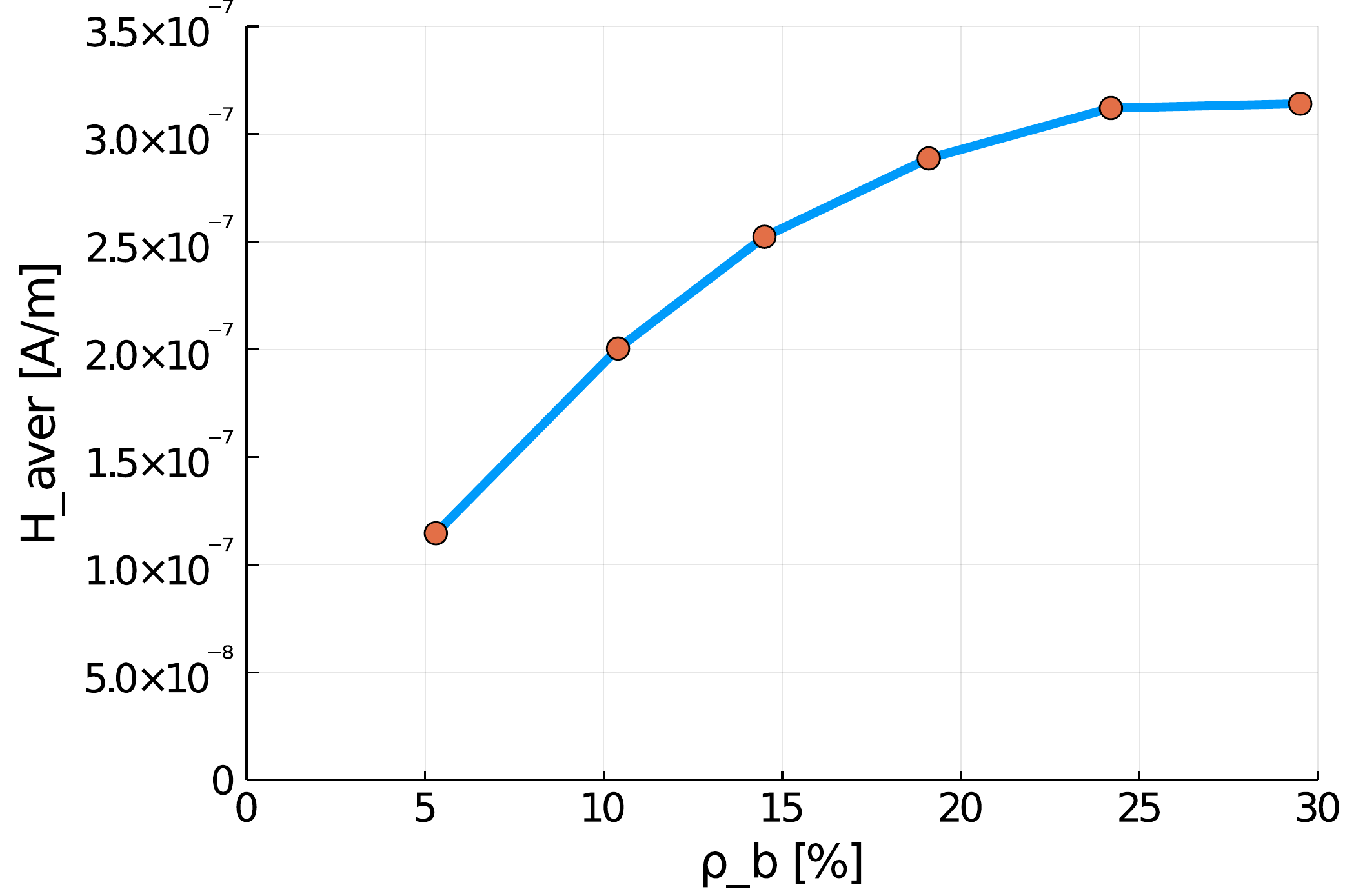}
\caption{Average magnetic field stength for the different RVEs at $t=50$.}
\label{tab7}
\end{figure}

\begin{figure}[h]
\centering 		 
	\includegraphics[width=0.42\textwidth]{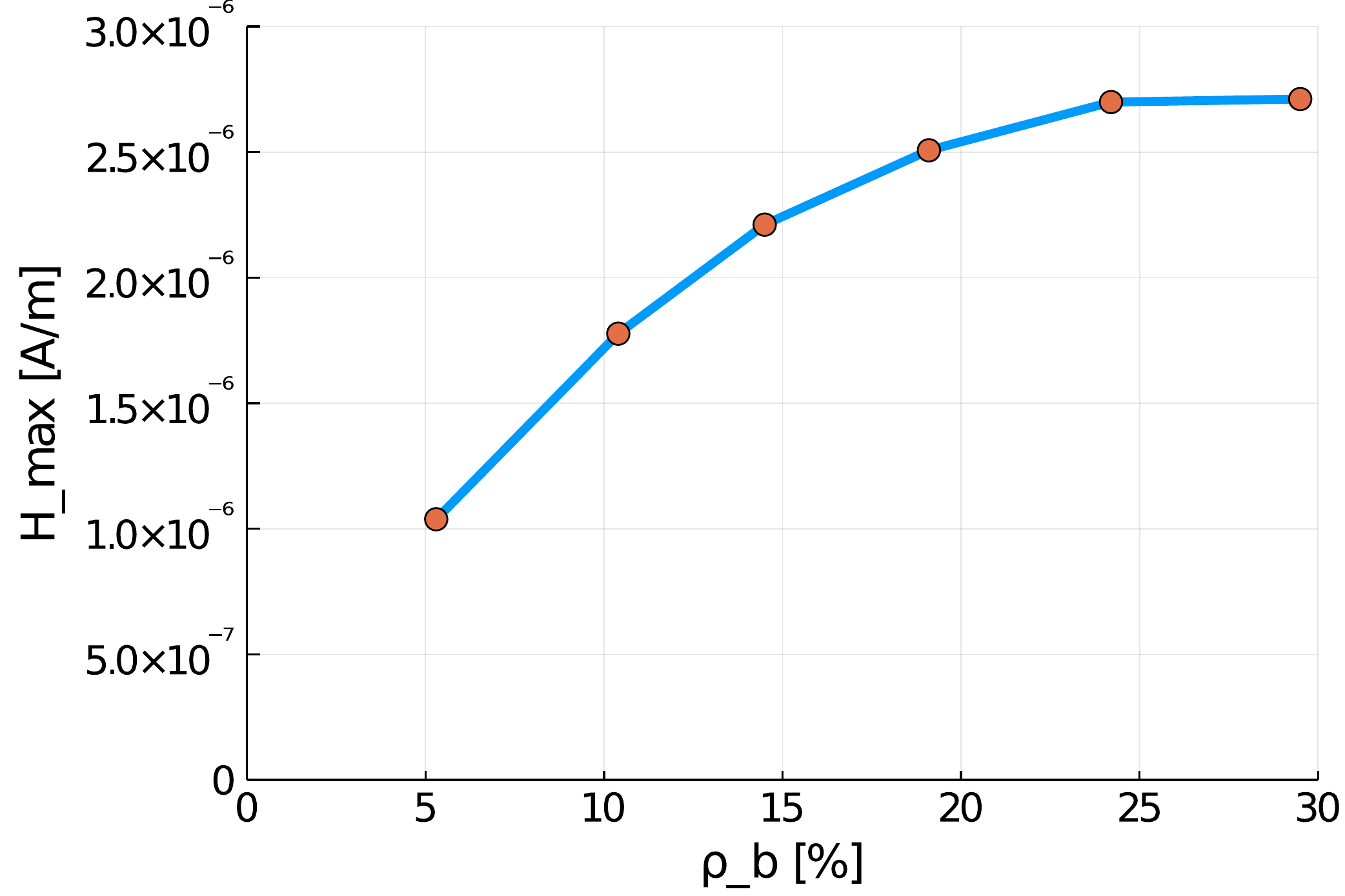} 
\caption{Maximum magnetic field stength for the different RVEs at $t=50$.}
	\label{tab7a}
\end{figure}

Again, the simulations show qualitatively similar results, but a significant increase for all quantities the higher the cortical bone volume fraction is. Additionally, the difference between the RVEs is smaller the higher the volume fraction of cortical bone is. Thus, both the functionality of the bone and the results of the sonography are only slightly affected at earlier stages of osteoporosis, but significantly at later ones. This confirms the disease as being often imperceptible for many subjects at earlier stages. This is especially important regarding the magnetic field strength $\bfH$, as it is the quantity measured at sonography-aided early detection. To further examine the results, we calculate the average and maximum magnetic field strength $\bfH$ at time step $t=50$ for the different RVEs. The results are shown in Figures \ref{tab7} and \ref{tab7a}.

Here, for both quantities a similar behavior can be observed. While there is nearly no reduction between the two RVEs with the highest volume fraction of cortical bone, the difference between the single RVEs increases for lower volume fractions of cortical bone. The average magnetic field strength reduces for the ill bone ($\rho_{\mathrm{b}}=5.3\%$) to $36.5\%$ compared to the healthy bone ($\rho_{\mathrm{b}}=29.5\%$), from $3.14 \cdot 10^{-7} \; \text{A}/\text{m}$ to $1.15 \cdot 10^{-7} \; \text{A}/\text{m}$. The maximum magnetic field strength for the healthy bone is $2.711 \cdot 10^{-6} \; \text{A}/\text{m}$, while the maximum for the degenerated bone is only $1.038 \cdot 10^{-6} \; \text{A}/\text{m}$. This equals a reduction to $38.2\%$. These results show the order of magnitude to be expected for the results of experimental research. For advanced stages of osteoporosis, sonography should measure a magnetic field strength, whose magnitude is only about one third compared to a healthy bone.

%%%%%%%%%%%%%%%%%%%%
\section{Conclusion and outlook}
In this contribution, we present a fully coupled multiscale model for cancellous bone considering mechanical, electrical and magnetic effects. We model bone as a two-phase material with the cortical bone phase assumed as a piezoelectric, insulating solid and the bone marrow phase described as a viscoelastic, conducting solid. Electrical and magnetic effects are coupled via the Maxwell equations. Based on energy methods in mechanics, we establish a thermodynamically consistent material model and derive the weak and strong form of the microscale boundary value problem. 

In order to solve the macroscale problem, we create an RVE and apply the FEM to solve the problem numerically. For the time integration of the FEM, we use a JWH-$\alpha$-scheme \cite{kadapagenalpha}. The numerical simulations on the microscale show mesh independence and quadratic convergence. Additionally, we show that the effective Young's modulus of the RVE depends strongly on the volume fraction of the different phases. Here, we find a reduction by $43\%$ for the degenerated bone ($\rho_{\mathrm{b}}=5.3\%$) compared to the healthy bone ($\rho_{\mathrm{b}}=29.5\%$), achieving similar results than in \cite{Ilic2010}.

For the multiscale calculations, we use $\text{FE}^2$ and apply periodic boundary conditions and volume averaging for the transition between the scales. We apply a time-dependent displacement boundary condition. The macroscopic cylinder model again shows quadratic convergence. To compare different stages of osteoporosis with a healthy bone, we create six different RVEs with different volume fractions of cortical bone phase and run the simulations for all RVEs. The simulations show a strong reduction of all quantities with decreasing volume fraction of cortical bone phase. The differences between the healthy bone RVE ($\rho_{\mathrm{b}}=29.5\%$) and a slightly degenerated bone ($\rho_{\mathrm{b}}=24.2\%$) are very small, while the differences in the later stages of the illness, ($\rho_{\mathrm{b}}=10.4\%$ compared to $\rho_{\mathrm{b}}=5.3\%$), increase drastically. To examine the interaction between the time derivative of the electric displacement field and the electric current density in the Maxwell equation, we perform a parameter study regarding the electric conductivity parameter $\kappa_1$. Here, the results show a slight decrease of the magnetic field strength with increasing $\kappa_1$, suggesting that both parts evolve in the same way, but in opposite directions.

As a final example, we apply our model to a true to scale model of a human femur bone. Here, the results show again a similar behavior for all quantities. Between the two RVEs with the highest volume fraction of cortical bone phase, nearly no reduction of the magnetic field strength can be observed. With decreasing $\rho_{\mathrm{b}}$, the differences grow increasingly larger. Compared to the healthy bone ($\rho_{\mathrm{b}}=29.5\%$), the bone with late stage osteoporosis ($\rho_{\mathrm{b}}=5.3\%$) shows a drastic reduction of the magnetic field strength by nearly two thirds. These results show, in which order of magnitude differences between healthy and degenerated bones can be expected, when performing experimental research and sonography for the purpose of early detection of osteoporosis.

For future research, the comparison of experimental with simulations results could provide further insights. It is of great importance to obtain accurate material parameters from the experiments, which then can be used for the simulations. Additionally, we aim to solve the inverse problem by using an Artificial Neural Network to predict simulation outputs. Here, the network should recover the distribution of cortical bone phase in the macroscopic model from the magnetic field data, thus diagnosing either a healthy or a degenerated bone. Another important aspect is to investigate the microscale behavior for RVEs which differ in size and structure of the phases. Finally, wave propagation in cancellous bone will be investigated in more detail.
%%%%%%%%%%%%%%%%%%%%
%Bibliography
\bibliographystyle{unsrt}
\bibliography{references}
%%%%%%%%%%%%%%%%%%%%
\appendix
\section{Calculation of the strong form for the micro problem}
To calculate the strong form of the problem, we use the reduced weak form:
\begin{align}
\int\limits_{\Omega} ( \bfsigma \cdot \delta \bfvarepsilon - \bfD \cdot \delta \bfE + \bfH \cdot \delta \bfB - \bfJ \cdot \delta \bfA & \nonumber \\
 + \gamma (\nabla \cdot \bfA) \cdot \delta (\nabla \cdot \bfA) ) \dd V - \delta W_{\mathrm{ext}} &= 0 \quad \forall \delta \bfu \text{,} \; \delta \varphi \text{,} \; \delta \bfA \; \text{.} \label{eqwfreda}
\end{align} % weak form reduced
Now, we apply partial integration to each term followed by the use of a surface-volume integral rule:
\begin{align}
\int\limits_{\Omega} \bfsigma \cdot \delta \bfvarepsilon \; \dd V &= \int\limits_{\Omega} \nabla \cdot (\bfsigma \cdot \delta \bfu) \; \dd V - \int\limits_{\Omega} (\nabla \cdot \bfsigma) \cdot \delta \bfu \; \dd V \nonumber \\
& = \int\limits_{\partial \Omega} \bfsigma \cdot \bfn \cdot \delta \bfu \; \dd A - \int\limits_{\Omega} (\nabla \cdot \bfsigma) \cdot \delta \bfu \; \dd V \text{,}
\end{align} % part Int sigma
\begin{align}
\int\limits_{t_0}^{t_1} \int\limits_{\Omega} -\bfD \cdot \delta \bfE \dd V \dd t &= \int\limits_{t_0}^{t_1} \int\limits_{\Omega} (\bfD \cdot \delta(\nabla \varphi) + \bfD \cdot \delta \dot{\bfA}) \; \dd V \dd t \nonumber \\
= \int\limits_{t_0}^{t_1} \int\limits_{\Omega} \nabla \cdot (\bfD \delta \varphi) \dd V \dd t &- \int\limits_{t_0}^{t_1} \int\limits_{\Omega} (\nabla \cdot \bfD) \delta \varphi \dd V \dd t \nonumber \\
- \int\limits_{t_0}^{t_1} \int\limits_{\Omega} \dot{\bfD} \cdot \delta \bfA \; \dd V \dd t &+ \underbrace{\int\limits_{\Omega} [\bfD \cdot \delta \bfA]_{t_0}^{t_1} \; \dd V}_{=0} \nonumber \\
= \int\limits_{t_0}^{t_1} \int\limits_{\partial \Omega} \bfD \cdot \bfn \delta \varphi \dd A \dd t &- \int\limits_{t_0}^{t_1} \int\limits_{\Omega} (\nabla \cdot \bfD) \delta \varphi \dd V \dd t \nonumber \\
&- \int\limits_{t_0}^{t_1} \int\limits_{\Omega} \dot{\bfD} \cdot \delta \bfA \; \dd V \dd t \text{,} \label{eqpid}
\end{align} % part Int D
\begin{align}
\int\limits_{\Omega} \bfH \cdot \delta \bfB \; \dd V &= \int\limits_{\Omega} \nabla \cdot (\delta \bfA \times \bfH) \; \dd V + \int\limits_{\Omega} \nabla \times \bfH \cdot \delta \bfA \; \dd V \nonumber \\
&= \int\limits_{\partial \Omega} \bfH \times \bfn \cdot \delta \bfA \; \dd A + \int\limits_{\Omega} \nabla \times \bfH \cdot \delta \bfA \; \dd V \text{,} \label{eqpih}
\end{align} % part Int H
\begin{align}
\int\limits_{\Omega} \gamma (\nabla \cdot \bfA) \cdot \delta(\nabla \cdot \delta \bfA) \; \dd V & \nonumber \\
=\int\limits_{\Omega} \gamma \nabla \cdot ((\nabla \cdot \bfA) \cdot \delta \bfA) \; \dd V &- \int\limits_{\Omega} \gamma \nabla (\nabla \cdot \bfA) \delta \bfA \; \dd V \nonumber \\
= \int\limits_{\partial \Omega} \gamma (\nabla \cdot \bfA) \cdot \bfn \cdot \delta \bfA \; \dd A &- \int\limits_{\Omega} \gamma \nabla (\nabla \cdot \bfA) \delta \bfA \; \dd V \text{.}
\end{align} % part Int gauge
Here, $\bfn$ is the normal vector pointing outwards. It should be noted that the term $\bfD \cdot \delta \bfA$ in Eq. \eqref{eqpid} vanishes, as the test function $\delta \bfA$ does not change between the time $t_0$ and $t_1$. For the magnetic energy variation Eq. \eqref{eqpih}, the sign is different compared to the other equations because of the cross product rule. Additionally, the resulting triple product in the surface integral allows cyclic permutation without changing the result. Inserting these identities in the reduced weak form Eq. \eqref{eqwfreda} yields
\begin{align}
\int\limits_{\Omega} \bigl(-\nabla \cdot \bfsigma - \bff \bigr) \cdot \delta \bfu \dd V &+ \int\limits_{\partial \Omega} \bigl(\bfsigma \cdot \bfn - \bft \bigr) \cdot \delta \bfu \dd A \nonumber \\
+ \int\limits_{\Omega} \bigl(-\nabla \cdot \bfD + q_{\mathrm{v}}\bigr) \cdot \delta \varphi \dd V &+ \int\limits_{\partial \Omega} \bigl(\bfD \cdot \bfn  + q_{\mathrm{s}}\bigr) \cdot \delta \varphi \dd A \nonumber \\ 
+ \int\limits_{\Omega} \bigl(\nabla \times \bfH -\dot{\bfD} - \bfJ &- \gamma \nabla (\nabla \cdot \bfA) - \bfj_{\mathrm{v}} \bigr) \cdot \delta \bfA \dd V \nonumber \\
+ \int\limits_{\partial \Omega} \bigl( \bfH \times \bfn - \bfj_{\mathrm{s}} + \gamma (\nabla \cdot \bfA) \cdot \bfn \bigr) & \cdot \delta \bfA \dd A = 0 \; \forall \delta \bfu \text{,} \delta \varphi \text{,} \delta \bfA \text{.}
\end{align}
Then, we calculate the strong form by using the arbitrariness of the test functions and by splitting the volume and surface of the area, resulting in Eq. \eqref{eqstrong} and the matching conditions Eq. \eqref{eqmc}.
%%%%%%%%%%%%%%%%%%%%%%%%%
\section{Derivation of the matrix form for the FEM}
Inserting the FEM ansatz into the reduced weak form of the micro problem Eq. \eqref{eqwfreda} yields
\begin{align}
\int\limits_{\Omega_{\mathrm{b}}} \bigl( (\dsC_\mathrm{b} \bfB_\mathrm{u} \boldface{\hat{u}} - \bfe_\mathrm{b}^T [- \bfB_\mathrm{grad} \boldface{\hat{\varphi}} - \bfN_\mathrm{A} \boldface{\dot{\hat{A}}}]) \bfB_\mathrm{u} \delta \boldface{\hat{u}} \nonumber \\
- [\bfe_\mathrm{b} \bfB_\mathrm{u} \boldface{\hat{u}} + \bfxi_\mathrm{b} (- \bfB_\mathrm{grad} \boldface{\hat{\varphi}} - \bfN_\mathrm{A} \boldface{\dot{\hat{A}}})] [- \bfB_\mathrm{grad} \delta \boldface{\hat{\varphi}} - \bfN_\mathrm{A} \delta \boldface{\dot{\hat{A}}}] \nonumber \\
+ \mu_{\mathrm{b}}^{-1} \bfB_\mathrm{curl} \boldface{\hat{A}} \bfB_\mathrm{curl} \delta \boldface{\hat{A}} + \gamma \bfB_\mathrm{div} \bfB^T_\mathrm{div} \delta \boldface{\hat{A}} \bigr) \dd V \nonumber \\
+ \int\limits_{\Omega_{\mathrm{m}}} \bigl( \bfsigma_\mathrm{m} \bfB_\mathrm{u} \delta \boldface{\hat{u}} - [\bfxi_\mathrm{m} (- \bfB_\mathrm{grad} \boldface{\hat{\varphi}} - \bfN_\mathrm{A} \boldface{\dot{\hat{A}}})] [- \bfB_\mathrm{grad} \delta \boldface{\hat{\varphi}} \nonumber \\ 
- \bfN_\mathrm{A} \delta \boldface{\dot{\hat{A}}}] + \mu_{\mathrm{m}}^{-1} \bfB_\mathrm{curl} \boldface{\hat{A}} \bfB_\mathrm{curl} \delta \boldface{\hat{A}} + \gamma \bfB_\mathrm{div} \bfB^T_\mathrm{div} \delta \boldface{\hat{A}} \nonumber \\
- \bfkappa_\mathrm{m} [- \bfB_\mathrm{grad} \boldface{\hat{\varphi}} - \bfN_\mathrm{A} \boldface{\dot{\hat{A}}}] \bfN_\mathrm{A} \delta \boldface{\hat{A}} \bigr)\dd V \nonumber \\
- \biggl( \int\limits_{\Omega} \bfN_\mathrm{u}^T \bff \dd V + \int\limits_{\partial \Omega} \bfN_\mathrm{u}^T \bft \dd A \biggr) \delta \boldface{\hat{u}} \nonumber \\
+ \biggl( \int\limits_{\Omega} \bfN_{\varphi}^T q_\mathrm{v} \dd V + \int\limits_{\partial \Omega} \bfN_{\varphi}^T q_\mathrm{s} \dd A \biggr) \delta \boldface{\hat{\varphi}} \nonumber \\
- \biggl( \int\limits_{\Omega} \bfN_{A}^T \; \bfj_\mathrm{v} \dd V + \int\limits_{\partial \Omega} \bfN_{A}^T \; \bfj_\mathrm{s} \dd A \biggr) \delta \boldface{\hat{A}} = 0 \; \forall \delta \boldface{\hat{u}} \text{,} \delta \boldface{\hat{\varphi}} \text{,}  \delta \boldface{\hat{A}} \text{.} \label{eqfem}
\end{align}
Here, we use partial integration as follows:
\begin{align}
\int\limits_{t_0}^{t_1} \int\limits_{\Omega} \bfN_{\mathrm{A}}^T \bfe_{\mathrm{b}} \bfB_{\mathrm{u}} \; \boldface{\hat{u}} \; \delta \boldface{\dot{\hat{A}}} \; \dd V \dd t & \nonumber \\
= \underbrace{\int\limits_{\Omega} [\bfN_{\mathrm{A}}^T \bfe_{\mathrm{b}} \bfB_{\mathrm{u}} \; \boldface{\hat{u}} \; \delta \boldface{\hat{A}}]_{t_0}^{t_1} \dd V}_{=0} &- \int\limits_{t_0}^{t_1} \int\limits_{\Omega} \bfN_{\mathrm{A}}^T \bfe_{\mathrm{b}} \bfB_{\mathrm{u}} \; \boldface{\dot{\hat{u}}} \; \delta \boldface{\hat{A}} \; \dd V \dd t \\
\int\limits_{t_0}^{t_1} \int\limits_{\Omega} \bfN_{\mathrm{A}}^T \bfxi \bfB_{\mathrm{grad}} \; \boldface{\hat{\varphi}} \; \delta \boldface{\dot{\hat{A}}} \; \dd V \dd t & \nonumber \\
= \underbrace{\int\limits_{\Omega} [\bfN_{\mathrm{A}}^T \bfxi \bfB_{\mathrm{grad}} \; \boldface{\hat{\varphi}} \; \delta \boldface{\hat{A}}]_{t_0}^{t_1} \dd V}_{=0} &- \int\limits_{t_0}^{t_1} \int\limits_{\Omega} \bfN_{\mathrm{A}}^T \bfxi \bfB_{\mathrm{grad}} \; \boldface{\dot{\hat{\varphi}}} \; \delta \boldface{\hat{A}} \; \dd V \dd t \\
\int\limits_{t_0}^{t_1} \int\limits_{\Omega} \bfN_{\mathrm{A}}^T \bfxi \bfN_{\mathrm{A}} \; \boldface{\dot{\hat{A}}} \; \delta \boldface{\dot{\hat{A}}} \; \dd V \dd t & \nonumber \\
= \underbrace{\int\limits_{\Omega} [\bfN_{\mathrm{A}}^T \bfxi \bfN_{\mathrm{A}} \; \boldface{\dot{\hat{A}}} \; \delta \boldface{\hat{A}}]_{t_0}^{t_1} \dd V}_{=0} &- \int\limits_{t_0}^{t_1} \int\limits_{\Omega} \bfN_{\mathrm{A}}^T \bfxi \bfN_{\mathrm{A}} \; \boldface{\ddot{\hat{A}}} \; \delta \boldface{\hat{A}} \; \dd V \dd t
\end{align} %part Int FEM
Again, the nodal test function $\delta \boldface{\hat{A}}$ does not change in time, so the corresponding terms vanish. Inserting this into Eq. \eqref{eqfem} yields
\begin{align}
\int\limits_{\Omega_{\mathrm{b}}} \bigl( (\dsC_\mathrm{b} \bfB_\mathrm{u} \boldface{\hat{u}} - \bfe_\mathrm{b}^T [- \bfB_\mathrm{grad} \boldface{\hat{\varphi}} - \bfN_\mathrm{A} \boldface{\dot{\hat{A}}}]) \bfB_\mathrm{u} \delta \boldface{\hat{u}} \nonumber \\ 
- [\bfe_\mathrm{b} \bfB_\mathrm{u} \boldface{\hat{u}} + \bfxi_\mathrm{b} (- \bfB_\mathrm{grad} \boldface{\hat{\varphi}} - \bfN_\mathrm{A} \boldface{\dot{\hat{A}}})] [- \bfB_\mathrm{grad} \delta \boldface{\hat{\varphi}}] \nonumber \\
+ [\bfe_\mathrm{b} \bfB_\mathrm{u} \boldface{\dot{\hat{u}}} + \bfxi_\mathrm{b} (- \bfB_\mathrm{grad} \boldface{\dot{\hat{\varphi}}} - \bfN_\mathrm{A} \boldface{\ddot{\hat{A}}})] [- \bfN_\mathrm{A} \delta \boldface{\hat{A}}] \nonumber \\
+ \mu_{\mathrm{b}}^{-1} \bfB_\mathrm{curl} \boldface{\hat{A}} \bfB_\mathrm{curl} \delta \boldface{\hat{A}} + \gamma \bfB_\mathrm{div} \bfB^T_\mathrm{div} \delta \boldface{\hat{A}} \bigr) \dd V  \nonumber \\
+ \int\limits_{\Omega_{\mathrm{m}}} \bigl(\bfB^T_\mathrm{u} \dsC_\mathrm{tang} \bfB_\mathrm{u} \delta \boldface{\hat{u}} 
- [\bfxi_\mathrm{m} (- \bfB_\mathrm{grad} \boldface{\hat{\varphi}} - \bfN_\mathrm{A} \boldface{\dot{\hat{A}}})] 
[- \bfB_\mathrm{grad} \delta \boldface{\hat{\varphi}}] \nonumber \\
+ [\bfxi_\mathrm{m} (- \bfB_\mathrm{grad} \boldface{\dot{\hat{\varphi}}} - \bfN_\mathrm{A} \boldface{\ddot{\hat{A}}})]  [- \bfN_\mathrm{A} \delta \boldface{\hat{A}}] + \mu_{\mathrm{m}}^{-1} \bfB_\mathrm{curl} \boldface{\hat{A}} \bfB_\mathrm{curl} \delta \boldface{\hat{A}} \nonumber \\
+ \gamma \bfB_\mathrm{div} \bfB^T_\mathrm{div} \delta \boldface{\hat{A}} - \bfkappa_\mathrm{m} [- \bfB_\mathrm{grad} \boldface{\hat{\varphi}} - \bfN_\mathrm{A} \boldface{\dot{\hat{A}}}] \bfN_\mathrm{A} \delta \boldface{\hat{A}} \bigr)\dd V \nonumber \\
- \biggl( \int\limits_{\Omega} \bfN_\mathrm{u}^T \bff \dd V + \int\limits_{\partial \Omega} \bfN_\mathrm{u}^T \bft \dd A \biggr) \delta \boldface{\hat{u}}  \nonumber \\
+ \biggl( \int\limits_{\Omega} \bfN_{\varphi}^T q_\mathrm{v} \dd V + \int\limits_{\partial \Omega} \bfN_{\varphi}^T q_\mathrm{s} \dd A \biggr) \delta \boldface{\hat{\varphi}}  \nonumber \\
- \biggl( \int\limits_{\Omega} \bfN_{A}^T \; \bfj_\mathrm{v} \dd V + \int\limits_{\partial \Omega} \bfN_{A}^T \; \bfj_\mathrm{s} \dd A \biggr) \delta \boldface{\hat{A}} = 0 \; \forall \delta \boldface{\hat{u}} \text{,} \delta \boldface{\hat{\varphi}} \text{,} \delta \boldface{\hat{A}} \text{.} 
\end{align}
This equation can be split by the variations. By using the arbitrariness of the test functions and introducing the generalized nodal forces as
\begin{align}
\begin{pmatrix}
\boldface{\hat{f}} \\ \boldface{\hat{q}} \\ \boldface{\hat{j}}
\end{pmatrix} := 
\begin{pmatrix} \int\limits_{\Omega} \bfN_\mathrm{u}^T \bff \dd V + \int\limits_{\partial \Omega} \bfN_\mathrm{u}^T \bft \dd A \\ 
- \int\limits_{\Omega} \bfN_{\varphi}^T q_\mathrm{v} \dd V - \int\limits_{\partial \Omega} \bfN_{\varphi}^T q_\mathrm{s} \dd A \\
\int\limits_{\Omega} \bfN_{A}^T \; \bfj_\mathrm{v} \dd V + \int\limits_{\partial \Omega} \bfN_{A}^T \; \bfj_\mathrm{s} \dd A 
\end{pmatrix} \; \text{,}
\end{align} % forces
the residual is then $\bfR = \begin{pmatrix}
	\bfR_\mathrm{u} & \bfR_\mathrm{\varphi} & \bfR_\mathrm{A}
\end{pmatrix}^T = $
\be
\begin{pmatrix}\boldface{\hat{f}} + \int\limits_{\Omega} -\bfB_\mathrm{u}^T \boldface{\hat{\sigma}} \dd V \\
\boldface{\hat{q}} + \int\limits_{\Omega} -\bfB_{\mathrm{grad}}^T \boldface{\hat{D}} \dd V \\
\boldface{\hat{j}} + \int\limits_{\Omega} ( -\bfB_{\mathrm{curl}}^T \boldface{\hat{H}} + \bfN_\mathrm{A}^T (\dot{\hat{\bfD}}+\hat{\bfJ}) - \gamma \bfB_{\mathrm{div}} \bfB^T_{\mathrm{div}} \boldface{\hat{A}} ) \dd V \end{pmatrix} 
\ee % residual
By writing this equation in matrix form, we recover the equation system Eq. \eqref{eqfemmat}.
\end{document}